\documentclass[reprint, superscriptaddress, amsmath,amssymb,aps,pre floatfix]{revtex4-2}

\usepackage{graphicx}% Include figure files
\usepackage{dcolumn}% Align table columns on decimal point
\usepackage{bm}% bold math
\usepackage{xcolor}
\usepackage{empheq}

\usepackage[normalem]{ulem}

\usepackage{subcaption}
\usepackage[english]{babel}
\usepackage{amsmath, amsfonts, amssymb, amstext, amscd, amsthm, makeidx, graphicx, hyperref, url, mathtools}
\usepackage{setspace}
\usepackage{physics}
\usepackage{color, soul}
\usepackage{float}

\begin{document}

\preprint{APS/123-QED}

\title{The Dynamics of Inducible Genetic Circuits}
\thanks{Corresponding authors:\\
Rebecca J. Rousseau (rroussea@caltech.edu) and\\
Rob Phillips (phillips@pboc.caltech.edu).}

\author{Zitao Yang}
\thanks{These authors contributed equally to this work.}
\affiliation{Department of Physics, California Institute of Technology, Pasadena, CA 91125}
\author{Rebecca J. Rousseau}
\thanks{These authors contributed equally to this work.}
\affiliation{Department of Physics, California Institute of Technology, Pasadena, CA 91125}
\author{Sara D. Mahdavi}
\thanks{These authors contributed equally to this work.}
\affiliation{Division of Biology and Biological Engineering, California Institute of Technology, Pasadena, CA 91125}
\author{Hernan G. Garcia}
\affiliation{Biophysics Graduate Group, University of California, Berkeley, CA 904720}
\affiliation{Department of Physics, University of California, Berkeley, CA 94720}
\affiliation{Institute for Quantitative Biosciences-QB3, University of California, Berkeley, CA 94720}
\affiliation{Department of Molecular and Cell Biology, University of California, Berkeley, CA 94720}
\affiliation{Chan Zuckerberg Biohub–San Francisco, San Francisco, CA 94158}
\author{Rob Phillips}
\affiliation{Department of Physics, California Institute of Technology, Pasadena, CA 91125}
\affiliation{Division of Biology and Biological Engineering, California Institute of Technology, Pasadena, CA 91125}

\date{\today}% It is always \today, today,
             %  but any date may be explicitly specified

\begin{abstract}
Genes are connected in complex networks of interactions where often the product of one gene is a transcription factor that alters the expression of another. Many of these networks are based on a few fundamental motifs leading to switches and oscillators of various kinds. And yet, there is more to the story than which transcription factors control these various circuits. These transcription factors are often themselves under the control of effector molecules that bind them and alter their level of activity. Traditionally, much beautiful work has shown how to think about the stability of the different states achieved by these fundamental regulatory architectures by examining how parameters such as transcription rates, degradation rates and dissociation constants tune the circuit, giving rise to behavior such as bistability. However, such studies explore dynamics without asking how these quantities are altered in real time in living cells as opposed to at the fingertips of the synthetic biologist's pipette or on the computational biologist's computer screen.  In this paper, we make a departure from the conventional dynamical systems view of these regulatory motifs by using statistical mechanical models to focus on endogenous signaling knobs such as effector concentrations rather than on the convenient but more experimentally remote knobs such as dissociation constants, transcription rates and degradation rates that are often considered. We also contrast the traditional use of Hill functions to describe transcription factor binding with more detailed thermodynamic models. This approach provides insights into how biological parameters are tuned to control the stability of regulatory motifs in living cells, sometimes revealing quite a different picture than is found by using Hill functions and tuning circuit parameters by hand. 
\end{abstract}

%\tableofcontents

\maketitle

\section{\label{sec:intro}Introduction}

The first half of the twentieth century was a time in which many of the great mysteries of nineteenth century physics were resolved~\cite{Kelvin1901,Gamow1985,Segre2007c,Longair2020}.  Though perhaps less well known, the study of living organisms had enormous mysteries of its own including the molecular and cellular basis of the laws of heredity~\cite{Judson1996}. One key puzzle centered on a
phenomenon known at the time as ``enzymatic adaptation''~\cite{Novick1957,Barnett2004}.  Those words refer to the apparent induction of enzyme action as a result of changes in the metabolic or physiological state of cells, such as those that occur upon shifting from one carbon source to another~\cite{Barnett2004}.  In the nineteenth century, Louis Pasteur had noted that yeasts behave differently under different growth conditions.  Fr\'{e}d\'{e}ric Dienert followed up on those observations with great foresight by doing  experiments that quantitatively characterized the phenomenon~\cite{Barnett2004}.  Jacques Monod made these studies a fine art through the use of bacterial growth curves ``as a method for the study of bacterial physiology and biochemistry''~\cite{Monod1949} (see Figure 9 of Monod's paper for a compelling example of the induction phenomenon). 

\begin{figure*}
    \centering
    \includegraphics[width=0.75\linewidth]{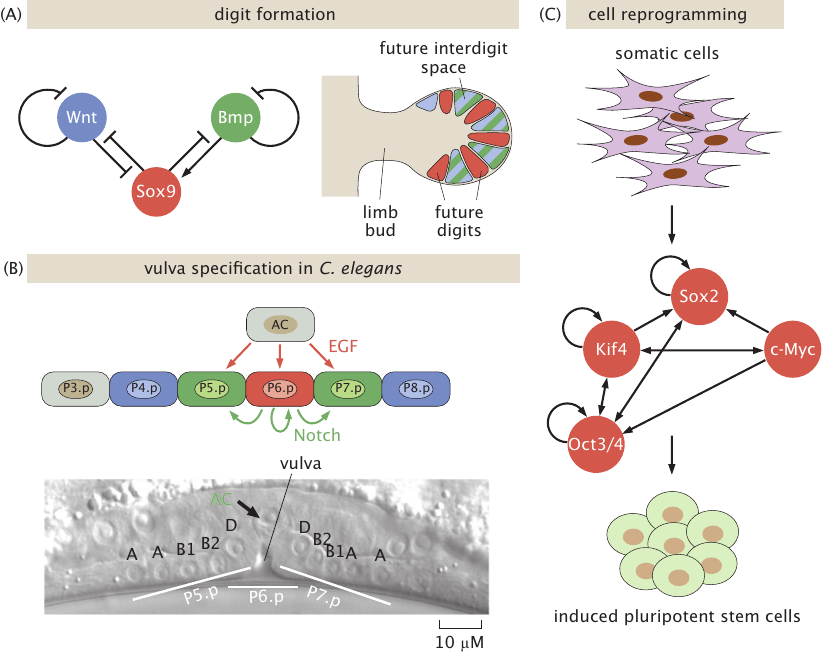}
    \caption{Gallery of examples of  regulatory circuits participating in the genetic decisions of animal development. (A) A three-node network thought to be relevant to the control of digit formation, adapted from \cite{zuniga2014turing}. (B) An example involving vulval development in {\it C. elegans}, where epidermal growth factor (EGF) and Notch induce cells toward one of three possible fates, adapted from \cite{corson2012geometry} and \cite{schindlermorphogenesis}. (C) Transcription factors compete and maintain cell pluripotency unless sufficiently induced to reprogram a cell to a differentiated fate, adapted from \cite{liu2008yamanaka}.}
    \label{fig:ExampleCircuits}
\end{figure*}

As a result of studies like these, in the early 1960s  Jacob and Monod shook the world of biology by showing that there are genes whose job it is to control other genes~\cite{Judson1996,Echols2001}, culminating in their repressor-operator model which showed how proteins could bind to DNA and repress the expression of nearby genes~\cite{Jacob1961,Muller-Hill1996,Judson1996,Echols2001}. Their original work  was extended and amplified through the discovery of architectures that were mediated not only by repression, but by activation as well~\cite{Englesberg1965}, and even by combinations of activators and repressors~\cite{Schleif1993Book}. In the late 1960s, the vision was considerably broadened through the generalization of these ideas from their first context in bacteria to the much broader set of regulatory problems associated with animal development such as those schematized in Fig.~\ref{fig:ExampleCircuits}~\cite{Britten1969}. The study of the lysis-lysogeny decision in bacteriophage lambda became a paradigm for the genetic switch~\cite{Ptashne2004,Ptashne2002}, and in the time since then those ideas have been generalized, realized, and exploited across biology. 

The repressor-operator model of Jacob and Monod provided not only a successful conceptual vision for gene expression writ large, but also served as the basis of mathematical models of transcription based upon the precepts of statistical mechanics~\cite{Ackers1982,Shea1985,Buchler2003a}. These models provided a quantitative description of a variety of different regulatory contexts in which the strengths of binding sites, the repressor copy number and  DNA loop length were altered, illustrating how genetic circuits could be tuned directly and quantitatively~\cite{Oehler1994,Muller1996,Vilar2003a,Vilar2003b,Kuhlman2007,saiz2008ab, vilar2011control, vilarSaiz2013}.  Interestingly, these pioneering studies became a jumping off point for the construction of a number of synthetic variants that when combined with fluorescent reporters made it possible to watch synthetic switches and oscillators in real time in single cells~\cite{Gardner2000,Elowitz2000,ozbudak2004multistability}.

In addition to the seminal discoveries of the existence of gene circuits themselves, a parallel set of discoveries unfolded which added a second layer of regulatory control to the original repressor-operator model and its 
subsequent generalizations and elaborations. Specifically, the mystery of induction required another insight into biological feedback and control.
 Enzymatic adaptation, the idea that somehow enzymes that were latent would become active in the presence of the right substrate~\cite{Novick1957}, led to the discovery of allostery, a concept that Monod himself referred to as the ``second secret of life''~\cite{Ullmann2011}.
 In the context of gene regulation, this idea implies that transcription factors themselves are subject to control through the binding of effector molecules that alter their activity~\cite{Gerhart1962, Monod1963,Monod1965,Martins2011,Marzen2013,Changeux2013,Gerhart2014,Phillips2020,o1980equilibrium, DalyMatthews1986exptind}. Writ large,
these insights now fall under the general heading of allosteric transitions, a phenomenon in which proteins of all types undergo conformational changes that alter their activity.  This idea applies broadly to ion channels, enzymes, the respiratory protein hemoglobin, membrane receptors mediating chemotaxis and quorum sensing, and of course, to the main subject of our paper, transcription factors~\cite{Phillips2020}.

\begin{figure*}
    \centering
    \includegraphics[width=\linewidth]{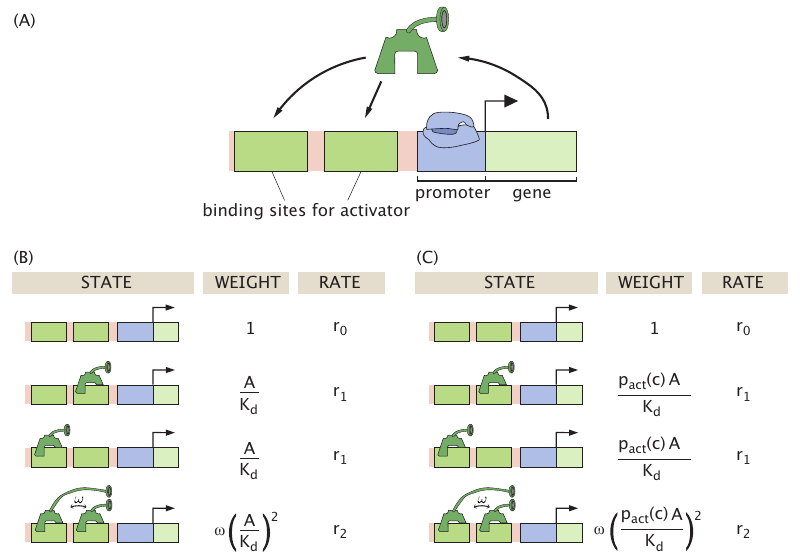}
    \caption{The auto-activation regulatory circuit.  (A) Schematic of the operation of the circuit. Polymerase binding at the promoter (blue) transcribes the gene (encoded in the light green region), producing a protein that can activate its own expression at a sufficient concentration. In our model, an activator can bind at one of two possible sites to enhance gene transcription.  (B) Thermodynamic states, weights, and rates for the circuit in the traditional model without induction. The parameter $\omega$ denotes the cooperative strength of two activators binding. (C) Thermodynamic states, weights and rates for the case in which the effector tunes the fraction of active activators. Note that in both of these cases the parameters $K_d$, $\omega$, $r_0$, $r_1$ and $r_2$ are {\it effective} parameters that have hidden dependence upon the number of polymerases and the strength with which it binds the promoter.  The explicit definitions of these effective parameters are worked out in Appendix~\ref{Section:CoarseGraining}.}
    \label{fig:AutoActivation}
\end{figure*}

The mathematical analysis of genetic circuits is its own fascinating enterprise, using the tools of dynamical systems to explore the stability of switches and oscillations~\cite{Goodwin1963, Cherry2000,Alon2020, Covert2015,Ferrell2022}. The idea for describing some circuit involving $n$ different proteins is to write dynamical equations of the form
\begin{equation}
    \frac{d [TF_i]}{dt}= f_i(\{[TF_j]\}),
\end{equation}
where there is one such equation for each transcription factor (for which we will often use the shorthand notation). The function on the right side acknowledges that the dynamics of the $i^{th}$ TF can depend upon the concentrations of all the others, represented here by the notation $\{TF_j\}$ signifying ``the set of all $n$ TFs.'' Perhaps the simplest such example we will discuss as our first case study is the auto-activation switch as shown in Fig.~\ref{fig:AutoActivation} and described by an equation of the form

\begin{equation}
{dA \over dt}= -\gamma A+ {r_0 + r_1 2{A \over K_d} + r_2 \omega ({A \over K_d})^2 \over 1+ 2{A \over K_d} + \omega ({A \over K_d})^2},
\label{eqn:AutoActivationThermoModel}
\end{equation}
where $A$ is the number of activators, $K_{d}$ is the dissociation constant for $A$ in its interaction with DNA, and $\omega$ is the cooperativity between two activators bound to the gene promoter along the DNA. The first term on the right captures the degradation of activator at rate $\gamma$, and the second term characterizes protein production with a basal level of production $r_{0}$ and a saturating level $r_{2}$.

The production rate in Eqn.~\ref{eqn:AutoActivationThermoModel}, and throughout this work, is modeled using a thermodynamic framework relating promoter occupancy to output~\cite{TLHill1977,TLHill1989,Vilar2003b,Buchler2003a,bintu2005transcriptional,bintu2005transcriptional2,Sherman2012}.
We note that often instead of adopting the full thermodynamic model to treat promoter occupancy,  it is convenient to use Hill functions as an approximation to describe the probability of promoter binding~\cite{Cherry2000,Alon2020, Covert2015,Ferrell2022, bottani2017hill, rogers2015synthetic, tkavcik2011information, tkavcik2012optimizing,walczak2010optimizing}. The auto-activation example will serve as our first foray into the problem of induction of genetic circuits by effector molecules as well as an opportunity to bring some critical scrutiny to the use of Hill functions to describe the physics of occupancy.

\begin{figure*}
    \centering
    \includegraphics[width=\linewidth]{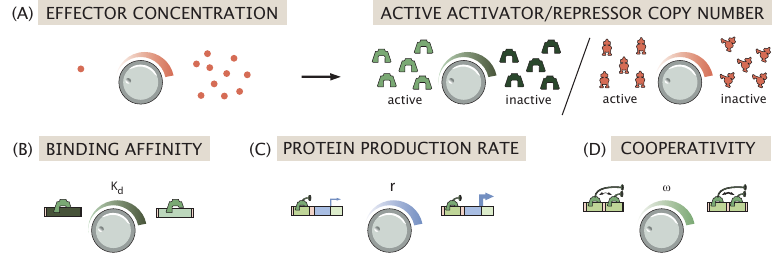}
    \caption{Tuning genetic circuits.  The schematic shows different knobs which are available to the cell, the theorist and the experimentalist, namely (A) effector concentration (and by consequence the number of active activator or repressor molecules), (B) binding affinity $K_{d}$, (C) protein production rate $r$, and (D) cooperativity $\omega$.}
    \label{fig:TuningInducible}
\end{figure*}

Typically, the exploration of the stability behavior of these circuits is based upon varying theoretically accessible parameters such as dissociation constants ($K_d$), transcription rates ($r_i$) and degradation rates ($\gamma$) as featured in Eqn.~\ref{eqn:AutoActivationThermoModel} and  in Fig.~\ref{fig:TuningInducible}(B-D), without reference to how such parameters are themselves controlled by living cells~\cite{Phillips2020}. In fact, often the underlying response is dictated by the presence and absence of  experimentally accessible effector molecules that alter the balance between inactive and active forms of key regulatory molecules such as transcription factors, as shown in Fig.~\ref{fig:TuningInducible}(A). Thus, while all of the tunable parameters in Fig.~\ref{fig:TuningInducible}   make it possible to systematically tune the level of gene expression, some of these parameters are more conveniently accessible to the experimentalist and to the cell itself as it rapidly tunes its behavior in response to stimuli.  Our goal is to use explicit statistical mechanical models of the induction phenomenon to explore the behavior of genetic circuits as a function of the presence or absence of effectors. 

In the next few sections, we work our way through an array of increasingly sophisticated gene regulatory circuits and leverage the statistical mechanical framework to uncover how the presence of effectors dictates complex gene expression dynamics. We envision that the predictions and systematic analysis stemming from our work will make it possible to better understand how cells exploit these genetic circuits to regulate their decision making processes, as well as enable the predictive design of synthetic circuits with prescribed functions in response to input effector dynamics.

\section{The Statistical Mechanics of Induction} \label{sec:statmechind}

We now undertake a systematic analysis of a number of different regulatory circuits from an allosteric perspective, building upon earlier work in which biological parameters are tuned by hand rather than by effectors~\cite{Gardner2000,Cherry2000,Alon2020,Phillips2012,Sokolik2015,ozbudak2004multistability}. Although some previous studies also vary inducer concentration, they typically either address different regulatory contexts than ours, map inducer concentrations to protein activity without modeling the allosteric transition explicitly, or consider different types of dynamical systems analyses than we do~\cite{yagil1971relation,Novick1957,Jacob1961,ozbudak2004multistability,saiz2008ab,vilar2011control,vilar2013systems, walczak2010optimizing, landman2017self, michel2010transcription, daber2011thermodynamic, daber2009one, rogers2015synthetic}.

In particular, we argue that often the number of active transcription factors $TF_{\text{act}}$ is given by
\begin{equation}
    TF_{\text{act}}=p_{\text{act}}(c)TF_{\text{tot}},
\end{equation}
where $TF_{\text{tot}}$ is the total number of transcription factors and $p_{\text{act}}(c)$ is the probability that the transcription factor is active as a function of the effector concentration $c$. For example, effector binding can render a protein inactive, decreasing $p_\text{act}$ as effector concentration $c$ increases. We will show that the fraction of transcription factors that are active can be given by the Monod-Wyman-Changeux (MWC) model, which can be used to compute $p_{\text{act}}(c)$ using statistical mechanics~\cite{o1980equilibrium,Gerhart1962, Monod1963,Monod1965,Martins2011,Marzen2013,Changeux2013,Gerhart2014,Phillips2020}. Note that while we invoke the MWC model to describe allostery, we could just as well use the KNF model or even the phenomenological Hill functions to capture the role of the effector~\cite{Phillips2020,o1980equilibrium}. The KNF model, for example, would describe effector binding as sequential and local. In this case, transcription factor structure would change at the subunit level, rather than through a single coordinated change in the protein's quartenary conformation as described by the MWC interpretation \cite{Phillips2012, einav2018theoretical}.

\begin{figure*}
    \centering
    \includegraphics[width=0.5\linewidth]{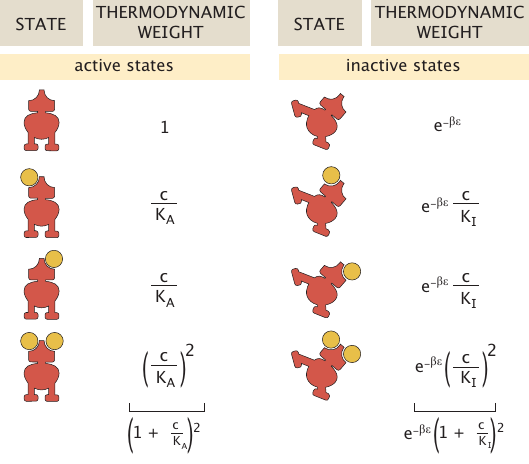}
    \caption{States and weights for an allosteric transcription factor with an effector that can bind at two sites on
    the protein. Effectors can bind to both the active and inactive forms of the transcription factor with different dissociation constants $K_A$ and $K_I$, which determine whether the effector stabilizes the protein more strongly in its active or inactive configuration. The sum of the thermodynamic weights for the active and inactive conformations are shown at the bottom.}
    \label{fig:MWCStatesWeights}
\end{figure*}

In the MWC model, we consider an inactive state and an active state of the transcription factor with energy difference $\varepsilon = \varepsilon_i - \varepsilon_a$. In this setting, an effector can bind to a transcription factor in both its active and inactive forms with different dissociation coefficients. The values of these constants then determine whether increasing effector concentration stabilizes the active or inactive form of the transcription factor. The states and weights for such an allosteric transcription factor with two binding sites is shown in Fig.~\ref{fig:MWCStatesWeights}. Appealing to these states and weights, the probability of a transcription factor being active is then of the form
\begin{align}
    p_\text{act}(c) = \frac{(1 + \frac{c}{K_A})^2}{(1 + \frac{c}{K_A})^2 + e^{-\beta \varepsilon}(1 + \frac{c}{K_I})^2},
    \label{eqn:MWC2site}
\end{align} 
where $c$ is the concentration of effector molecules. Here we define $\beta = 1/k_BT$, and $K_A$ and $K_I$ as the dissociation constants for the transcription factor in its active and inactive states, respectively.

\begin{figure}
    \centering
    \includegraphics[width=\linewidth]{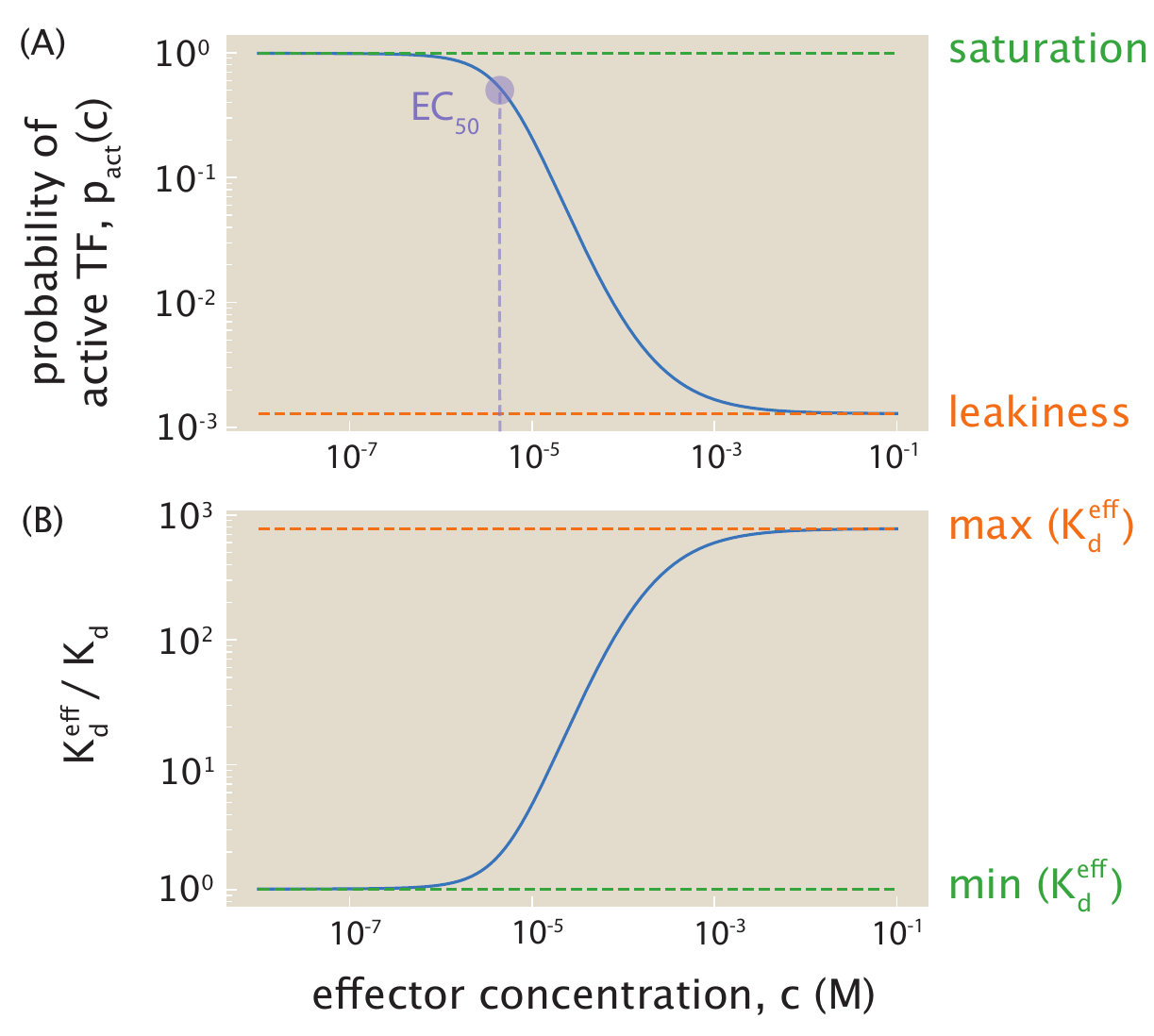}
    \caption{Activity of a transcription factor as a function of effector concentration $c$. Parameters used throughout the paper, unless otherwise stated, are: $K_A = 140 \, \mu M$, $K_I = 530 \, nM$, $\varepsilon = 4.5 \, k_BT$.
    (A) Probability of active transcription factor as a function of effector concentration, defined by Eqn.~\ref{eqn:MWC2site}. The half maximal effective concentration EC$_{50}$, defined as the effector concentration $c^*$ such that $p_{\text{act}}(c^*) = (p_{\text{act}}^{max} + p_{\text{act}}^{min})/2$, is plotted in purple. (B) The effective dissociation constant $K_d^{\text{eff}}$ (dimensionless with respect to $K_{d}$) as a function of effector concentration. Saturation of $p_\text{act}(c)$ corresponds to minimal $K_d^{\text{eff}}$, and leakiness of $p_\text{act}(c)$ corresponds to maximal $K_d^{\text{eff}}$.}
    \label{fig:pact}
\end{figure}

An alternative way of thinking about this approach is to express an effective dissociation constant $K_d^{\text{eff}}$ between transcription factors and DNA as 
\begin{equation}
K_d^{\text{eff}}=K_d/p_\text{act}(c), 
\label{eqn:EffectiveKd}
\end{equation}
where $K_d$ is the fixed physical dissociation constant and $p_\text{act}(c)$ modulates the activity of that transcription factor in an effector-concentration-dependent way. For a transcription factor with two effector binding sites, we can write the probability of being in the active state in the form given by Eqn.~\ref{eqn:MWC2site}. The activity of the transcription factor as a function of effector concentration is shown in Fig.~\ref{fig:pact}(A).  This input-output function has the typical sigmoidal behavior of $p_\text{act}(c)$. It is worth noting that here we show the behavior of a transcription factor for which the effector renders the proteins inactive. By tuning the relative values of the active and inactive dissociation constants, however, we can also generate situations in which the activity of the transcription factor increases with effector concentration.

\begin{figure}
    \centering
    \includegraphics[width=\linewidth]{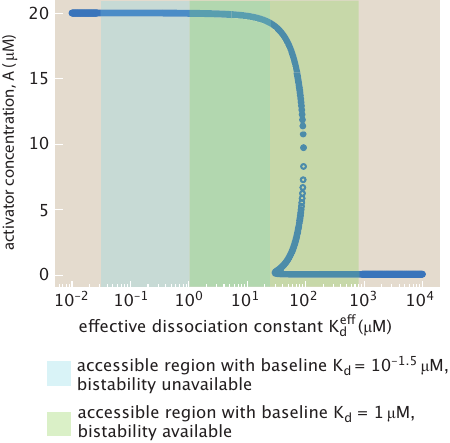}
    \caption{Allosteric tuning restricts the range of accessible $K_d^{\text{eff}}$. The dark blue curve shows the steady-state values of activator concentration $A$, with filled circles indicating stable fixed points and open circles indicating unstable ones. The light blue and light green shaded regions represent two example ranges of accessible $K_d^{\text{eff}} = K_d / p_{\text{act}}(c)$ values, obtained by varying the effector concentration $c$ for two proteins with different DNA binding affinities (and thus different $K_d$ values). For the blue region, $K_d =3.2 \times 10^{-2}$ $\mu$M $=10^{-1.5}$ $\mu$M; for the green region, $K_d = 1$ $\mu$M. The parameters for the auto-activation stability curve are: $\gamma = 1/$min, $r_0 = 0.1$ $\mu$M$/$min, $r_1 = 1$ $\mu$M$/$min, $r_2 = 20$ $\mu$M$/$min, $\omega = 100$.}    \label{fig:ClassicalAutoActivation}
\end{figure}

Writing $K_d^{\text{eff}} = K_d/p_\text{act}(c)$ in gene regulatory network motifs is useful and informative in many ways. First, traditionally, theoretical models of the behavior of network motifs are studied by tuning $K_d^{\text{eff}}$, while experiments generally tune the effector concentration $c$. Incorporating $p_\text{act}(c)$ bridges theory and experiments by providing experimentally accessible ``knobs'' to control the behavior of the genetic circuit of interest. Second, since $p_\text{act}(c)$ is highly nonlinear in $c$, other model variables might react differently to varying $c$ than to varying $K_d^{\text{eff}}$. Third, $p_\text{act}(c)$ constrains the range of accessible $K_d^{\text{eff}}$ values. 
When discussing input-output curves, there are a variety of summary parameters that help us understand their character qualitatively. For example, in the case where effector binding renders the protein inactive, the leakiness is the amount of activity at saturating concentrations  of effector, namely, $p_\text{act}(\infty)$. Similarly, the maximal activity known as the saturation occurs in the zero-effector limit, namely, $p_\text{act}(0)$. These important summary variables  can be calculated as 
\begin{equation}
    p_{\text{act}}^{\max} = \lim_{c \to 0} p_{\text{act}} = \frac{1}{1 + e^{-\beta \varepsilon}}\label{eq:pact_limit1}
\end{equation}
and
\begin{equation}
    p_{\text{act}}^{\min} = \lim_{c \to \infty} p_{\text{act}} = \frac{1}{1+ e^{-\beta\varepsilon}\bar{K}_{c}^2},
    \label{eq:pact_limit2}
\end{equation}
where $\bar{K}_c=K_A/K_I$. With the parameters used in Fig.~\ref{fig:pact}(A), $p_\text{act}^{\min}$ and $p_\text{act}^{ \max}$ are separated by about three orders of magnitude, and thus $K_d^{\text{eff}}$ is also only tunable across three orders of magnitude, as shown in Fig.~\ref{fig:pact}(B). The restricted $K_d^{\text{eff}}$ range can have important consequences. For example, consider a bistable system with a stability curve such as that shown in Fig.~\ref{fig:ClassicalAutoActivation}, where the $x$-axis is $K_d^{\text{eff}}$ and the $y$-axis tracks the steady state concentration of some protein $A$. Tuning the parameter $c$ imposes constraints on the range of achievable $K_d^{\text{eff}}$ for a given $K_d$. The light blue range shown does not intersect the bistable regime, while the light green range does, illustrating that bistability may not be fully accessible due to the functional dependence of $K_d^{\text{eff}}$ on $p_{\text{act}}(c)$. This restriction on $K_d$ follows a tradition of relating thermodynamic binding interactions to experimentally measured biochemical parameters~\cite{saiz2012physics}.

\begin{figure*}
    \centering
    \includegraphics{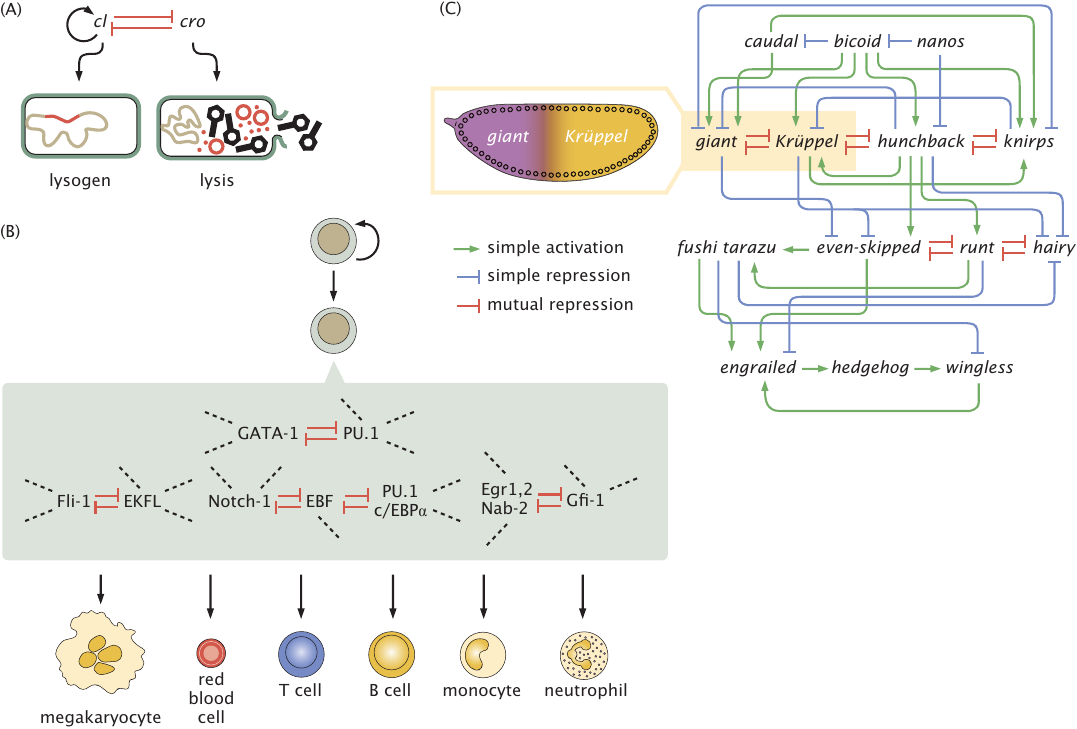}

\caption{Mutual repression is ubiquitous in cellular decision making.
(A) The bacteriophage lambda switch that mediates the phage decision to become a lysogen in the bacterium or engage in lysis through mutual repression of {\it cI} and {\it cro} (and some auto-activation of {\it cI} as well). (B)  In hematopoietic development,
mutual repression between different genes have been suggested to ensure the switch-like adoption of
alternate cellular fates. Diagram adapted from \cite{laslo2008gene} and \cite{chute2010minireview}.
(C) Mutual repression in the
early fruit fly gene network has been associated with the emergence of discrete domains of gene expression. Diagram adapted from \cite{carroll2001dna} and \cite{jaeger2011gap}. 
\label{fig:MutualRepressionExamples}}
\end{figure*}

Our fundamental goal is to reconsider the classic stability analysis for a broad array of different regulatory architectures in light of the MWC model for transcription factor activity described above. For example, as seen in Fig.~\ref{fig:ClassicalAutoActivation}, tuning the $K_d$ for the simple auto-activation circuit yields two stable fixed points.  In the first section, we examine this circuit by modulating the concentration of effector molecules, which tunes the concentration of active transcription factors. From this simple genetic circuit, we then turn to the
ubiquitous mutual repression switch, well known not only as a key part of the repertoire of natural genetic circuits as shown in 
Fig.~\ref{fig:MutualRepressionExamples}, but also as one of the
classic examples of synthetic biology~\cite{Gardner2000}. Both auto-activation and mutual repression can exhibit bistable dynamics~\cite{Gardner2000,Cherry2000,Phillips2012,Sokolik2015}. We examine the conditions for bistability as well as the relaxation dynamics in each. Finally, we turn to three-gene feed-forward loops, in which an input gene regulates expression of another both directly and indirectly through regulation of an intermediary. Depending on the precise architecture, these circuits exhibit unique time-varying behavior in response to pulsing effector signals \cite{mangan2003structure, mangan2003coherent, mangan2006incoherent}. 

\section{\label{sec:bistability}Bistability in Genetic Circuits}
Different gene networks serve different biological functions. Among the most important classes of networks are those that yield bistability.  Bistability refers to the situation in which, for a given set of parameters, a system can exist in one of two possible stable steady states. Such a feature is biologically significant. For example, often the expression level of one protein determines a cell's fate. To obtain cells with different functions, there might be some cells with high concentrations of the protein of interest and some cells with low concentrations, requiring a bistable system regulating the protein in question. Broad computational surveys have shown that switch-like, bistable behavior can emerge from a wide range of simple biochemical network architectures~\cite{siegal2011emergence}. 
 Here, we analyze the two simplest and most ubiquitous gene circuits that produce bistability~\cite{Covert2015}, auto-activation and mutual repression, now through the new lens of how effector molecules modulate the dissociation constants of regulatory proteins binding to DNA. Although there are examples in the literature that also use this knob in other regulatory or modeling contexts (as noted in Section \ref{sec:statmechind}), the approach considered here offers a contrast to the conventional setting often employed in studying auto-activation and mutual repression. In such cases, the tuning strategy is typically to modulate the binding parameters within thermodynamic models~\cite{Cherry2000,Alon2020,Covert2015,Phillips2012}, rather than those parameters being naturally tuned through the action of signaling molecules. For simplicity, the following sections do not consider transcription factor oligomerization, either in solution or when bound to DNA, which could give rise to more complex configurations such as DNA looping. We discuss modest generalizations to this effect in Appendix~\ref{app:oligo}.

\subsection{\label{subsec:selfact}The auto-activation regulatory motif}

Auto-activation circuits, in which a gene product enhances its own transcription, are among the simplest genetic regulatory motifs and are capable of generating bistable behavior ~\cite{Cherry2000}. Such motifs have been studied extensively in both synthetic and natural biological systems. In vitro synthetic networks have demonstrated robust switching between high and low expression states under controlled biochemical conditions \cite{subsoontorn2011bistability}, and positive feedback loops have been implicated in natural processes like cell differentiation, where they convert graded input signals into binary gene expression responses \cite{becskei2001positive}. The inducible auto-activation circuit has also been explored with Hill function formulations and with $p_{act}$ as an input parameter \cite{hermsen2011speed}. Motivated by this work, we now present our theoretical analysis of the auto-activation switch that incorporates a full thermodynamic model and explicit treatment of induction. The architecture we study is shown schematically in Fig.~\ref{fig:AutoActivation}, where a transcription factor activates its own production, forming a feedback loop. Given the states, weights, and rates in Fig.~\ref{fig:AutoActivation}, we can now write the kinetic equation governing the dynamics of the auto-activation system as
\begin{align}
\begin{split}
    \frac{dA}{dt} = -\gamma A &+ \frac{r_0 + r_1 (2p_\text{act}(c)\frac{A}{K_d}) + r_2 \omega (p_\text{act}(c)\frac{A}{K_d})^2}{1+2p_\text{act}(c)\frac{A}{K_d} + \omega (p_\text{act}(c)\frac{A}{K_d})^2},
\end{split}
\label{eqn:AutoActivationNoPolymerase}
\end{align}
where $\gamma$ is the degradation rate of protein $A$, $r_{i}$ is the production rate when $i$ activators are bound, $\omega$ is the cooperativity of activator binding, and $K_d$ is the biophysical dissociation constant specific to a gene and a transcription factor. The effect of allosteric regulation is included in $p_\text{act}(c)$, as we introduced earlier. This probability modifies the active transcription factor concentration from $A$ to $p_\text{act}(c)A$. 

Note that we could alternatively describe this auto-activation switch by explicitly considering all possible regulatory states with bound and unbound RNA polymerase (RNAP), and explicitly defining an energy for interaction between activators and RNAP. Appendix~\ref{appendix:polymerase_autoact} demonstrates, however, that this representation is equivalent to Eqn.~\ref{eqn:AutoActivationNoPolymerase}, with the rates $r_{i}$, dissociation constant $K_{d}$, and cooperativity $\omega$ implicitly dependent on polymerase concentration, the strength of polymerase binding to the DNA, and the strength of interaction between polymerase and activator. The discussion throughout this paper will thus remain in the equivalent coarse-grained realm, as depicted in Fig.~\ref{fig:AutoActivation}, and effectively focus on polymerase-bound states.

It is helpful to write our dynamical equation in
dimensionless form.  To do so, we non-dimensionalize the system by using $1/\gamma$ as the unit of time and $K_d$ as the unit of concentration. Within the $p_\text{act}$ paradigm, we can write the dynamical equation in dimensionless form as
\begin{align}
    \frac{d\bar{A}}{d\bar{t}} = - \bar{A} + \frac{\bar{r}_0 + \bar{r}_1 (2p_\text{act}\bar{A}) + \bar{r}_2 \omega (p_\text{act}\bar{A})^2}{1+2p_\text{act}\bar{A} + \omega (p_\text{act}\bar{A})^2},
    \label{eq:auto_activation_master_equation}
\end{align}
where $\bar{t} = \gamma t$, $\bar{A} = A/K_d$ and  $\bar{r}_i = r_i/\gamma K_d$.

\begin{figure}[b]
    \centering
    \includegraphics[width=\linewidth]{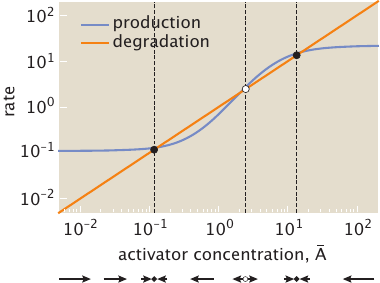}
    \caption{Plot of production and degradation rates for an auto-activation switch as a function of activator concentration. This figure illustrates the competition between the production and degradation terms for a system with rate constants $\bar{r}_{0} = 0.1$, $\bar{r}_{1} = 1$, $\bar{r}_{2} = 20$, and cooperativity $\omega = 10$ at effector concentration $c = 25 \; \mu$M. Intersections of the curve denote stable (filled nodes at low concentration $\bar{A}_{\text{low}}$ and high concentration $\bar{A}_{\text{high}}$) and unstable (unfilled node $\bar{A}_{\text{unstable}}$) fixed points. The vectors denote a phase portrait representing the direction and magnitude of change in activator concentration as a function of activator concentration itself.}
    \label{fig:autoactphaseportrait}
\end{figure}

At a given effector concentration, we can represent the gene expression dynamics that unfold through a phase portrait as in Fig.~\ref{fig:autoactphaseportrait}. The points of intersection of the production and degradation curves correspond to steady state activator concentrations. Fig.~\ref{fig:autoactphaseportrait} highlights a system that can stabilize to one of two possible states with a high ($\bar{A}_{\rm high}$) or low ($\bar{A}_{\rm low}$) activator concentration. Depending on the initial concentration of activator protein, the system will converge to one of these stable points. At the unstable steady state $\bar{A}_{\rm unstable}$, only a small perturbation is needed for the system to evolve toward one of the two stable steady states.

\begin{figure*}
    \centering
    \includegraphics[width=0.9\linewidth]{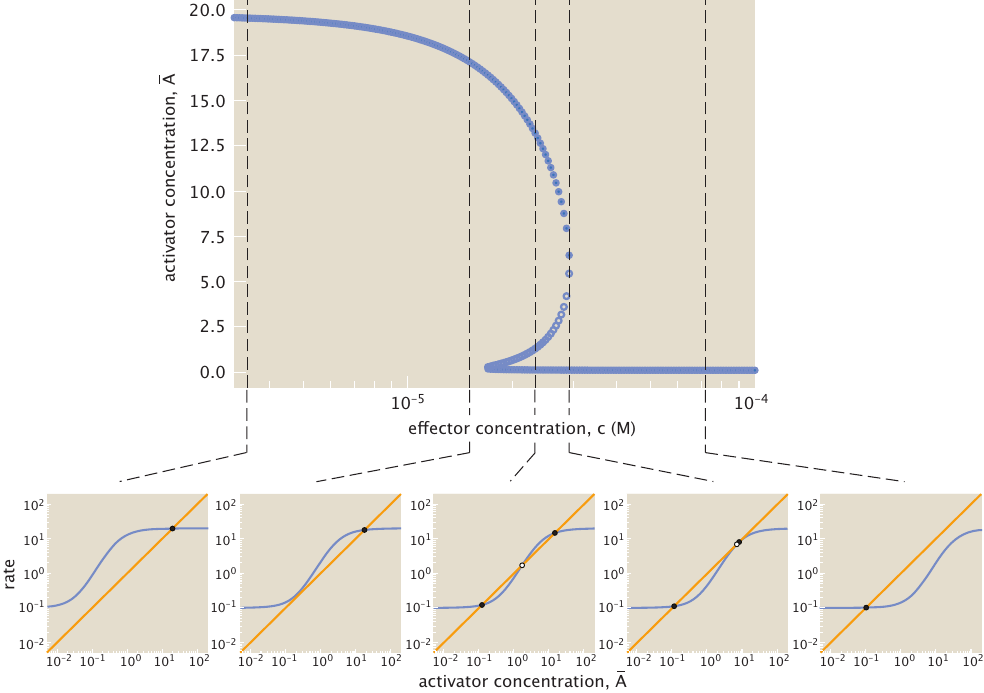}
    \caption{Bifurcation diagram for auto-activation system with rate constants $\bar{r}_{0} = 0.1$, $\bar{r}_{1} = 1$, $\bar{r}_{2} = 20$, and cooperativity $\omega = 10$. This curve plots all stable (filled nodes) and unstable (unfilled nodes) fixed points, demonstrating a shift at intermediate effector concentrations from one to three fixed points. Each black dashed line denotes a specific effector concentration, pointing to a plot of the corresponding production (blue) and degradation (orange) rates as a function of activator concentration. Curve intersections mark the stable (filled nodes) and unstable (unfilled nodes) fixed points found at the given effector concentration.}
    \label{fig:autoactbifurc+phase}
\end{figure*}

We can now qualitatively visualize how the dynamics of auto-activation transform at different effector concentrations. Specifically, as we will see explicitly in the next section, for each effector concentration $c$ we generate a phase portrait analogous to that shown in Fig.~\ref{fig:autoactphaseportrait}. We then determine the number of stable and unstable fixed points and their corresponding activator concentrations. Performing this analysis as we tune effector concentration yields the bifurcation curve shown in Fig.~\ref{fig:autoactbifurc+phase}. At a low effector concentration, activators are more likely to be found in their active configurations, enhancing gene expression such that the system always stabilizes to a state with a high concentration of activator. The magnitude of this concentration approaches a maximal value defined by the rate $\bar{r}_{2}$ for activated protein production. As the effector concentration increases, the system becomes bistable, allowing a bimodal distribution in protein concentrations for an ensemble of cells~\cite{Sokolik2015}.

Ultimately, at sufficiently high effector concentration, activators are sequestered in their inactive configuration such that the system can only stabilize to a state with low activator concentration. The magnitude of activator expressed is then largely defined by the basal rate of production without bound activator, i.e., $\bar{r}_{0}$. Tracking the system's corresponding production and degradation curves through a series of snapshots in Fig.~\ref{fig:autoactbifurc+phase}, we observe that these shifts between bistable and monostable dynamics emerge because the increasing effector concentration shifts the system production curve toward higher activator concentration. Expressed differently, as effector concentration increases, a higher activator concentration is necessary to achieve a given production rate.

\begin{figure}[t]
    \centering
    \includegraphics[width=\linewidth]{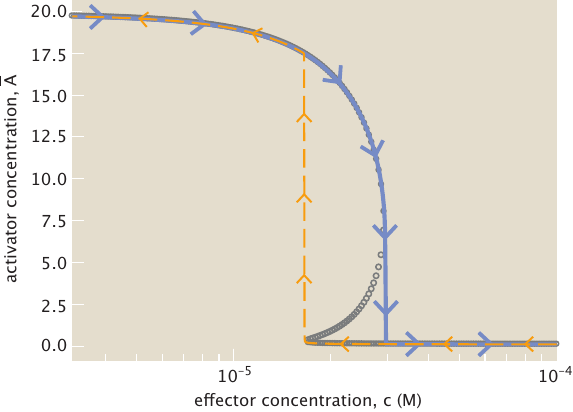}
    \caption{Plot of stable state evolution, exhibiting hysteresis under different trends for effector concentration. The blue curve plots how the stable state evolves with increasing effector concentration from an initial high concentration of activator. The orange dashed curve plots the change in stable activator concentration starting from an initially low level as effector concentration decreases. Comparing to the bifurcation diagram (gray) previously shown in Fig.~\ref{fig:autoactbifurc+phase}, the history of steady-state evolution determines the threshold concentration of effector at which the system switches state, highlighting hysteretic behavior.}
    \label{fig:autoacthysteresis}
\end{figure}

Note that the threshold at which the system switches from one stable state to another differs when increasing and decreasing the effector concentration. If the system initially contains a high concentration of activator before tuning, a higher concentration of effector is necessary to switch to the low activator state than is required when decreasing effector concentration for a system with initially low activator concentration. Considering Fig.~\ref{fig:autoactbifurc+phase} again, the higher threshold corresponds to the maximum effector concentration at which the system is bistable, and the lower threshold to the minimum concentration at which the system is bistable. Fig.~ \ref{fig:autoacthysteresis} illustrates this phenomenon of hysteresis more explicitly, overlaying the previously discussed bifurcation diagram, and showing how the threshold at which the blue stable equilibrium trajectory switches from high to low activator concentration differs from the threshold for the orange trajectory tracing the switch in reverse.

We are particularly interested in characterizing the conditions for bistability to occur at effector concentrations $c$, as well as how the bistable regime (if it exists) responds to changes in parameters such as production rates and cooperativity. While previous studies have analyzed bistability in auto-activation circuits using thermodynamic models without effectors~\cite{Cherry2000, laxhuber2020theoretical}, our aim is to extend this to include effector dependence explicitly and systematically explore how the bistable region evolves across a broader parameter space.

\begin{figure*}
    \centering
    \includegraphics[width=\linewidth]{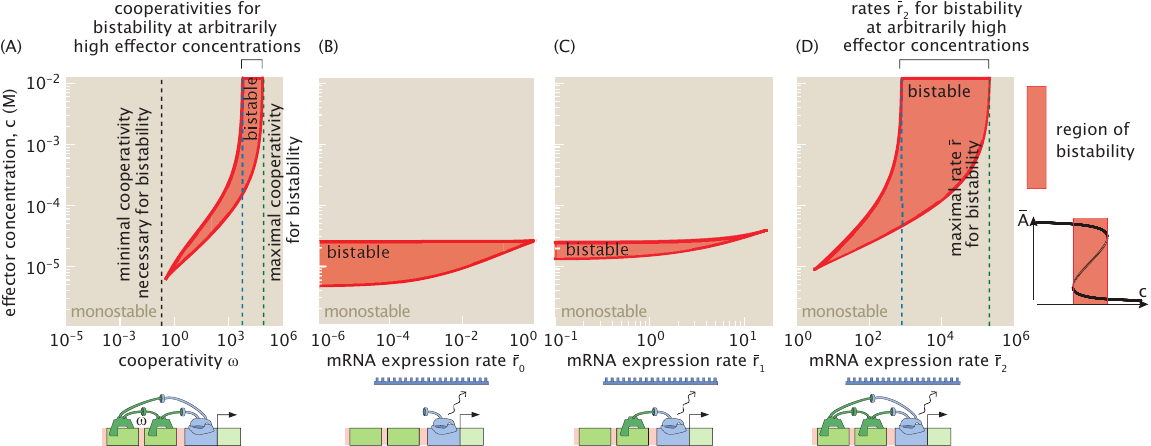}
    \caption{Range of effector concentrations for which the system exhibits bistability. The red shaded region indicates bistability. Outside this region, the system is monostable. Unless otherwise specified, parameters are fixed at: $\omega = 7.5$, $\bar{r}_0 = 0.1$, $\bar{r}_1 = 1$, and $\bar{r}_2 = 20$. (A) Bistable concentration range as a function of cooperativity $\omega$, varied over the interval $\omega \in [10^{-6}, 10^6]$, corresponding to interaction energies between the two activators from approximately $+14,k_BT$ to $-14,k_BT$, since $\omega = e^{-\beta \varepsilon_{\text{int}}}$. The dotted line shows the analytical lower bound on the minimal cooperativity required for bistability. (B) Bistable concentration range as a function of $\bar{r}_0$, sampled over the interval $\bar{r}_0 \in [10^{-6}, \bar{r}_1]$. (C) Bistable concentration range as a function of $\bar{r}_1$, sampled over the interval $\bar{r}_1 \in [\bar{r}_0, \bar{r}_2]$. (D) Bistable concentration range as a function of $\bar{r}_2$, sampled over the interval $\bar{r}_2 \in [\bar{r}_1, 10^6]$. The rate parameters are varied under the constraint of the auto-activation condition $\bar{r}_0 \leq \bar{r}_1 \leq \bar{r}_2$, which ensures that the production rate increases with the number of bound activators.}
\label{fig:autoactivation_ccmin_ccmax_sweep}
\end{figure*}

\subsubsection{Regimes of bistability}
We now analyze the ways in which bistability emerges in the context of effector-mediated genes. One could begin with simply a fixed set of parameters and asking whether the system is bistable. In other words, can the concentration of gene product $A$ settle at either a high or low steady state depending on the initial condition, thereby functioning as a binary switch?

The question becomes more nuanced, however, when taking the role of effector molecules into account. Some parameters are more intrinsic and less readily tunable than others. For instance, molecular constants such as binding affinities or cooperativity are typically encoded in the system's molecular architecture. In contrast, cells can regulate the concentration of effector molecules relatively easily and rapidly, either through controlled expression, import/export mechanisms, or degradation pathways. Given this, the most relevant questions to ask are (\textit{i}) under what constraints on the system’s intrinsic parameters can bistability be achieved for at least some range of effector concentration $c$, and (\textit{ii}) how wide is that range? Indeed, even if the effector concentration is a more accessible control knob, the cell is unlikely to operate at a single precise value, if only due to molecular noise and environmental fluctuations. It is therefore biologically relevant to assess not just whether bistability is possible, but whether it is robust to such fluctuations in effector levels.

We observe that even under the idealized assumption that the cell could freely choose its effector concentration, bistable behavior arises only within a restricted range of parameter values. This is shown numerically in the red regions of Fig.~\ref{fig:autoactivation_ccmin_ccmax_sweep}. 
To investigate the conditions under which multiple steady states are possible, we also derive analytical bounds in parameter space in Appendix~\ref{appendix:autoact_derivation_of_descartes_bound}. By setting $d\bar{A}/d\bar{t} = 0$ and rewriting the resulting expression in standard polynomial form, we can infer the number of positive real roots using classical results (i.e., Descartes' rule of signs) that relate sign changes in the coefficients of a polynomial to the number of positive roots. This leads to a necessary condition on the parameters of the system: for bistability to be possible for at least one value of the effector concentration, the system must satisfy the inequalities
\begin{eqnarray}
\displaystyle 
\frac{\omega \bar{r}_2}{2} &>& 1 + e^{-\beta \varepsilon},\label{eq:bistabconds_final1}\\
2\bar{r}_{1} &<& 1 + e^{-\beta \varepsilon}\bar{K}_c^2,\label{eq:bistabconds_final2} \\
 \omega \bar{r}_2 &>& 4\bar{r}_1.
 \label{eq:bistabconds_final3}
\end{eqnarray}
Notably, these conditions do not depend on $\bar{r}_0$.

In Fig.~\ref{fig:autoactivation_ccmin_ccmax_sweep}, we systematically modulate the four parameters of the auto-activation system: the strength of cooperative binding $\omega $, the protein production rate $\bar{r}_{0}$ in the absence of activator binding, the protein production rate $\bar{r}_1$ when one activator is bound to the DNA, and the protein production rate $\bar{r}_2$ when two activators are bound. Fig.~\ref{fig:autoactivation_ccmin_ccmax_sweep}(A) shows the consequences of varying the cooperativity parameter $\omega$, sampling values from $10^{-6}$ to $10^{6}$ to represent systems with positive cooperativity ($\omega >1$), systems that are non cooperative ($\omega=1$), and systems with negative cooperativity ($\omega <1$). Recall that cooperativity describes the energy of interaction between two bound activators, $\varepsilon_{\text{int}}$, and therefore can be written as $\omega = e^{-\beta \varepsilon_{\text{int}}}$. The range of cooperativity shown in the figure thus corresponds to interaction energies ranging from $\varepsilon_{\text{int}} \approx -14k_BT$ to $\varepsilon_{\text{int}} \approx 14k_BT$, encompassing a broad and biologically relevant spectrum of interaction strengths~\cite{bintu2005transcriptional,bintu2005transcriptional2,Phillips2012}. 

Fig.~\ref{fig:autoactivation_ccmin_ccmax_sweep}(B), (C), and (D) explore the effects of varying the rates $\bar{r}_0$, $\bar{r}_1$ and $\bar{r}_2$, respectively. To do so, we impose constraints on these parameters to ensure that the system remains within the auto-activation regime, as defined in Appendix~\ref{appendix:condition_r1_r2_autoact}. Specifically, we require that the production term in Eqn.~\ref{eq:auto_activation_master_equation} remains a monotonically increasing function of $\bar{A}$, such that $\bar{A}$ consistently acts as an activator across all concentrations. This condition imposes the inequality $\bar{r}_0 \leq \bar{r}_1 \leq \bar{r}_2$, which we enforce throughout our analysis when varying $\bar{r}_0$, $\bar{r}_1$ and $\bar{r}_2$. 
 
We note that the necessary criteria for bistability derived in Eqns.~\ref{eq:bistabconds_final1} -~\ref{eq:bistabconds_final3} do not impose any effective constraint when tuning either $\bar{r}_0$, $\bar{r}_1$, or $\bar{r}_2$ for the set of parameters chosen in Fig.~\ref{fig:autoactivation_ccmin_ccmax_sweep}. Specifically, the predicted lower bound on $\bar{r}_1$ exceeds the upper limit on $\bar{r}_1$ allowed by the auto-activation constraint (i.e., $\bar{r}_1 \leq \bar{r}_2$). Similarly, the threshold for $\bar{r}_{2}$ above which bistability is possible lies below $\bar{r}_1$, meaning the system no longer behaves as a strictly auto-activating unit.  Indeed, we observe that for parameter values consistent with the auto-activation regime, when either $\bar{r}_1$ or $\bar{r}_2$ is varied individually in Fig.~\ref{fig:autoactivation_ccmin_ccmax_sweep}(C) and (D), the system fails to exhibit bistability only in a narrow region where $\bar{r}_1$ approaches $\bar{r}_2$. Note, however, that this behavior is not universal. For different parameter values, the relative positioning of these thresholds may change, and the necessary conditions specified by Eqns.~\ref{eq:bistabconds_final1} -~\ref{eq:bistabconds_final3} could become more explicitly informative. The conclusions drawn here are therefore specific to the parameter set used in this analysis. 
 
 We show that the parameters $\bar{r}_2$ and $\omega$ play similar roles in shaping the system’s ability to exhibit bistability. As seen in Fig.~\ref{fig:autoactivation_ccmin_ccmax_sweep}(A), there exists a critical threshold of cooperativity $\omega$ below which the system is strictly monostable, indicating that a minimal level of nonlinearity is required for bistability. The analytical bounds derived in Eqns.~\ref{eq:bistabconds_final1} and~\ref{eq:bistabconds_final3} accurately capture this threshold. Likewise, Fig.~\ref{fig:autoactivation_ccmin_ccmax_sweep}(D) demonstrates that $\bar{r}_2$ must exceed a minimum value to support bistability; below this threshold, the system remains monostable for all effector concentrations. 
 
 Increasing $\bar{r}_2$ strengthens the contrast between states with high and low activator steady state concentrations in a positively cooperative system ($\omega > 1$), and as a result, the system becomes unstable for a broader range of effector concentrations. Therefore, as either $\omega$ or $\bar{r}_2$ increases, the region of effector concentrations that supports bistability broadens. This leads to a wider hysteresis zone and expands the range over which the system can toggle between high and low steady states under a fixed set of parameters. The region of bistability in effector concentration space is displaced toward higher concentrations, where the activation probability $p_\text{act}(c)$ approaches the leakiness limit. This reflects the fact that the effector destabilizes the activator by decreasing its effective DNA binding affinity. In this sense, increasing effector concentration counteracts the effect of high $\omega$ and $\bar{r}_2$, which tend to promote high expression levels of $A$. These two opposing effects—activation-driven amplification and effector-driven destabilization—create a balance that enables bistability. 
 
 Due to the system's leakiness, complete inactivation of the activator is never achieved, even at high effector concentrations. As a result, for sufficiently large values of $\omega$ and $\bar{r}_2$, bistability can occur for all concentrations above a finite lower bound $c_{\min}(\omega, \bar{r}_0, \bar{r}_1, \bar{r}_2)$. This behavior corresponds to a limited region in parameter space where the system remains bistable at large $c$, as illustrated by the blue and green dotted lines in Fig.~\ref{fig:autoactivation_ccmin_ccmax_sweep}(A) and (C). In Appendices~\ref{appendix:descart_for_c_auto_act} and~\ref{appendix:auto_act_no_bistab_high_coop}, we derive analytical bounds that predict the onset and breakdown of this upper-unbounded bistable regime.
 
 Beyond a certain point and with fixed values of effector concentrations, further increases in either parameter have the opposite effect. When $\omega$ becomes too large, activators bind excessively tightly to the DNA, effectively locking the system into a high-expression state. Similarly, if $\bar{r}_2$ becomes too large, the system favors high levels of gene expression, and bistability is again lost (we analyze this transition using two-dimensional numerical sweeps and supporting analytical arguments in Appendix \ref{appendix:auto_act_no_bistab_high_coop}). This behavior is a direct consequence of how the effector enters the model. If, instead of varying the effector concentration, we varied an effective dissociation constant $K_d^{\text{eff}}$, the bistable region would always remain bounded within a finite range of values.
 
This behavior contrasts with the effect of increasing $\bar{r}_0$ or $\bar{r}_1$. As shown in Fig.~\ref{fig:autoactivation_ccmin_ccmax_sweep}(B) and (C), higher values of either parameter reduce the width of the bistable region, and beyond a critical threshold, bistability disappears entirely. We can therefore infer that, in a positively cooperative system ($\omega > 1$), elevated values of $\bar{r}_0$ and $\bar{r}_1$ undermine the system’s ability to function as a bistable switch.

Interestingly, we observe bistability for values of the cooperativity parameter $\omega$ that are less than one, as seen in Fig.~\ref{fig:autoactivation_ccmin_ccmax_sweep}(A). This can be reconciled in two complementary ways. First, we can define an effective cooperativity for the system, given by $\omega_{\text{eff}} = \omega \cdot \bar{r}_2 / 2$. From the necessary conditions for bistability derived in Eqns.~\ref{eq:bistabconds_final1} and~\ref{eq:bistabconds_final3}  we find that $\omega_{\text{eff}} > 1$ is required for bistability. While this condition is necessary but not sufficient, it suggests that $\omega_{\text{eff}}$ captures the functional cooperativity of the system more accurately than $\omega$ alone, as it incorporates both the interaction between bound activators and the maximal rate of activator production. We can also reconcile our bistable results containing values of $\omega$ less than one by examining the effective Hill coefficient of the steady-state input–output function. As shown in Appendix \ref{appendix:auto_act_bistab_low_coop}, bistability is observed only when the effective Hill coefficient exceeds one, as anticipated from theoretical considerations ~\cite{griffith1968mathematics}. This reinforces the idea that the system can exhibit bistability even when $\omega < 1$, provided the overall nonlinearity—quantified either by $\omega_{\text{eff}}$ or the effective Hill coefficient—is sufficiently strong.

\subsubsection{Comparing Hill function and thermodynamic formulations}
\label{sec:hill_v_thermo}
Thus far our paper has employed a thermodynamic formulation of the auto-activation switch, rooted in the principles of statistical mechanics. In previous discussions of gene circuits, however, a phenomenological Hill function is the predominant method to model these dynamics (e.g., \cite{Gardner2000,Elowitz2000}).
We note the fascinating origins of the Hill function in the work of Archibald Hill.
More than 100 years ago, Hill wrote down a
description of the binding between oxygen and hemoglobin that we now know as the Hill function, which he wrote as
\begin{equation}
p_{\text{bound}}(x)={\left({x \over K}\right)^n \over 1+ \left({x \over K}\right)^n },
\label{eqn:HillEquation}
\end{equation}
where $x$ is the concentration of $O_2$ and $K$ is its allied dissociation constant.

As Hill himself tells us, this functional form provides a summary of the occupancy of hemoglobin (the example he used, though it has been applied much more broadly).
If we think of the huge topic of input-output functions in biology, then the kind of
characteristics embodied in the Hill approach include a representation of leakiness (the amount of output even in the absence of input, $p_{\text{bound}}(0)$), dynamic range, 
$EC_{\text{50}}$ (the input concentration at which the output reaches half its maximum, $EC_{\text{50}}=K$) and the sensitivity as measured by the slope of the input-output curve (usually in
logarithmic variables) at the midpoint.  It is instructive to hear Hill commenting on his thinking: ``My object was rather to see whether an equation {\it of this type} can satisfy all the observations, than to base any direct physical meaning on $n$ and $K$~\cite{Hill1910}.''   He goes further in his 1913 paper noting~\cite{Hill1913} ``In point of fact $n$ does not turn out to be a whole number, but this is due simply to the fact that aggregation is not into one particular type of molecule, but rather into a whole series of different molecules: so that equation (1) (our Eqn.~\ref{eqn:HillEquation}) is a rough mathematical expression for the sum of several similar quantities with $n$ equal to 1, 2, 3, 4 and possibly higher integers.'' We think it is important to remember that the Hill function is a phenomenological
description of equilibrium binding that assumes certain states of occupancy are irrelevant, or alternatively, that bunches all of the states of occupancy into one effective non-integer state of occupancy~\cite{Sherman2012}.

In comparing and contrasting the two treatments of transcription factor binding, we find that they can yield different results. Hill functions approximate away some thermodynamic states and describe the dynamics as
\begin{align}
\begin{split}
    \frac{dA}{dt} = -\gamma A &+ \frac{r_0+r_2(p_\text{act}(c)\frac{A}{K_d^\text{eff}})^n}{1 + (p_\text{act}(c)\frac{A}{K_d^\text{eff}})^n},
    \label{eq:autoactdim}
\end{split}
\end{align}
where $n$ is the Hill coefficient and $K_d^\text{eff}$ is an effective dissociation constant. Appendix~\ref{appendix:hill_v_thermo_sup} demonstrates explicitly how this Hill function form derives from the high cooperativity limit of the thermodynamic model, removing the single activator-bound state as has been reported previously in the literature \cite{Sherman2012}.

\begin{figure*}
    \centering
    \includegraphics[width=\linewidth]{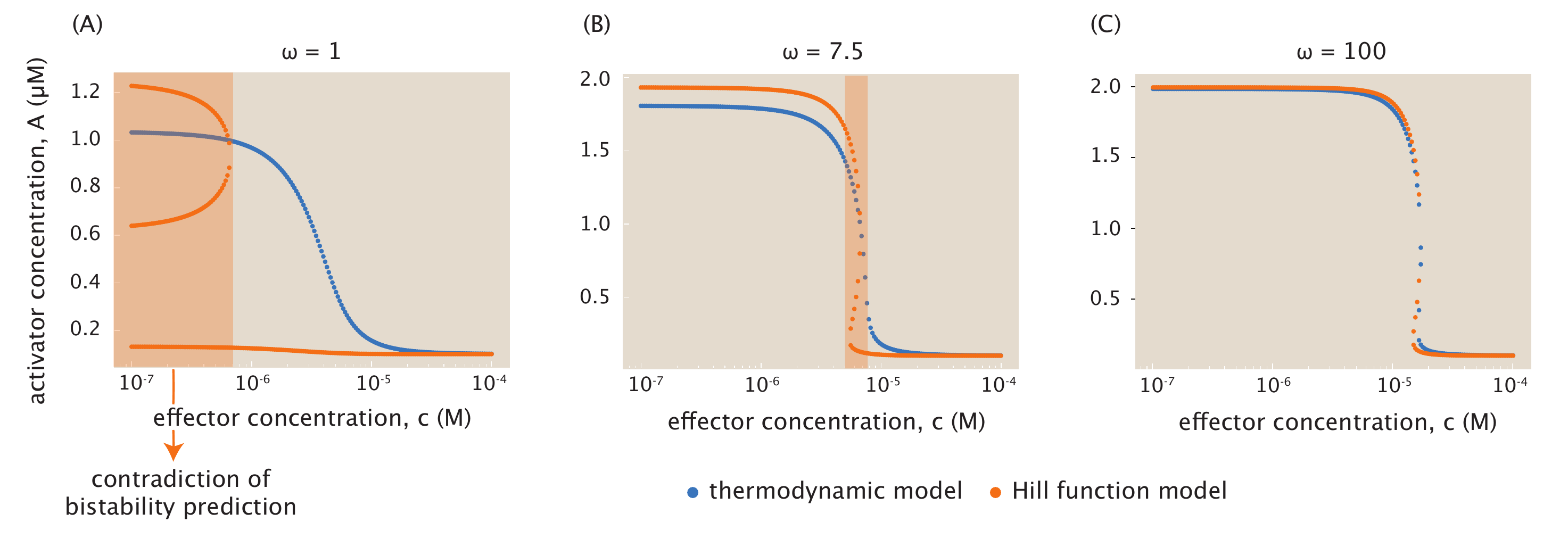}
    \caption{Bifurcation diagrams generated by the Hill function model and the thermodynamic model differ at small to intermediate cooperativity but converge at large cooperativity. The blue dots are fixed points in the thermodynamic model, and the orange dots are the fixed points in the Hill function model. The orange shaded region tracks where conflicting bistability predictions are made between the two models. The shared parameters are $r_0 = 0.1 \mu$M/min, $r_2=2 \mu$M/min, $\gamma = 1$/min, $K_d = 1 \mu$M. The thermodynamic model has $r_1 = 1\mu$M/min across the three panels. The cooperativities are (A) $\omega = 1$, (B) $\omega = 7.5$, (C) $\omega = 100$.}
    \label{fig:hill_v_thermo_revised}
\end{figure*}

Reconciling the Hill and thermodynamic models for $n = 2$ in the limit of large $\omega$ requires an effective dissociation constant of $K_d^\text{eff} = K_d/\sqrt{\omega}$, as discussed in Appendix~\ref{appendix:hill_v_thermo_sup}. Since this means that $K_d^\text{eff}\neq K_d$ when $\omega \neq 1$, the natural concentration scales differ in the Hill and thermodynamic settings, and we must thus exercise caution in comparison. The remaining analysis of this subsection therefore employs the dimensional forms of each model to allow direct comparison.

When cooperativity $\omega$ is small or moderate, the Hill function model can provide different predictions from the thermodynamic model. There are two ways to compare the models. One compares only the probability of state occupancy, as done in Appendix~\ref{appendix:hill_v_thermo_sup}, and the other compares the bifurcation curves of the full dynamical system when each production term is used. In the first case, the two models often yield similar end results, with their state occupancy probabilities agreeing to some degree even for small $\omega$. In contrast, when the production terms are embedded in the dynamical system, the models can diverge substantially. Figure~\ref{fig:hill_v_thermo_revised}(A) shows this for the low cooperativity case with $\omega=1$. Most notably, at low effector concentrations the Hill model shows bistability while the thermodynamic model does not. In the intermediate cooperativity regime of Fig.~\ref{fig:hill_v_thermo_revised}(B), the bifurcation diagrams resemble each other more closely. Nevertheless, they still make conflicting predictions for bistability in a range of effector concentration between $10^{-6}$ M to $10^{-5}$ M. In the high cooperativity regime of Fig.~\ref{fig:hill_v_thermo_revised}(C), the two models ultimately agree and produce identical bifurcation profiles.

Given the relatively light computational requirements for the model systems discussed here, as opposed to more complex and high-dimensional gene-interaction networks \cite{Rousseau2023,RousseauPhillips2025}, we will proceed with the thermodynamic formulation at high cooperativity for all other gene circuits considered in the paper. Since our analysis addresses the existence of bistability across all three cooperativity regimes, using the full thermodynamic model also ensures that the discussion captures all essential dynamical features across varying cooperativities.

\subsubsection{Timescale for stabilization}

\begin{figure*}
    \centering
    \includegraphics[width=0.9\linewidth]{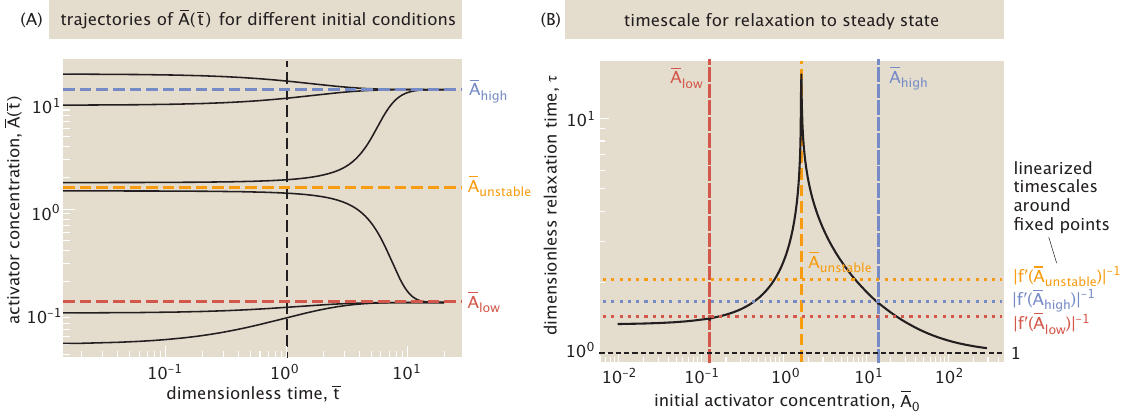}
    \caption{Dynamics of the auto-activation switch. The parameters of the system are fixed at: $\omega = 7.5$, $\bar{r}_0 = 0.1$, $\bar{r}_1 = 1$, $\bar{r}_2 = 20$, and $c=2 \cdot 10^{-5}\text{ M}$. (A) Time evolution of the activator concentration for various initial conditions. Each  black curve represents a trajectory $\bar{A}(\bar{t})$ starting from a different initial condition $\bar{A}_0$. Dashed horizontal lines indicate the stable ($\bar{A}_{\rm high}$ and $\bar{A}_{\rm low}$) and unstable ($\bar{A}_{\rm unstable}$) fixed points. (B) Relaxation timescale obtained from exponential fits to the trajectory $\bar{A}(\bar{t})$ as a function of the initial concentration $\bar{A}_0$. Vertical lines indicate the positions of the steady states, while horizontal dashed lines mark the timescales associated with small perturbations around each fixed point, computed from the inverse slope $|f'(\bar{A})|^{-1}$ of the function $f$ defined in Eqn.~\ref{eq:auto_act_def_f}. The solid horizontal black line corresponds to the reference timescale $\bar{t} = 1$.}
\label{fig:autoactivation_dynamics}
\end{figure*}

In addition to the question of steady states, it is interesting to examine the timescale for an inducible genetic circuit such as the auto-activation switch to reach steady state. In doing this analysis we remind the reader that our treatment  assumes separation of time scales between the dynamics of effector binding and allosteric transitions, and the dynamics of the relaxation to steady state. As a result, the binding of effector and the allosteric state of the activator rapidly equilibrate as the activator concentration changes dynamically. 

As shown in Fig.~\ref{fig:autoactivation_dynamics}(A), the relaxation timescale depends strongly on the initial activator concentration $\bar{A}_0$. 
In particular, as seen in the figure,  the relaxation timescale to steady state increases as the initial value of activator concentration, $\bar{A}_0$, gets closer to the unstable fixed point. The increased time to reach steady state near the unstable point reflects the typical behavior of positive feedback systems near bifurcations, where the system lingers near the threshold before switching ~\cite{tyson2003sniffers}. However, for large initial concentrations of $A$, the dimensionless relaxation timescale approaches 1, as shown by the line at $\bar{t}=1$ in Fig.~\ref{fig:autoactivation_dynamics}(A). This corresponds to the protein degradation timescale $1/\gamma$, which serves as the unit of time in our non-dimensionalization. 
This also corresponds to the relaxation timescale of a simply regulated gene, activated solely by an upstream transcription factor. We find that auto-activation consistently delays the response compared to simple regulation, although the two converge in the high-$A$ limit. Within the Hill-function framework, positive autoregulation has likewise been shown to slow gene circuit response times~\cite{mitrophanov2010positive}, in contrast to the accelerating effect of negative feedback~\cite{rosenfeld2002negative}.

To further interpret these trends, we apply linear stability analysis to Eqn.~\ref{eq:auto_activation_master_equation}, as is commonly done in the study of dynamical systems ~\cite{Strogatz2015}. Linearizing around a point $\bar{A}_i$ with $\bar{A}(\bar{t}) = \bar{A}_i + \delta \bar{A}(\bar{t})$ and expanding the right-hand side, that we denote $f(\bar{A})$, to first order yields
\begin{equation}
    f(\bar{A}(t)) \approx f(\bar{A}_i) + f'(\bar{A}_i)\, \delta\bar{A}(t)
\end{equation}
for sufficiently small $\delta \bar{A}(\bar{t})$, where $f'(\bar{A})$ is a derivative with respect to $\bar{A}$ rather than a time derivative. The function $f$ is given by
\begin{equation}
    f(\bar{A}) = -\bar{A} + \frac{\bar{r}_0 + \bar{r}_1(2p_\text{act}\bar{A}) + \bar{r}_2\omega(p_\text{act}\bar{A})^2}{1 + (2p_\text{act}\bar{A}) + \omega(p_\text{act}\bar{A})^2}.
    \label{eq:auto_act_def_f}
\end{equation}
Therefore, the linearized equation governing the evolution of a small perturbation $\delta \bar{A}(t)$ around the point $\bar{A}_i$ becomes
\begin{equation}
    \frac{d\delta \bar{A}}{d\bar{t}} = f(\bar{A}_i) + f'(\bar{A}_i)\delta \bar{A}.
    \label{eq:autoact_taylor_ai}
\end{equation}
If $\bar{A}_i$ is a steady state, then $f(\bar{A}_i) = 0$, and the equation reduces to exponential relaxation with dimensionless timescale $|f'(\bar{A}_i)|^{-1}$. Stable fixed points, denoted $\bar{A}_{\text{low}}$ and $\bar{A}_{\text{high}}$, satisfy $f'(\bar{A}_{\text{low/high}}) < 0$, so perturbations decay over time. For unstable fixed points, where $f'(\bar{A}_\text{unstable}) > 0$, small deviations grow exponentially and drive the system away from the fixed point.

However, if $\bar{A}_i$ is not a fixed point, then $f(\bar{A}_i) \ne 0$ and the solution to Eqn.~\ref{eq:autoact_taylor_ai} does not represent convergence to a steady state. Instead, it predicts that $\delta \bar{A}$ asymptotes to a finite offset $f(\bar{A}_i)/|f'(\bar{A}_i)|$, breaking the assumption of linearity ($\delta \bar{A}(\bar{t})\to 0$). In this case, the derivative $f'(\bar{A}_i)$ still encodes the local response to small perturbations, but only describes the dynamics while deviations remain small. Therefore, the timescale $|f'(\bar{A})|^{-1}$ is most meaningful in the vicinity of fixed points, even if it can still provide qualitative insights elsewhere.

These insights are reflected in Fig.~\ref{fig:autoactivation_dynamics}(B), where we numerically compute the relaxation timescale as a function of the initial condition $\bar{A}_0$. For initial values near the stable fixed points, the timescale closely follows the linear prediction $|f'(\bar{A}_{\text{low/high}})|^{-1}$. In contrast, when the initial condition lies near the unstable fixed point, the system initially diverges exponentially  and only later relaxes nonlinearly to a stable state. 
 This leads to an overall increase in the time to reach steady state, consistent with the trajectories displayed in Fig.~\ref{fig:autoactivation_dynamics}(A) (see Appendix~\ref{appendix:auto_act_timescale} for the precise numerical procedure used to compute these timescales). Together, these results reveal that the stabilization timescale in auto-activation circuits is not fixed but depends sensitively on initial conditions, particularly near unstable fixed points—highlighting the importance of considering nonlinear transient dynamics when predicting response times in bistable gene networks.

\subsection{\label{subsec:mutrep}The mutual repression regulatory motif}
\begin{figure*}
    \centering
    \includegraphics[width=\linewidth]{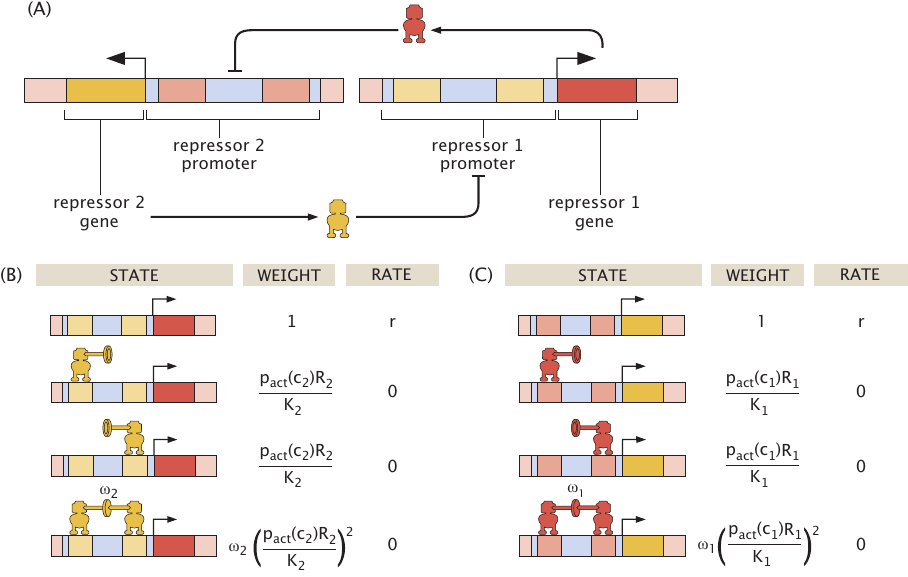}
    \caption{The mutual repression regulatory circuit.  (A) Schematic of the operation of the circuit. When the gene for repressor 1 is expressed, the resulting protein downregulates the expression of the gene for repressor 2. Repressor 2, in turn, downregulates the expression of the gene for repressor 1.
     (B) Thermodynamic states, weights, and rates for expression of repressor 1. In our model, a repressor can bind non-exclusively at one of two possible sites within the target promoter region to suppress gene transcription. The parameter $\omega_2$ denotes the cooperative strength between two bound repressors $R_{2}$.  (C) Thermodynamic states, weights, and rates for expression of repressor 2. The states and weights for the regulation of the promoter responsible for the production of repressor 2 are analogous to those shown for repressor 1. However, the dissociation constant of repressor 1 in this case is given by $K_1$, and the cooperativity term for the interaction of two repressor 1 molecules bound to the DNA is $\omega_1$. }
    \label{fig:MutRep}
\end{figure*}

\begin{figure*}
    \centering
    \includegraphics[width=0.78\linewidth]{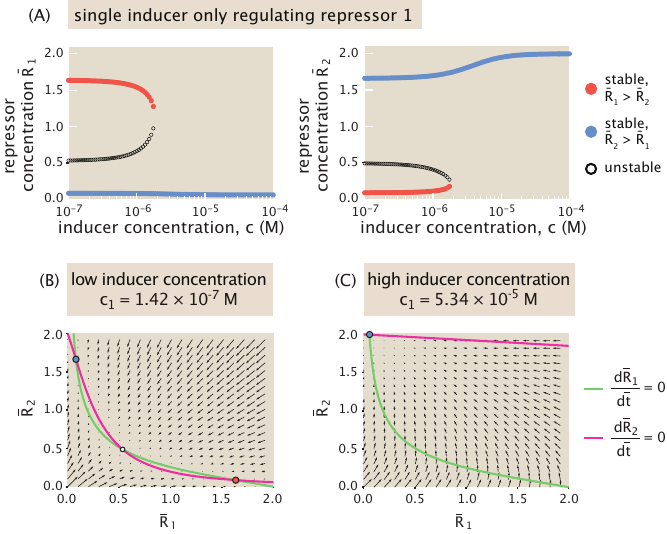}
    \caption{Bifurcation diagrams and phase portraits for mutual repression in the presence of a single inducer regulating the activity of repressor $R_{1}$ (with fixed parameters $\bar{K} = K_{1}/K_{2} = 1$, $\bar{r} = 2$ and $\omega_1 = \omega_2 = 7.5$). (A) Bifurcation diagrams tracking steady-state $\bar{R}_{1}$ and $\bar{R}_{2}$ expression as inducer concentration increases. The red curves track expression in the steady state for which $\bar{R}_{1} > \bar{R}_{2}$, the blue curves track expression in the steady state for which $\bar{R}_{2} > \bar{R}_{1}$, and the unfilled nodes denote unstable saddle points. (B) Phase portrait at a low inducer concentration, demonstrating bistable dynamics between two possible stable states favoring either $\bar{R}_{1}$ (red) or $\bar{R}_{2}$ (blue). Steady states occur at the intersections of the nullclines as shown. (C) Phase portrait at a high inducer concentration, for which the system is monostable to favor $\bar{R}_{2}$.}
    \label{fig:mutual_repression_singleinducer}
\end{figure*}

Many regulatory circuits in prokaryotes and eukaryotes alike are mediated by the interaction between two genes that repress each other as shown in Fig.~\ref{fig:MutRep}~\cite{Huang2007,laslo2008gene,chute2010minireview}. Indeed, one of the signature achievements of the early days of synthetic biology was the consideration of a mutual repression switch in bacteria whose behavior was reported by the use of fluorescent proteins and controlled by a small molecule inducer~\cite{Gardner2000}. As the name suggests, two genes $R_1$ and $R_2$ mutually repress each other. To simplify our analysis we assume that the two genes share the same degradation rate $\gamma$ and production rate $r$.

By inspecting the states and weights diagrams of Fig.~\ref{fig:MutRep}, we can express the dynamics of repressor 1 expression as
\begin{equation}
\label{eq:R1SwtichDynamics}
    {{\rm d} R_1 \over {\rm d}t} = - \gamma R_1 +{r \over 1 + 2 {p_\text{act}(c_2) R_2 \over K_2} + \left({p_\text{act}(c_2) R_2 \over K_2} \right)^2 \omega_2},
\end{equation}
where we have defined an effector with concentration $c_2$ that regulates the activity of repressor 2. Similarly, 
the dynamics for $R_2$ expression are described by an equation analogous to Eqn.~\ref{eq:R1SwtichDynamics} but swapping $R_1$ and $R_2$, given by
\begin{equation}
\label{eq:R2SwtichDynamics}
    {{\rm d} R_2 \over {\rm d}t} =- \gamma R_2 + {r \over 1 + 2  {p_\text{act}(c_1) R_1 \over K_1} + \left(
 {p_\text{act}(c_1)  R_1 \over K_1} \right)^2 \omega_1},
\end{equation}
where we have also defined an effector with concentration $c_1$ that regulates the activity of repressor 1. The production term attributes a rate $r$ to the state with no bound repressors as shown in
Fig.~\ref{fig:MutRep}. As in auto-activation, this analysis only accounts for the presence of RNA polymerase implicitly, which we discuss further in Appendix~\ref{appendix:polymerase_mut_rep}. Note that having two effectors means that the control knob becomes a 2-dimensional vector $(c_1,c_2)$. The bistable region becomes an area, and $(c_1,c_2)$ can be tuned following arbitrary trajectories in the inducer concentration space. We will explore these threads in this section.

To continue the analysis, we can write dimensionless forms of these kinetic equations by transforming $\bar{t} = \gamma t$, $\bar{R}_{i} = R_{i}/K_{2}$, and $\bar{r} = r/\gamma K_2$, and obtain
\begin{small}
\begin{eqnarray}
\label{eq:mutRep1}
\frac{d\bar{R}_{1}}{d\bar{t}} &=& - \bar{R}_{1} +\frac{\bar{r}}{1 + 2p_\text{act}(c_2)\bar{R}_{2} + \omega_2 \Big[p_\text{act}(c_2)\bar{R}_{2} \Big]^2}, \\
 \frac{d\bar{R}_{2}}{d\bar{t}} &=& - \bar{R}_{2} + \frac{\bar{r}}{1 + 2p_\text{act}(c_1)\frac{\bar{R}_{1}}{\bar{K}} + \omega_1 \Big[p_\text{act}(c_1)\frac{\bar{R}_{1}}{\bar{K}}\Big]^2}.
\end{eqnarray}
\end{small}
As a reminder, $\omega_1$ and $\omega_2$ are the cooperativity for $\bar{R}_{1}$ and $\bar{R}_{2}$, respectively, $c_1$ and $c_2$ are the inducer concentrations for each repressor, and $\bar{K} = K_1 / K_2$ is the ratio of dissociation constants. Note that the equations above assume that each repressor responds to induction with the same inducer-protein binding chemistry, obeying the same activation probability function $p_\text{act}$, but responding to potentially different inducer concentrations $c_1$ and $c_2$. In the most general case, however, the two probability functions could differ.

\begin{figure*}
    \centering
    \includegraphics[width=0.9\linewidth]{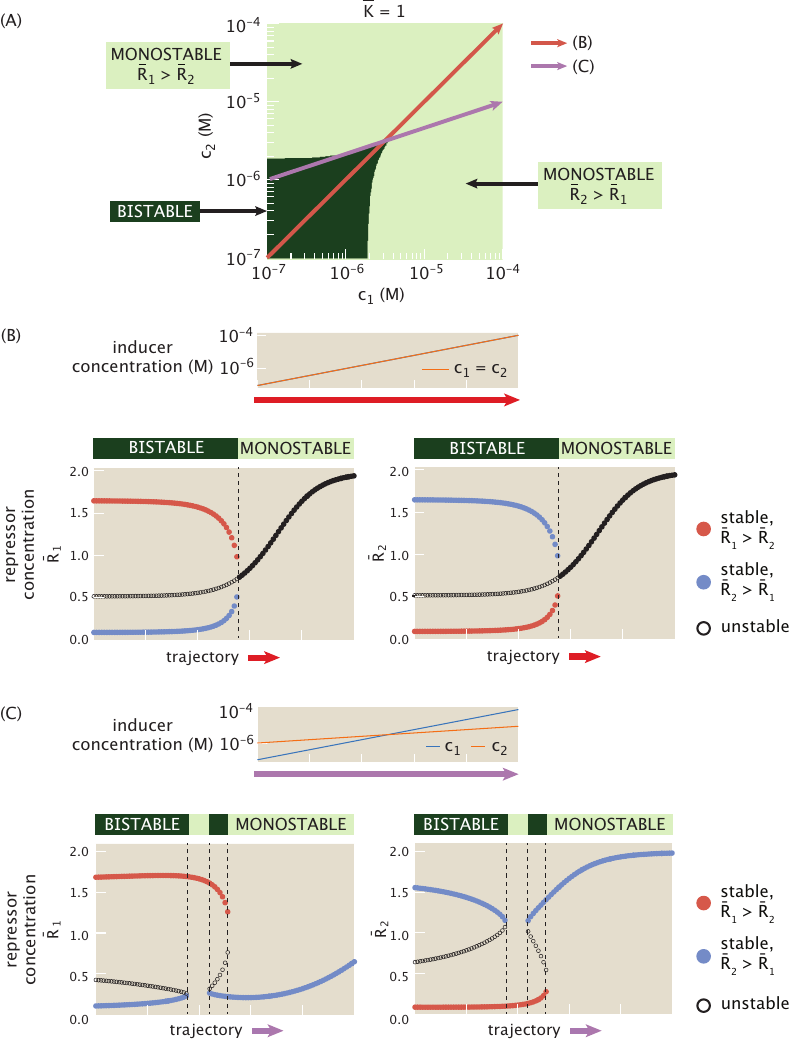}
    \caption{Phase diagram and bifurcation diagrams for mutual repression when each repressor's activity is regulated by an inducer (with fixed parameters $\bar{K} = K_1/K_2 = 1$, $\bar{r} = 2$ and $\omega_1 = \omega_2 = 7.5$.) (A) Phase diagram in which the dark green region denotes the inducer concentration regime for which the system has bistable dynamics. Trajectories $(B)$ and $(C)$ follow the shifts in dynamics moving through phase space along different paths, and are shown in the corresponding panels. (B) Bifurcation diagrams for $\bar{R}_{1}$ and $\bar{R}_{2}$ steady-state expression as the inducer concentrations controlling each repressor increase at the same rate. (C) Bifurcation diagrams as inducer concentrations increase at different rates.}
    \label{fig:mutual_repression}
\end{figure*}

This formulation now provides a two-dimensional knob for tuning the concentrations of the inducers corresponding to $\bar{R}_{1}$ and $\bar{R}_{2}$. The system may then be tuned to follow arbitrary trajectories in the two-dimensional parameter space spanned by $c_1$ and $c_2$. Frequently, this tuning generates non-trivial bifurcation curves.

To build intuition about this system, we first consider a scenario in which only one of the repressors may be induced. Fig.~\ref{fig:mutual_repression_singleinducer}(A) plots the resulting bifurcation diagrams for $\bar{R}_{1}$ and $\bar{R}_{2}$ steady-state expression. At low inducer concentrations, the system is bistable and can either evolve to favor $\bar{R}_{1}$ (red) or $\bar{R}_{2}$ (blue) expression. Fig.~\ref{fig:mutual_repression_singleinducer}(B) shows an example phase portrait within this low induction regime that depicts vector field flows through expression space toward these fixed points. At a sufficiently high inducer concentration, $\bar{R}_{1}$ expression can no longer be maintained, and the system transitions to a monostable regime in which only the state favoring $\bar{R}_{2}$ expression survives. A higher inducer concentration amplifies the expression of $\bar{R}_{2}$ up to its production limit. Fig.~\ref{fig:mutual_repression_singleinducer}(C) depicts a phase portrait in this regime.

\begin{figure*}
    \centering
    \includegraphics[width=0.7\linewidth]{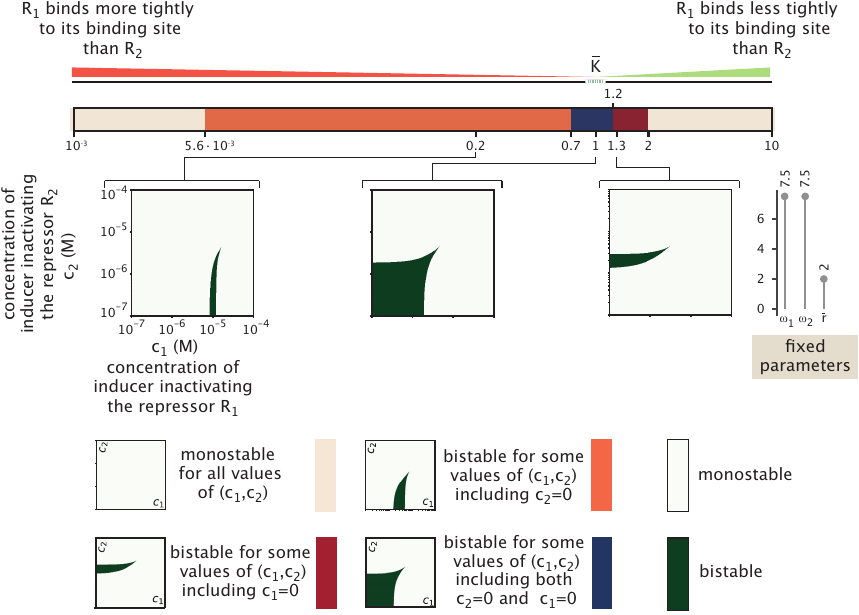}
    \caption{ Bistability regimes in mutual repression as a function of the relative DNA-binding affinities of repressors $R_1$ and $R_2$. Colored regions denote distinct bistable phase space geometries, defined by whether bistability occurs for a finite window of $c_{1}$, a finite window of $c_{2}$, or the presence of both at sufficiently small concentrations. Otherwise, the system is never bistable for any concentration $(c_1,c_2)$ in the interval $[10^{-7} \text{M},10^{-4} \text{M}]$ considered. These geometries evolve as the binding affinity ratio $\bar{K}=K_1/K_2$ is varied, with $\omega_1 = \omega_2 = 7.5$ and $\bar{r} = 2$ held constant.}
    \label{fig:color_bar_phase_space}
\end{figure*}

We now turn to a two-dimensional setting for allosteric regulation, with distinct inducers downregulating the activity of each repressor. Fig.~\ref{fig:mutual_repression}(A) plots the phase diagram for the dynamics of mutual repression at different combinations of inducer concentrations. The dark green region denotes the region of parameter space with bistable dynamics, with monostable behavior elsewhere. Note the symmetry of this phase diagram with respect to the diagonal $c_1=c_2$. This is unsurprising given the condition $\bar{K} = 1$ which amounts to saying that the two different repressors bind with the same  dissociation constant. This condition will be discussed further in the following subsection.

We can now consider how dynamics evolve as inducer concentrations change at different rates. 
Fig.~\ref{fig:mutual_repression}(A) considers different ``protocols'' for simultaneously varying the concentration of each inducer. For example, in Fig.~\ref{fig:mutual_repression}(B), we examine a protocol in which the inducer concentration for each gene increases at the same constant rate (corresponding to the red trajectory shown in Fig.~\ref{fig:mutual_repression}(A)). We then plot the corresponding bifurcation diagram. As inducer concentration increases, the scope of the bistable switch shrinks in expression space, with the stable states continuously approaching each other. At %$c_1 = c_2\approx 10^{-5.5}\:\text{M}$  
$c_1 = c_2\approx 3.2 \cdot 10^{-6}\:\text{M}$, the system then undergoes a pitchfork bifurcation to monostable expression, stabilizing at increasingly high concentrations of both $\bar{R}_{1}$ and $\bar{R}_{2}$.

We could follow an alternative trajectory (denoted by the purple arrow in Fig.~\ref{fig:mutual_repression}(A)) through parameter space such that the inducer concentrations evolve at different rates, in this case with $c_{1}$ increasing more rapidly than $c_{2}$. This purple trajectory then passes in and out of the green bistable region several times. Fig.~\ref{fig:mutual_repression}(C) plots the corresponding bifurcation diagram tracking stable and unstable steady states as the inducer concentrations increase, and demonstrates the switches between bistability and monostability. Note that while the intermediate monostable regime favors $\bar{R}_{1}$ expression, the monostable regime at later times favors $\bar{R}_{2}$ instead, reflecting the swap in the dominant inducer concentration that occurs between these time periods. Thus, by modulating the induction dynamics of each repressor we can access a broad range of dynamical responses in repressor concentrations.

\subsubsection{Conditions for bistability}

We now study how the size, shape, and symmetry of the bistable region observed in the phase diagram of Fig.~\ref{fig:mutual_repression}(A) varies with system parameters. Specifically, in Fig.~\ref{fig:color_bar_phase_space} we first identify three distinct geometries for the bistable region in the $(c_1, c_2)$ plane, each reflecting different limiting behaviors of the inducers. 

The first geometry (marked orange in the legend of Fig.~\ref{fig:color_bar_phase_space}) corresponds to a situation where bistability is present only when $c_1$ lies within a finite interval $[c_1^{\min}, c_1^{\max}]$ for a given $c_2$. In this case, both bounds of the interval are strictly positive and finite, and $c_2$ must be smaller than a certain threshold. The limiting factor is therefore $c_1$, which must be finely tuned to enable bistability, while $c_2$ simply needs to remain below a critical level. Nevertheless, when decreasing $c_2$ the bistable interval in $c_1$ broadens, showing that lower $c_2$ expands the range of $c_1$ values supporting bistability. 

A second geometry (marked red in the legend of Fig.~\ref{fig:color_bar_phase_space}) mirrors the first, but with the roles of $c_1$ and $c_2$ reversed. Here, bistability is present only when $c_2$ lies within a finite interval, while $c_1$ must remain below a threshold. In this case, $c_2$ becomes the more constrained parameter to tune. In contrast, a third geometry (marked blue in the legend of Fig.~\ref{fig:color_bar_phase_space}) arises when bistability is supported broadly for small enough values of both $c_1$ and $c_2$, with no lower bound required for either parameter. Although the bistable region remains upper-bounded, neither inducer is particularly limiting, with a broad range of concentrations allowing for bistability as long as neither becomes too large.

Fig.~\ref{fig:color_bar_phase_space} illustrates how these phase space geometries for bistability depend on system parameters. We vary the relative DNA-binding affinity of the repressors by tuning $\bar{K} = K_1 / K_2$, while keeping the basal production $\bar{r}$ fixed, and the cooperativities fixed and equal ($\omega_1 = \omega_2$). We observe that the most permissive bistable region—broad in both $c_1$ and $c_2$, for the range of concentration studied—occurs when $\bar{K} \approx 1$, corresponding to a symmetric system where both repressors bind with comparable affinity. As $\bar{K}$ decreases (i.e., $R_1$ binds more tightly than $R_2$), the phase space becomes increasingly constrained in $c_1$. If $c_1$ is too low, $R_1$ remains fully bound and strongly represses $R_2$, suppressing bistability. Conversely, if $c_1$ is too high, $R_1$ becomes fully unbound, leaving $R_2$ unrepressed and again eliminating bistability. Only an intermediate range of $c_1$ supports bistability in this regime, while $c_2$ simply needs to be small enough. A similar scenario occurs when $\bar{K}$ becomes large (i.e., $R_2$ binds more tightly than $R_1$), but with $c_1$ and $c_2$ effectively reversed. Eventually, $c_2$ is no longer sufficient to counteract the tight binding of $R_2$, and above a critical value of $\bar{K}$, bistability disappears entirely from the parameter space.
Tuning the cooperativity parameter $\omega_2$ produces similar qualitative changes in the bistability phase space as varying the relative binding affinity $\bar{K}$. 

In both cases, we observe the same sequence of transitions in the structure of the bistable region, as shown in Appendix~\ref{appendix:mutrep_colorbar_coop}. High values of $\omega_2$ reflect strong cooperative binding of repressor $R_2$ to the DNA, meaning that binding becomes more favorable when two repressors are present. This effect mirrors what happens when increasing $\bar{K}$: if $\bar{K} > 1$, then $K_1 > K_2$, implying that higher concentrations of $R_1$ are required for effective DNA binding compared to $R_2$. As a result, increasing $\bar{K}$ effectively enhances the influence of $R_2$, analogous to increasing $\omega_2$. Conversely, tuning $\omega_1$ affects the system similarly, but with the roles of $c_1$ and $c_2$ reversed. The effects of cooperativity are examined in greater detail in Appendix~\ref{appendix:mutrep_colorbar_coop}.

\begin{figure}
    \centering
    \includegraphics[width=\linewidth]{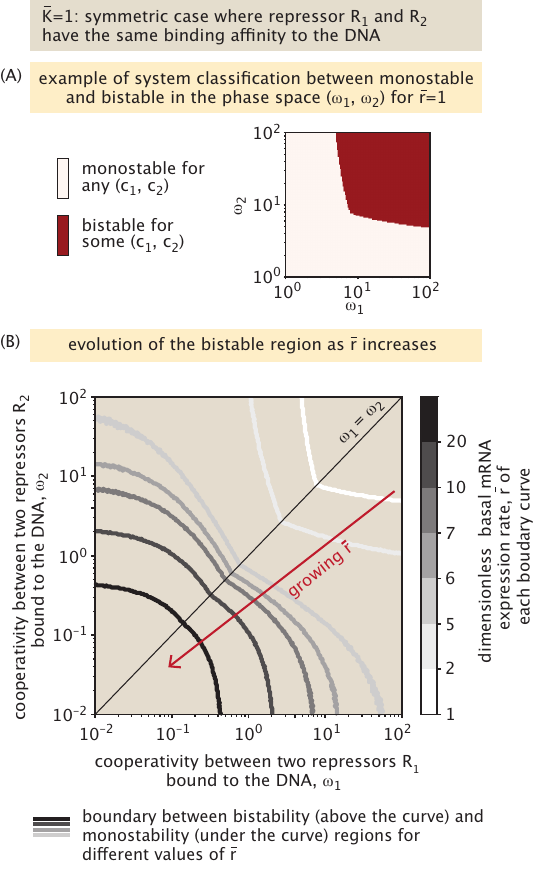}
    \caption{Bistability regions in mutual repression as a function of cooperativity and production rate. In both panels, the two repressors have equal DNA-binding affinities, with $\bar{K} = 1$. We determine the boundary in $(\omega_1, \omega_2)$ space that separates regions where bistability is possible for some $(c_1, c_2)$ from those where it is not possible. (A) Example for $\bar{r}=1$. (B) Boundary curves in the $(\omega_1,\omega_2)$ plane showing how the region of possible bistability expands as $\bar r$ increases. Each curve corresponds to fixed $\bar r$ values and separates parameter combinations that are monostable for all $(c_1,c_2)$ (below the curve) from combinations that are bistable for some $(c_1,c_2)$ (above the curve). Colors indicate the value of $\bar r$ used to compute each boundary curve.}
\label{fig:mutrep_bistability_region_different_r}
\end{figure}

From Fig.~\ref{fig:color_bar_phase_space}, we note that extreme (low or high) values of $\bar{K}$ tend to suppress bistability, as they strongly favor one repressor over the other across all concentrations. In contrast, high values of $\omega_1$ or $\omega_2$ amplify repression mostly when the corresponding repressor is present at high concentration. As a result, the system requires finely tuned inducer concentrations to counteract this cooperative imbalance and sustain bistability, effectively constraining the range of inducer concentrations for which the system can be bistable---as shown in Appendix \ref{appendix:mutrep_colorbar_coop}.

We next explore how the interplay between cooperativity and production rate controls the presence and extent of bistability in mutual repression systems, focusing on the symmetric case $\bar{K}=1$ shown in Fig.~\ref{fig:mutrep_bistability_region_different_r}. We note that, in contrast to parameters like $\omega_1$, $\omega_2$, or $K$, tuning the rate parameter $\bar{r}$ does not break the symmetry between the two genes, as it controls the production rate of both repressors equally. 

In Fig.~\ref{fig:mutrep_bistability_region_different_r}(A), we classify parameter combinations in the $(\omega_1, \omega_2)$ plane for $\bar{r}=1$ according to whether the system exhibits bistability for any inducer concentrations $(c_1, c_2)$. Fig.~\ref{fig:mutrep_bistability_region_different_r}(B) shows how the boundary between monostable and bistable regimes shifts with $\bar{r}$, separating regions where bistability is either inaccessible or achievable for at least some inducer pairs. We observe that the production rate, when coupled to cooperativity, plays a critical role in enabling bistability—much like in the auto-activation system, where the product $\omega \bar{r}_2$ must exceed a threshold to generate bistability. We quantify this cooperative relationship between $\omega_i$ and $\bar{r}$ further in Appendix~\ref{appendix:mutrep_bistab_bounds} by deriving a necessary condition for bistability where
\begin{equation}
    \bar{r}>\frac{1}{p^{\text{max}}_{\text{act}}-p^{\text{min}}_{\text{act}}+\min(\omega_2,\omega_1)p^{\text{max}}_{\text{act}}/2}\label{eq:necessary_condition_max_w1w2final_maintext},
\end{equation}
with $p^{\text{max}}_{\text{act}}$ and $p^{\text{min}}_{\text{act}}$ defined in Eqns.~\ref{eq:pact_limit1} and~\ref{eq:pact_limit2}.

Increasing $\bar{r}$ systematically expands the range of cooperativity values that can support bistability in the range of $\bar{r}$ swept in Fig.~\ref{fig:mutrep_bistability_region_different_r}. We observe from the figure that at low production rates, bistability only arises when both repressors exhibit stabilizing cooperative binding to DNA ($\omega_i > 1$). As $\bar{r}$ increases, this constraint relaxes: bistability becomes possible even without cooperativity ($\omega_1 = \omega_2 = 1$), and for sufficiently high $\bar{r}$, bistability can occur even in the presence of destabilizing interactions between the two repressors ($\omega_i < 1$). Intermediate production rates typically require at least one positively cooperative repressor. 

In auto-activation systems, the effective Hill coefficient can vary above or below one, but bistability only occurs when it exceeds one. In contrast, for mutual repression, the effective Hill coefficients for $R_1$ and $R_2$ vary with parameters but remain strictly greater than one when $\omega_1, \omega_2 > 0$ (Appendix~\ref{appendix:mutrep_bistab_bounds}), making the Hill coefficient less informative about the existence of bistability. Even in models using empirical Hill functions to describe the production terms, Hill coefficients above one are not always sufficient for bistability, and computational studies show that extended network interactions can yield bistability even with coefficients below one~\cite{pal2016functional}.

\subsubsection{Timescale for stabilization}

\begin{figure*}
    \centering
    \includegraphics[width=0.7\linewidth]{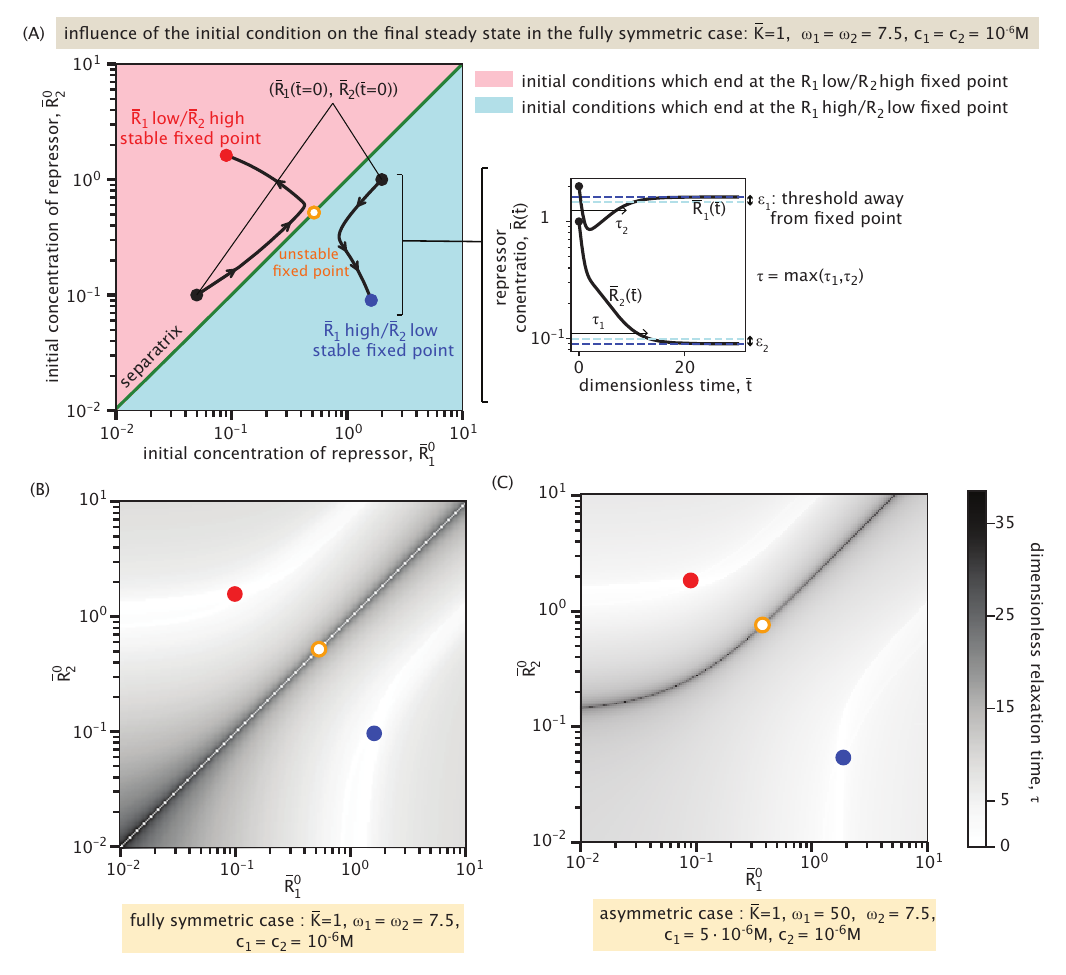}
    \caption{Dynamics and relaxation time in a mutual repression system.
(A) Influence of the initial condition on the final steady state for a symmetric case ($\bar{K}=1$, $\omega_1 = \omega_2 = 7.5$, $c_1 = c_2 = 10^{-6} \ \text{M}$). Two stable steady states and one unstable steady state are shown. The colored regions correspond to sets of initial conditions that converge to each respective stable steady state. The green curve separating these regions is the separatrix. Dimensionless relaxation times $\tau_1$ and $\tau_2$ are defined as the time required for the system to approach within thresholds $\epsilon_1$ and $\epsilon_2$ of the stable fixed points, where these thresholds correspond to $95\%$ of the respective steady-state values. The global relaxation time of the system, denoted $\tau$, is then defined as the maximum of the two: $\tau = \max(\tau_1, \tau_2)$. (B-C) Dimensionless relaxation time as a function of initial concentrations of repressors $R_1$ and $R_2$. Panel (B) corresponds to the symmetric case, whereas panel (C) plots an asymmetric case ($K=1$, $\omega_1 = 50$, $\omega_2 = 7.5$, $c_1 = 5 \cdot 10^{-6}\ \text{M}$, $c_2 =10^{-6}\ \text{M}$).}
    \label{fig:mutrep_dynamics}
\end{figure*}
Beyond identifying the final steady state of the mutual repression system, it is important to characterize the time required for the system to reach steady state starting from different initial conditions. We define the relaxation time $\tau$ as the maximum of the times $\tau_1$ and $\tau_2$ taken by the two trajectories, $\bar{R}_1(\bar{t})$ and $\bar{R}_2(\bar{t})$, to reach $95\%$ of their respective steady states. Fig.~\ref{fig:mutrep_dynamics}(A) illustrates the influence of the initial condition on the final steady state in a symmetric system, where $\bar{K} = 1$, $\omega_1 = \omega_2 = 7.5$, and $c_1 = c_2 = 10^{-6}\ \text{M}$. The figure reveals that the phase space is divided into two basins of attraction, each leading to one of the two stable steady states. The separatrix, plotted in green, denotes the boundary between these basins, and is derived analytically in Appendix~\ref{appendix:sep_mut_rep}.

Fig.~\ref{fig:mutrep_dynamics}(B) and (C) quantify the relaxation time $\tau$ across the phase space $(\bar{R}_1^0,\bar{R}_2^0)$ for both symmetric and asymmetric parameter regimes. In both cases, the relaxation time is shorter when the initial condition lies far from the separatrix and near the final stable steady state. In contrast, initial conditions close to the separatrix result in significantly longer relaxation times, as the system evolves slowly near the unstable fixed point before diverging toward a stable state. Exactly on the separatrix, the relaxation to steady state is more difficult to quantify because the system may remain indefinitely near an unstable manifold without converging to either stable fixed point. In practice, however, even minimal noise in a real system will eventually drive the system away from this unstable region toward a stable state. For this reason, we disregard the white line—signifying artificially short relaxation times—observed exactly on the separatrix in Fig.~\ref{fig:mutrep_dynamics}(B). This feature is absent in Fig.~\ref{fig:mutrep_dynamics}(C), as the separatrix has a more complex shape and the numerical sweep over initial conditions does not sample it precisely.

In the asymmetric case (Fig.~\ref{fig:mutrep_dynamics}(C)), with $\omega_1 = 50$, $\omega_2 = 7.5$, and distinct inducer concentrations ($c_1 = 5 \cdot 10^{-6}\ \text{M}$, $c_2 = 10^{-6}\ \text{M}$), the phase space becomes skewed. The separatrix, corresponding to the ridge of maximal relaxation time in the grayscale colormap, delineates the boundary between the two basins of attraction. We do not overlay it explicitly, as doing so would interfere with the visualization of the relaxation times, which are particularly sensitive near this boundary. Compared to Fig.~\ref{fig:mutrep_dynamics}(B), where the separatrix coincides with the diagonal $\bar{R}_1^0 = \bar{R}_2^0$ due to the symmetry of the system, we observe that the separatrix is now deformed, and the relative sizes of the basins of attraction have shifted.  Despite these geometric changes, the maximal relaxation times across both regimes remain comparable. This indicates that, although asymmetry reshapes the phase space and can affect which attractor is reached, it does not substantially alter the overall timescale required for the system to stabilize. The main contribution to long relaxation times remains the initial conditions--and how close they are to the separatrix--regardless of symmetry.

This whole section has had as its primary ambition to carefully consider the famed mutual repression genetic switch from the new perspective in which the two repressors are controlled separately by different effector molecules.  We have seen that the steady-states and the  dynamics in this case are extremely rich, making it clear that there is much freedom in the biological context to exploit different kinds of behavior.

\section{\label{sec:FFL} Kinetics and Time Delays in Feed-Forward Loops}

\begin{figure*}
    \centering
    \includegraphics[width=\linewidth]{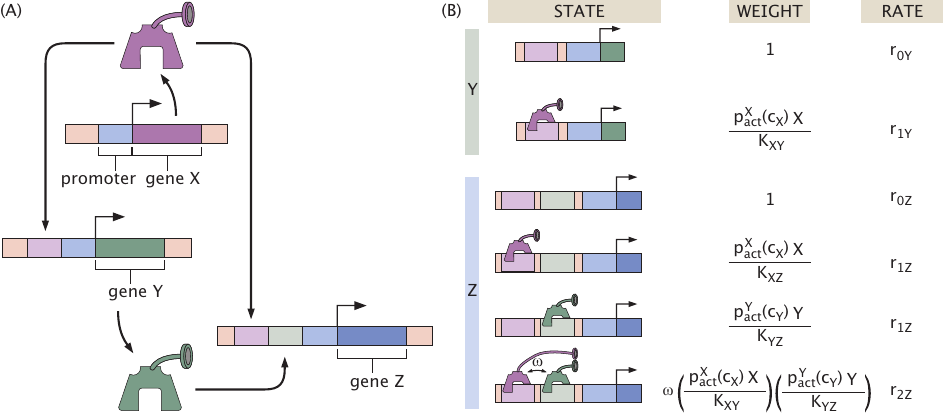}
    \caption{The coherent feed-forward loop. (A) Schematic representation of the coherent feed-forward loop. Expression of output protein $Z$ is controlled by expression of protein $X$, either by direct activation or indirectly, first activating expression of $Y$ which in turn activates $Z$. The regulatory circuit is coherent because both pathways have the same activating effect on $Z$. (B) Thermodynamic states, weights, and rates for expression of activator $Y$ and output protein $Z$. Note that the model assumes both $X$ and $Y$ can bind together to activate expression at their respective binding sites, interacting with cooperativity $\omega$.}
    \label{fig:FFLcoherent}
\end{figure*}

In this section, we consider gene circuits whose functionality appears in their dynamics rather than in their steady state responses. Specifically, we focus on a three-gene circuit called the feed-forward loop shown schematically in Fig.~\ref{fig:FFLcoherent}(A) \cite{mangan2003structure}. Here, we  denote the input genes as $X$ and $Y$, and the output gene as $Z$. In a feed-forward loop, $X$ regulates $Y$, and $X$ and $Y$ together regulate $Z$. $X$ thus controls expression of output $Z$ through both direct and indirect regulatory paths. This network typically features a sign-sensitive delayed or accelerated response (depending on architecture) to a step-wise change in the effector concentration for protein $X$ \cite{mangan2003structure}. That is, while the qualitative nature of the response (delay or acceleration) remains fixed, its magnitude depends on the sign (an increase or a decrease) of the input change. This delayed or accelerated response has been hypothesized to have important biological consequences, particularly in information-processing systems that filter noisy inputs~\cite{mangan2003structure}. Beyond their role in shaping temporal responses, coherent feed-forward loops have been shown to attenuate input noise, thereby enhancing the reliability of gene expression~\cite{fflnoise_chakravarty2021systematic,fflnoise_gui2016noise,fflsoie_ghosh2005noise}.

Interestingly, there are various architectures of feed-forward loop depending on whether $X$ and $Y$ work together or at cross purposes. We will largely focus on the particular architecture where $X$ activates $Y$ and $Z$, and $Y$ also activates $Z$, which is the so-called type I coherent feed-forward loop, with the word ``coherent'' attached to this architecture since $X$ and $Y$ alter the expression of $Z$ in a coherent manner.  We consider this particular motif primarily because at the time of the most recent census of regulatory architectures in {\it E. coli}, this version of the feed-forward loop appeared the most frequently~\cite{Milo2002, shen2002network}. The logic of our analysis  can be applied to any of the other feed-forward architectures as well.   

Previous literature explores feed-forward loops from a dynamical systems perspective using Hill functions to model transcription factor-DNA interactions and considering the effector concentration for $X$ as a Boolean variable that is either fully on or fully off~\cite{mangan2003structure}. We build upon that earlier analysis also by making a systematic search for network parameters that gives rise to various functions. Our goal is to expand the theoretical understanding of the feed-forward loop architecture by incorporating the thermodynamic model to describe transcription factor binding to the DNA  and to explicitly include effector function. Specifically, we explore what gives rise to the dynamical features of the feed-forward loop, the robustness of such features, and the effect of continuously tuning the effector concentration.

As usual when writing the dynamical equations, we begin with the states, weights and rates for the regulatory architecture of interest. Fig.~\ref{fig:FFLcoherent}(B) provides the states, weights and rates for the coherent feed-forward loop architecture, where we assume one binding site per transcription factor. In light of the states and weights, we can write the time-evolution equations for the coherent feed-forward loop as 
\begin{equation}
    \frac{dY}{dt} = -\gamma Y + \frac{r_{0Y} + r_{1Y} \, p_\text{act}^X(c_X)\frac{X}{K_{XY}}}{1+p_\text{act}^X(c_X)\frac{X}{K_{XY}}}
\end{equation}    
for the regulation of Y by X and
\begin{widetext}
\begin{equation}
    \frac{dZ}{dt} = -\gamma Z + \frac{
 r_{0Z} + r_{1Z} \left(p_\text{act}^X(c_X)\frac{X}{K_{XZ}} + p_\text{act}^Y(c_Y)\frac{Y}{K_{YZ}}\right)
          \textstyle + r_{2Z} \, \omega \, p_\text{act}^X(c_X) \frac{X}{K_{XZ}}p_\text{act}^Y(c_Y)\frac{Y}{K_{YZ}}}
{%
{            1+p_\text{act}^X(c_X)\frac{X}{K_{XZ}} + p_\text{act}^Y(c_Y)\frac{Y}{K_{YZ}}}
          {+ \omega \, p_\text{act}^X(c_X)\frac{X}{K_{XZ}} p_\text{act}^Y(c_Y)\frac{Y}{K_{YZ}}}}
    \label{eq:fedForCoh}
\end{equation}
\end{widetext}
for the regulation of Z by both X and Y. We assume here that the proteins $Y$ and $Z$ have the same degradation rate $\gamma$. The production rates and dissociation constants are assumed to be different in general for each thermodynamic state. Specifically, $r_{iG}$ denotes the production rate of gene $G$ when $i$ number of transcription factor are bound. Further, $K_{G_1G_2}$ denotes the dissociation constant of transcription factor $G_1$ binding to gene $G_2$. $\omega$ is the cooperativity, which takes into account the extra interaction energy between $X$ and $Y$ when bound to the DNA. Finally, the probabilities $p_\text{act}^X(c_X)$ and $p_\text{act}^Y(c_Y)$ scale the activity of transcription factors $X$ and $Y$.  Note also that we consider different effectors $c_X$ and $c_Y$ for the two genes that can be varied independently. These functions may be distinct in principle, and the analytic discussions here make no assumption regarding their nature. For numerical results, however, we assume these probability functions, and thus effector activity functions, to be the same regardless of target transcription factor for simplicity.

We non-dimensionalize the equations by using $1/\gamma$ as our unit of time and $K_{XY}$ as our measure of concentration. In light of these conventions, we arrive at
\begin{equation}\label{eq:coh_ffl_nondim1}
    \frac{d\bar{Y}}{d\bar{t}} = - \bar{Y} + \frac{\bar{r}_{0Y} + \bar{r}_{1Y} \, p_\text{act}^X(c_X)\bar{X}}{1+p_\text{act}^X(c_X)\bar{X}},
\end{equation}
and
\begin{equation}\label{eq:coh_ffl_nondim2}
        \frac{d\bar{Z}}{d\bar{t}} = - \bar{Z} + \frac{\bar{r}_{0Z} + \bar{r}_{1Z} (\mathcal{X} + \mathcal{Y}) + \omega \bar{r}_{2Z} \mathcal{X}\mathcal{Y}}{1+\mathcal{X} +\mathcal{Y} + \omega \mathcal{X} \mathcal{Y}}.
\end{equation}
To make subsequent analysis less cumbersome, we further introduce $\displaystyle \mathcal{X}=p_\text{act}^X(c_X)\bar{X}/\bar{K}_{XZ}$ and $\displaystyle \mathcal{Y}=p_\text{act}^Y(c_Y)\bar{Y}/\bar{K}_{YZ}$ as simpler notation for the effective regulatory contributions of $\bar{X}$ and $\bar{Y}$ to $\bar{Z}$ expression. The bar indicates quantities where time is measured in units of $1/\gamma$, and where concentration and dissociation constants are measured in units of $K_{XY}$. Specifically, we define dimensionless dissociation constants $\bar{K}_{XZ}=K_{XZ}/K_{XY}$ and $\bar{K}_{YZ}=K_{YZ}/K_{XY}$. The rates are then in units of $\gamma K_{XY}$. Note that this model fixes the concentration of $\bar{X}$ for simplicity, such that its activity regulating $\bar{Y}$ and $\bar{Z}$ depends entirely on effector concentration $c_{X}$.

\subsection{Characterizing delay responses in coherent feed-forward loops}
\label{sec:ffl_charac_delay}
 
Previous work has shown that coherent feed-forward loops can delay a system's response to an input signal~\cite{mangan2003structure}. Here, we demonstrate from our thermodynamic modeling perspective the analytic origins of this delay. In particular, we rigorously define how the introduction of an indirect but coherent path for regulation of output $\bar{Z}$ affects its response.

Suppose that we keep the effector concentration $c_{Y}$ fixed, and that at time $\bar{t} = 0$ the effector concentration $c_{X}$ changes sharply from an initial concentration $c_{X}^{i}$ to a final concentration $c_{X}^{f}$ such that input $\bar{X}$ activity either increases (an ``ON'' step) or decreases (an ``OFF'' step). This then means that
\begin{equation}
    \mathcal{X} (\bar{t})= \begin{cases}
    \begin{aligned}
        \mathcal{X}_{i} = \frac{p_\text{act}^X(c_{X}^{i})\bar{X}}{\bar{K}_{XZ}} \quad & \text{if} \quad \bar{t}\leq 0,\\
        \mathcal{X}_{f} = \frac{p_\text{act}^X(c_{X}^{f})\bar{X}}{\bar{K}_{XZ}} \quad & \text{if} \quad \bar{t} > 0. \label{eq:mathcalXdef}
    \end{aligned}
    \end{cases}
\end{equation}
In the coherent feed-forward loop, since all the regulatory relations are activation, both $\bar{Y}$ and $\bar{Z}$ increase in response to an ON step, and decrease in response to an OFF step.

To understand intuitively how exactly the feed-forward loop regulatory structure responds to such a switch in input signal, let us first consider a simpler scheme in which $\bar{X}$ no longer regulates $\bar{Y}$, leaving $\bar{X}$ and $\bar{Y}$ to regulate $\bar{Z}$ independently with $\bar{Y}(\bar{t}) = \bar{Y}$ fixed at a value. Since we are keeping effector concentration $c_{Y}$ fixed constant, this then also means that $\mathcal{Y}$ is constant. We refer to this setting as ``simple regulation.'' 

Before the switch in $c_X$, the simple regulation system is at steady state, with initial output expression $\bar{Z}_s^i$ where subscript $s$ denotes simple regulation. After the switch, the system relaxes to a new steady state with final expression $\bar{Z}_s^f$. For $\bar{t} > 0$, the production term in Eqn.~\ref{eq:coh_ffl_nondim2} is simply a constant, yielding the following differential equation for $\bar{Z}$.
\begin{eqnarray}
    \frac{d\bar{Z}}{d\bar{t}} &=& - \bar{Z} + \frac{\bar{r}_{0Z} + \bar{r}_{1Z} (\mathcal{X}_{f} + \mathcal{Y}) + \omega \, \bar{r}_{2Z} \, \mathcal{X}_{f}\mathcal{Y}}{1+\mathcal{X}_{f} +\mathcal{Y} + \omega \mathcal{X}_{f} \mathcal{Y}} \nonumber\\
    &\equiv& -\bar{Z} + \bar{Z}_{s}^f.\label{eq:simpleregZ}
\end{eqnarray}
Integrating Eqn.~\ref{eq:simpleregZ}, we thus determine that under simple regulation, output $\bar{Z}$ evolves after the switch in input $\bar{X}$ signal by a standard exponential behavior defined as
\begin{equation}
    \bar{Z}_{s}(\bar{t}) = \bar{Z}_{s}^ie^{-\bar{t}} + \bar{Z}_{s}^f(1-e^{-\bar{t}}) \label{eq:simpleregZsol}.
\end{equation}
\begin{figure}
    \centering
    \includegraphics[width=\linewidth]{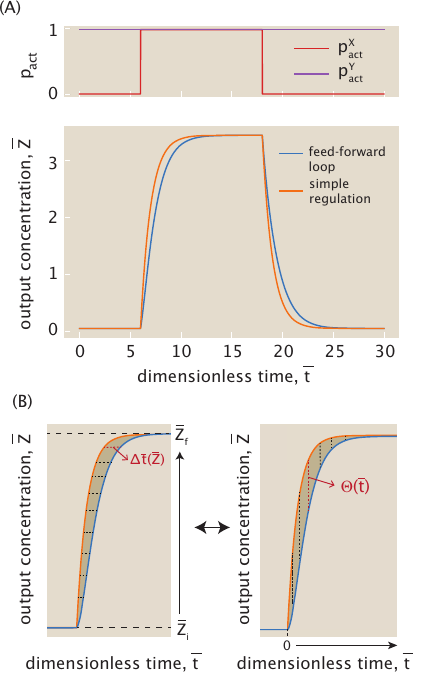}
    \caption{Delay in coherent feed-forward loop response compared to simple regulation. (A) The input signal is applied by tuning $c_X$ as a step function: from high ($10^{-4}$ M) to low ($10^{-7}$ M) at $\bar{t} = 6$, and back to high at $\bar{t} = 18$. The effector concentration $c_Y$ is held constant at $10^{-7}$ M. These inputs determine the activation probabilities $p_{\text{act}}^X$ and $p_{\text{act}}^Y$, shown in red and purple, respectively. The second panel plots the time evolution of the dimensionless output concentration $\bar{Z}(\bar{t})$ under feed-forward and simple regulation schemes, with $\bar{r}_{0Y} = \bar{r}_{0Z} = 0$, $\bar{r}_{1Y} = \bar{r}_{1Z} = 2$, $\bar{r}_{2Z} = 10$, $\omega = 1$, and $\bar{K}_{XZ} = \bar{K}_{YZ} = 1$. (B) Schematic demonstrating the two ways to equivalently quantify the delayed response of the feed-forward loop compared to simple regulation, captured by the shaded area between the two curves. One can either integrate over individual time delay measurements $\Delta\bar{t}(\bar{Z})$ as a function of $\bar{Z}$, or equivalently integrate the difference in responses $\Theta(\bar{t})$ as a function of $\bar{t}$.}
    \label{fig:FFL1}
\end{figure}

By contrast, in the coherent feed-forward loop, the concentration of $\bar{Y}$ directly depends on $\bar{X}$ as seen in Eqn.~\ref{eq:coh_ffl_nondim1}. As a result, additional time is needed for $\bar{Y}$ to evolve from its initial steady-state value $\bar{Y}_i$ to a new value $\bar{Y}_f$ following a change in $\bar{X}$.  Expressed mathematically, since $\bar{Y}$ itself is a function of time, the solution of $\bar{Z}$ to Eqn.~\ref{eq:coh_ffl_nondim2} is not strictly an exponential relaxation to steady state. More formally, output expression for the coherent feed-forward loop evolves by a function of the form
\begin{equation}
    \bar{Z}(\bar{t}) = \bar{Z}_{i}e^{-t} + \bar{Z}_{f}(1-e^{-t}) + \Theta(\bar{t}),\label{eq:ZFFL}
\end{equation}
where $\bar{Z}_i$ is the initial steady state in the feed-forward setting before the change in $c_X$, and $\bar{Z}_f$ is the final steady state after the change that $\bar{Z}(\bar{t})$ relaxes to eventually.
We derive Eqn.~\ref{eq:ZFFL} explicitly from Eqns.~\ref{eq:coh_ffl_nondim1} and~\ref{eq:coh_ffl_nondim2} in Appendix~\ref{appendix: step_func_signal}. Note that the sum of the first two terms in Eqn.~\ref{eq:ZFFL} describes behavior of the same form as simple regulation in Eqn.~\ref{eq:simpleregZsol}. Therefore, by rescaling Eqn.~\ref{eq:simpleregZsol} we can treat the exponential portion of Eqn.~\ref{eq:ZFFL} as an equivalent $\bar{Z}_{\text{simple}}$ with the same relaxation dynamics as observed for simple regulation. We can also then express output response for the feed-forward loop as
\begin{equation}
    \bar{Z}(\bar{t}) = \bar{Z}_{\text{simple}}(\bar{t}) + \Theta(\bar{t}).\label{eq:ZFFLfin}
\end{equation}
We thus observe that the feed-forward loop's output response differs from behavior in simple regulation by a function $\Theta(\bar{t})$. Analytically solving Eqns.~\ref{eq:coh_ffl_nondim1} and~\ref{eq:coh_ffl_nondim2} results in
\begin{align}
    \Theta(\bar{t}) = -\frac{\Phi \Delta \mathcal{Y}}{S^2} e^{-\bar{t}} \log \left( \frac{S e^{\bar{t}} - \Delta \mathcal{Y} (1 + \omega \mathcal{X}_{f})}{S - \Delta \mathcal{Y} (1 + \omega \mathcal{X}_{f})} \right).
    \label{eq:ffl_coh_full_sol}
\end{align}
Here, $\Theta(\bar{t})$ depends on the three quantities $\Delta\mathcal{Y}$, $\Phi$, and $S$. First,
the quantity
\begin{equation}
    \Delta\mathcal{Y} = \mathcal{Y}_{f} - \mathcal{Y}_{i} = \frac{p_\text{act}^Y(c_{Y})}{\bar{K}_{YZ}}\Big(\bar{Y}_{f} - \bar{Y}_{i}\Big)
\end{equation}
denotes the total change in quantity $\mathcal{Y}$ in response to the change in input effector concentration $c_{X}$. $\Delta\mathcal{Y}$ thus contains implicit information about regulation of $\bar{Y}$ by $\bar{X}$ from its evolution as defined in Eqn.~\ref{eq:coh_ffl_nondim1}. $\Theta(\bar{t})$ also depends on the coefficient
\begin{eqnarray}
    \Phi &=& \:\: \omega (\mathcal{X}_{f})^2 (\bar{r}_{2Z} - \bar{r}_{1Z}) + \omega \mathcal{X}_{f} (\bar{r}_{2Z} - \bar{r}_{0Z})\nonumber\\
&\quad& + (\bar{r}_{1Z} - \bar{r}_{0Z}),
\end{eqnarray}
which encodes how the rates and cooperativity regulate output $\bar{Z}$ expression as a function of input signal $\mathcal{X}_{f}$. Interestingly, we note that all types of feed-forward loops have the same solution as Eqn.~\ref{eq:ffl_coh_full_sol}, except with a potentially different $\Phi$. Appendix~\ref{appendix:ffl_other_networks} discusses this in more detail.
Finally, the quantity
\begin{equation}
    S = 1 + \mathcal{X}_{f} + \mathcal{Y}_{f} + \omega \mathcal{X}_{f} \mathcal{Y}_{f}
\label{eq:coh_full_sol2}
\end{equation}
is the sum of the dimensionless weights for all possible regulatory states with zero, one, or two transcription factors bound, at final concentrations of active $\bar{X}$ and $\bar{Y}$.

Fig.~\ref{fig:FFL1}(A) highlights the coherent feed-forward loop's delayed response to changes in input signal for a specific set of parameters. In our simulations, we set the rates $\bar{r}_{0G}$ to zero, indicating that this system requires activator(s) to be bound to express output $\bar{Z}$. In the top plot of Fig.~\ref{fig:FFL1}(A), the effector concentration $c_Y$ is fixed such that $p_\text{act}^Y(c_Y)$ is always at high activity, and the effector concentration $c_{X}$ jumps such that $p_\text{act}^X(c_X)$ reaches high activity as a step function (ON step), and then back to low activity (OFF step). The bottom plot of Fig.~\ref{fig:FFL1}(A) shows how the output $\bar{Z}$ evolves over time in response to a step function input $\bar{X}$ activity, evolving and stabilizing to a high $\bar{Z}$ value before the OFF step in $p_\text{act}^X(c_X)$ causes the output concentration to decay back down to steady state value zero. We observe that, compared to simple regulation, the change in output concentration $\bar{Z}$ is slower when responding to both an increase and a decrease in input activity, matching our expectation.

Fig.~\ref{fig:FFL1}(B) visualizes how to quantify this delay in the time it takes the feed-forward system to reach a given output concentration as it responds to an input pulse. The diagram on the left illustrates the response to the ON step, in which the output starts to increase from $\bar{Z}_{i}$ to $\bar{Z}_{f}$ at (dimensionless) time $\bar{t} = 0$. If we choose a value of $\bar{Z}$ in this time frame, we observe that it takes longer to reach this value on its way to steady state $\bar{Z}_{f}$ in the feed-forward loop setting than in simple regulation. We highlight one such horizontal distance between the two curves as the time delay $\Delta\bar{t}(\bar{Z})$. Explicitly, we define $\Delta \bar{t}(\bar{Z})$ to be the difference between the time it takes for simple regulation to reach a given $\bar{Z}$ and that for feed-forward loop. $\Delta \bar{t}(\bar{Z}) < 0$ signifies a delay and $\Delta \bar{t}(\bar{Z}) > 0$ signifies acceleration. 
We can then compute the average time difference observed between the two curves by integrating $\Delta\bar{t}(\bar{Z})$ over the range of output $\bar{Z}$, and normalizing by this range. Therefore, for a given step function change in input $\bar{X}$ activity, the average delay during the system's evolution toward its new final steady state is
\begin{align}
    \langle \Delta\bar{t} \rangle = \frac{1}{|\bar{Z}_{f} - \bar{Z}_{i}|} \int_{\bar{Z}_{i}}^{\bar{Z}_{f}} \Delta \bar{t} (\bar{Z}) d\bar{Z}.
    \label{eq:time_delay_physical}
\end{align}
Note, however, that it is not straightforward to derive an expression for $\Delta\bar{t}(\bar{Z})$. Instead, since this integral geometrically captures the area between the two curves, we can equivalently evaluate this area as shown in the second diagram of Fig.~\ref{fig:FFL1}(B) by integrating vertical slices through this shaded region. At a given time $\bar{t}$, the vertical dotted line corresponds to the difference in output response between the two curves, defined by the function $\Theta(\bar{t})$ previously derived in Eqn.~\ref{eq:ffl_coh_full_sol}.

From this description, we can then also derive the average time delay from the offset $\Theta(\bar{t})$, and thus arrive at the equivalent definition
\begin{equation}
    \langle \Delta\bar{t} \rangle = \:\frac{1}{\bar{Z}_{f}-\bar{Z}_{i}}\int_{0}^\infty\Theta(\bar{t})d\bar{t}.
    \label{eq:time_delay}
\end{equation}
Notice that the absolute value on $\bar{Z}_f - \bar{Z}_i$ is dropped to match the sign of $\langle \Delta\bar{t} \rangle$ in Eqn.~\ref{eq:time_delay_physical}.
Substituting $\Theta(\bar{t})$ from Eqn.~\ref{eq:ffl_coh_full_sol} and evaluating the integral, the average time delay becomes
\begin{align}
    \langle\Delta \bar{t} \rangle
    = \frac{\Phi(\bar{Z}_{f} - \bar{Z}_{i}) ^{-1}}{S(1+\omega \mathcal{X}_{f})} \log \Big( \frac{1 + \mathcal{X}_{f} + \mathcal{Y}_{i}+\omega \mathcal{X}_{f} \mathcal{Y}_{i}}{S}\Big).
    \label{eq:area_between_curves}
\end{align}
This result highlights that $\langle\Delta \bar{t}\rangle$ depends on both the initial state and the final state of $\mathcal{X}$ and $\mathcal{Y}$. In fact, it is this dual dependence that causes the delays in response to ON and OFF steps to differ in Fig.~\ref{fig:FFL1}(A). Switching from ON to OFF and vice versa simply swaps the initial and final expression states of $\mathcal{X}$ and $\mathcal{Y}$. Applying these transformations $\mathcal{X}_{i} \leftrightarrow \mathcal{X}_{f}$ and $\mathcal{Y}_{i} \leftrightarrow \mathcal{Y}_{f}$ in Eqn.~\ref{eq:area_between_curves} shows that the magnitude of the average delay $\langle\Delta \bar{t}\rangle$ is in general not conserved under the exchange of initial and final states. It is therefore this asymmetric dependence on initial and final states that directly leads to differences in output responses to ON and OFF steps in the coherent feed-forward loop. 

From Eqn.~\ref{eq:area_between_curves}, we can analytically deduce whether a feed-forward loop delays or accelerates output response from the sign of $\langle\Delta\bar{t}\rangle$. In Appendix~\ref{appendix: ffl_derive_delay}, we prove that in general $\langle\Delta\bar{t}\rangle \leq 0$ for the coherent feed-forward loop, leading to delay for both the ON and OFF steps as seen in the example of Fig.~\ref{fig:FFL1}(A).

\subsection{Robustness of delayed response for different coherent logic gates}\label{sec:FFLlogicsweep}

\begin{figure*}
     \centering
     \includegraphics[width=\linewidth]{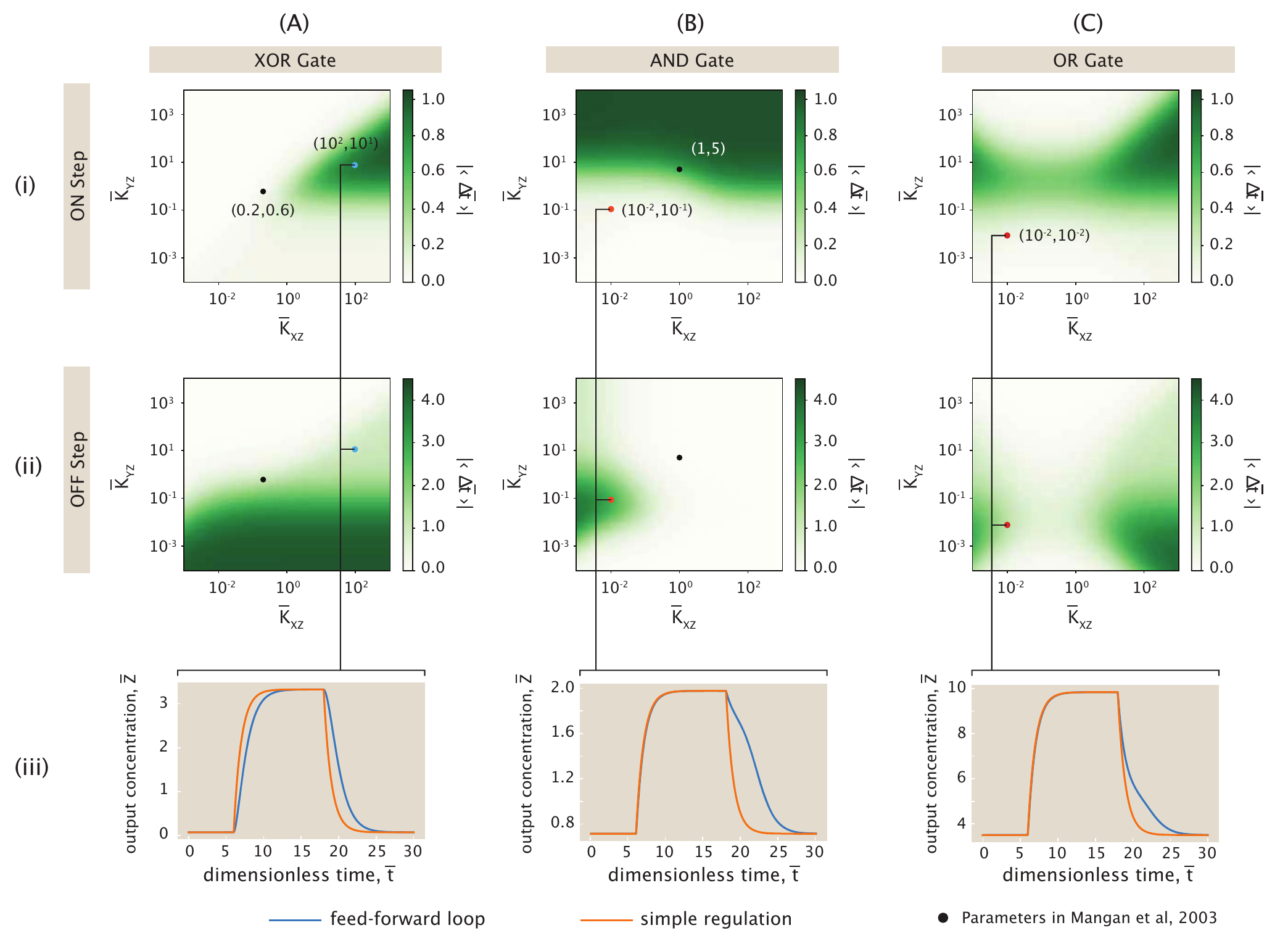}
     \caption{Magnitude of average time delay observed across parameter space in the ON and OFF steps of different coherent feed-forward logic gates. Each colorplot shows $\Delta \bar{t}$ as a function of $\bar{K}_{XZ}$ and $\bar{K}_{YZ}$. ON (row (i)) and OFF (row (ii)) steps are defined by the same $c_{X}$ step function as in Fig.~\ref{fig:FFL1}. Each panel represents a different logic gate --- (A) the XOR gate, (B) the AND gate, and (C) the OR gate. For each gate, we select a set of $(\bar{K}_{XZ}, \bar{K}_{YZ})$ that exhibit unexpected behaviors, and in row (iii) plot the corresponding feed-forward and simple regulation trajectories observed from numerical integration. The $c_X$ and $c_Y$ signal for these trajectories are the same as in Fig.~\ref{fig:FFL1}.
     The XOR gate parameters are $\bar{r}_{0Y} = \bar{r}_{0Z} = 0$, $\bar{r}_{1Y} = \bar{r}_{1Z} = 2$, and $\omega = 0$. The AND gate parameters are $\bar{r}_{0Y} = \bar{r}_{0Z} = \bar{r}_{1Z} = 0$, $\bar{r}_{1Y} = \bar{r}_{2Z} = 2$, and $\omega = 10$. 
     The OR gate parameters are $\bar{r}_{0Y} = \bar{r}_{0Z} = 0$, $\bar{r}_{1Y} = \bar{r}_{1Z} = 2$, $\bar{r}_{2Z} = 10$, and $\omega = 1$, which are the same as in Fig.~\ref{fig:FFL1}.} 
     \label{fig:ffl_func_param}
 \end{figure*}
 
While Fig.~\ref{fig:FFL1}(A) demonstrates delays in both the ON and OFF step of the feed-forward loop for an arbitrary choice of the dimensionless rates, dissociation constants, and cooperativity, distinct choices for this set of parameters lead to different magnitudes of delay. Returning to Fig.~\ref{fig:FFLcoherent}, the thermodynamic states and weights listed here are defined generally such that all of the states can contribute to transcription factor expression, and this is reflected in the example feed-forward loop shown in Fig.~\ref{fig:FFL1}(A). However, certain alternative choices for parameters can carry physical significance because they restrict the regulatory states allowing expression to only a subset of those depicted in Fig.~\ref{fig:FFLcoherent}. How would the feed-forward loop's behavior differ, for example, if expression could only be enhanced when both activators are bound? We use the framework of \emph{logic gates} to define such unique categories for regulatory conditions. 

Specifically, we highlight three commonly-encountered logic gates---the AND, XOR, and OR gates. We will assume here that all gates can express output $\bar{Z}$ at a basal level, as defined by rate $\bar{r}_{0Z}$ for the state. Each logic gate is then characterized by a different set of parameters for the states in which one or both activators can be bound, and these parameters determine whether a given state's expression is enhanced or remains unaffected at the basal level.

In the AND gate, $Z$ expression is enhanced only when both $X$ and $Y$ are bound. From the description in Fig.~\ref{fig:FFLcoherent}, this corresponds to systems in which cooperativity $\omega$ is nonzero and $\bar{r}_{2Z} > \bar{r}_{1Z} = \bar{r}_{0Z}$, such that $X$ and $Y$ have no activating effect on basal expression unless simultaneously bound. In the XOR gate, $X$ and $Y$ cannot be bound at the same time ($\omega = 0$), and single-activator bound states enhance $Z$ expression ($\bar{r}_{1Z} > \bar{r}_{0Z}$). Finally, the OR gate allows enhanced expression when either $X$, $Y$, or both are bound, and broadly applies to systems for which $\bar{r}_{2Z}\geq \bar{r}_{1Z} > \bar{r}_{0Z}$. Note that the expression rates when one or both transcription factors bind can differ.

In previous work, XOR and AND gates have been reported to exhibit sign-sensitive delays in response to a signal change in $c_X$: 
the XOR gate feed-forward loop delays the OFF step but not the ON step, while the AND gate delays the ON step but not the OFF step~\cite{mangan2003structure}. While these descriptions hold for certain parameter choices, it remains unclear how robust these patterns are across parameter space as the conditions governing regulatory interactions change. We will now examine the conditions under which the feed-forward loop response delays both ON and OFF steps, delays only one, or delays neither as we compare the different types of logic gates.

To evaluate the robustness of time delays across different types of logic gates, we sweep across parameter space and find regions with high average delay, $\langle \Delta \bar{t} \rangle$. For now, we choose to fix rate and cooperativity parameters and sweep across the two-dimensional space $(\bar{K}_{XZ}, \bar{K}_{YZ}$). The motivation is to find the suitable dissociation constants given a logic gate, in the case in which the production rates and cooperativity must satisfy certain requirements. For example, in the XOR gate, $\omega = 0$ is fixed and is not a tunable parameter.

In Fig.~\ref{fig:ffl_func_param}, we show a parameter sweep for three sets of rates and cooperativity parameters that correspond to three different logic gates. For each set of parameters, both ON and OFF steps are studied. These calculations inspire several observations. First, for a given step undergone by $p_\text{act}^X(c_X)$, we can computationally identify a maximum $\langle \Delta \bar{t} \rangle$ with respect to all other parameters. Further, this maximum is different for ON and OFF steps. For example, in Fig.~\ref{fig:ffl_func_param}(i), specific to the input step chosen here, we observe that in all logic gates the maximal $\langle \Delta \bar{t} \rangle$ achievable is about 1 for the ON step and 4.5 for the OFF step. 
To tie this back to units of time, we consider \textit{E. coli}. Here, proteins tend to be stable over the timescale of a cell cycle and hence the dilution resulting from cell division becomes the effective degradation rate. Taking a rate of degradation of order $\gamma=10^{-2}\text{min}^{-1}$, we then get a maximum delay of $500 \text{ min}$, which is more than $8 \text{ hours}$. 
The asymmetry between maximal $\langle \Delta \bar{t} \rangle$ the ON and OFF steps resonates with the analytic discussion in the previous section. We show computational evidence in Appendix~\ref{appendix:ffl_func_region} that sweeps across all other parameters indicate the existence of a global maximum $\langle \Delta \bar{t} \rangle$. Note that the exact value and existence of this maximum is a result specific to function $p_\text{act}(c_{X})$ and allosteric parameters we chose to describe effector activity in Fig.~\ref{fig:pact}.

Second, both the XOR and AND gates exhibit the expected sign-sensitive delay across a substantial portion of the $(\bar{K}_{XZ}, \bar{K}_{YZ})$ parameter space, though not uniformly. In particular, Fig.~\ref{fig:ffl_func_param}(A)(i,ii) shows that for the XOR gate, the lower half of this space yields negligible ON-step delay but a pronounced OFF-step delay. Conversely, Fig.~\ref{fig:ffl_func_param}(B)(i,ii) reveals that for the AND gate, the upper half of the space yields negligible OFF-step delay and a pronounced ON-step delay. 
Nevertheless, in specific regions of the $(\bar{K}_{XZ}, \bar{K}_{YZ})$ parameter space—namely, the upper left quadrant for the XOR gate and the lower right quadrant for the AND gate—neither the ON nor the OFF step exhibits any appreciable delay. More intriguingly, certain extreme choices of dissociation constants contradict the expected behavior: the XOR gate can show delay on both ON and OFF steps, and the AND gate can only delay the OFF step. Example trajectories corresponding to these atypical regimes are shown in Fig.~\ref{fig:ffl_func_param}(A,B)(iii).

Finally, the OR gate can produce sizable delays, but only for more extreme values of the dissociation constants. Near the region where $\bar{K}_{XZ} \approx \bar{K}_{YZ} \approx 1$—which corresponds to similar binding strengths between $X$, $Y$, and $Z$ with the DNA—the delays for both ON and OFF steps are minimal. This suggests that in some biologically relevant regimes, where dissociation constants are typically of the same order of magnitude, the OR gate is the least effective at generating a temporal delay.

We observe that the average delay $\langle \Delta \bar{t} \rangle$ in coherent feed-forward loops depends strongly on the dissociation constant $\bar{K}_{YZ}$, which sets the binding affinity of $Y$ to the promoter of $Z$—one of the interactions that distinguishes feed-forward loops from simple regulation. A clear trend emerges: ON steps (Fig.~\ref{fig:ffl_func_param}(A-C)(i)) exhibit stronger delays when $Y$ binding is weak (large $\bar{K}_{YZ}$), whereas OFF steps (Fig.~\ref{fig:ffl_func_param}(A-C)(ii)) show stronger delays when $Y$ binding is strong (small $\bar{K}_{YZ}$).

Since the feed-forward loop ultimately aims to control $Z$ activation, another relevant feature is the output amplitude $\Delta \bar{Z} = |\bar{Z}_f - \bar{Z}_i|$ for a given change in input $c_X$. A small $\Delta \bar{Z}$ would imply that $\bar{Z}$ remains nearly constant, making the response uninformative.  However, we show in Appendix~\ref{appendix:ffl_func_region} that $\Delta \bar{Z}$ scales linearly with the production rates. Although the dependence of $\Delta \bar{Z}$ on other parameters is non-trivial, globally the effect of the rates dominates. We thus reserve a detailed discussion of the optimization of $\Delta \bar{Z}$ values for Appendix~\ref{appendix:ffl_func_region} and remain focused here on the average delay $\langle \Delta \bar{t} \rangle$.

\subsection{Existence of pulse in incoherent feed-forward loop}
\label{sec:pulse}
\begin{figure*}
    \centering
    \includegraphics[width=\linewidth]{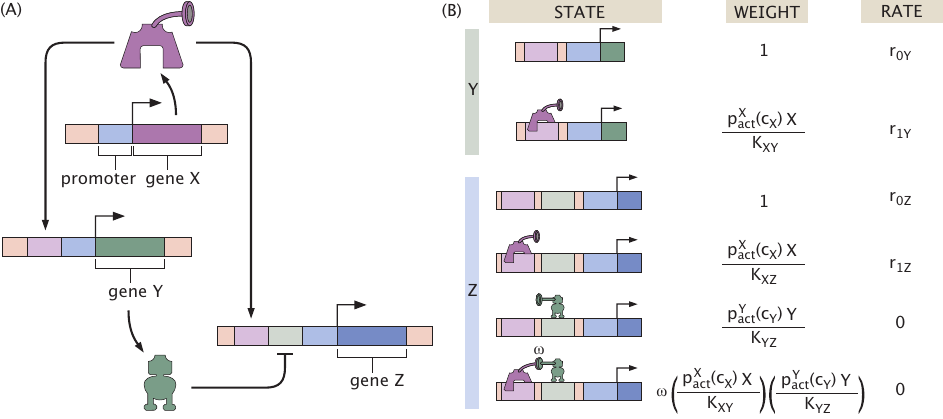}
    \caption{The incoherent feed-forward loop. (A) Schematic representation of the incoherent feed-forward loop. Expression of output protein $Z$ is controlled by expression of protein $X$, either by direct activation or indirectly, first activating expression of $Y$ which then represses $Z$. The regulatory circuit is incoherent because the pathways have opposing effects on $Z$. (B) Thermodynamic states, weights, and rates for expression of repressor $Y$ and output protein $Z$. $X$ and $Y$ interact with cooperativity $\omega$, but bound repressor suppresses expression regardless of activator presence.}
    \label{fig:FFLincoherent}
\end{figure*}

\begin{figure*}
    \centering
    \includegraphics[width=\linewidth]{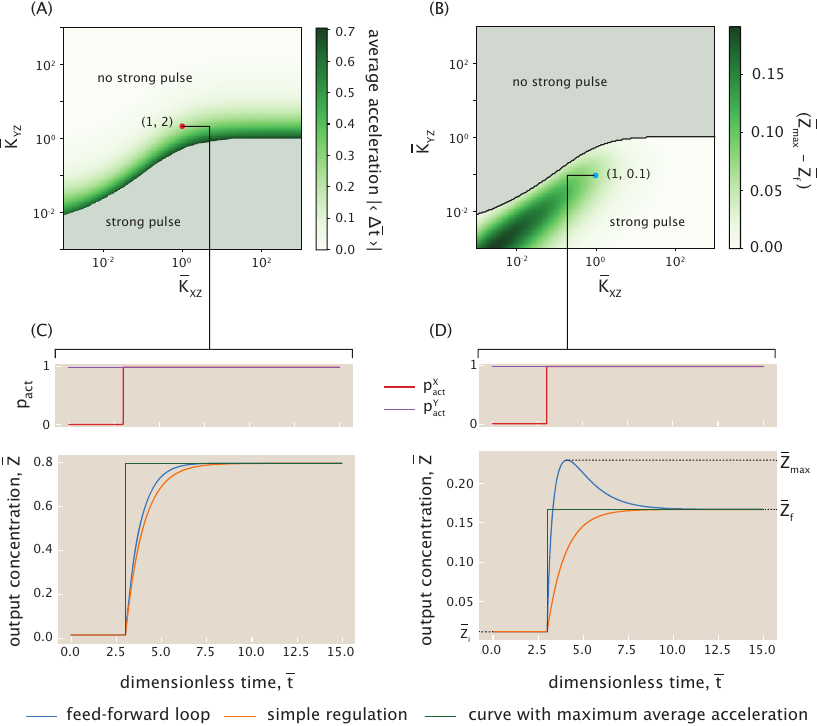}
    \caption{Existence of a pulse and acceleration in the incoherent feed-forward loop. Parameters used here are $\bar{r}_{0Y} = \bar{r}_{0Z} = 0$, $\bar{r}_{1Y} = \bar{r}_{1Z} = 2$, and $\omega = 0$.  (A) Average acceleration $|\langle\Delta \bar{t}\rangle|$ during ON step across the phase space $(\bar{K}_{YZ}, \bar{K}_{XZ})$, computed only for parameter sets where no strong pulse is observed. For the parameters we chose, the ON step in $p_\text{act}^X(c_X)$ coincides with $\bar{Z}(\bar{t})$ having an increasing response. (B) Pulse amplitude $\bar{Z}_{\max} - \bar{Z}_f$ across the same phase space, quantifying the transient overshoot above steady state. (C) Example trajectories corresponding to $(\bar{K}_{XZ}, \bar{K}_{YZ}) = (1, 2)$. No strong pulse is observed, but the feed-forward loop response is accelerated compared to the simple regulation response. The green curve is the trajectory with the largest $\langle \Delta\bar{t} \rangle$ without a pulse. The blue and orange curves are the feed-forward loop and simple regulation trajectories, respectively. (D) Example trajectories corresponding to $(\bar{K}_{XZ}, \bar{K}_{YZ}) = (1, 0.1)$. A strong pulse is observed.}
    \label{fig:ffl_pulse}
\end{figure*}

We now turn our attention to the incoherent feed-forward loop depicted in Fig.~\ref{fig:FFLincoherent}. This architecture  is a commonly observed motif in \textit{E.~coli}, where $X$ activates both $Y$ and $Z$, while $Y$ represses $Z$~\cite{Milo2002, shen2002network}. This motif gives rise to a qualitatively distinct dynamical behavior from the coherent feed-forward loop: a pulse in the output $Z$. A pulse is conventionally defined as a transient trajectory of $\bar{Z}(\bar{t})$ such that there exists a time $\bar{t}$ where $\bar{Z}(\bar{t}) > \bar{Z}_f$ if the response of $\bar{Z}(\bar{t})$ is increasing, or $\bar{Z}(\bar{t}) < \bar{Z}_f$ if the response of $\bar{Z}(\bar{t})$ is decreasing.
In incoherent feed-forward loops, an ON step signal where $p_\text{act}^X(c_X)$ increases does not necessarily produce an increasing response in $\bar{Z}(\bar{t})$, as activation induced by increasing $\bar{X}$ competes with repression induced by increasing $\bar{Y}$. This is precisely the incoherence in the name of such circuits.

Referring to the states, weights, and rates shown in Fig.~\ref{fig:FFLincoherent}, the non-dimensional dynamics of the system are described by
\begin{equation}\label{eq:incoh_ffl_nondim}
    \frac{d\bar{Y}}{d\bar{t}} = - \bar{Y} + \frac{\bar{r}_{0Y} + \bar{r}_{1Y} \, p_\text{act}^X\bar{X}}{1+p_\text{act}^X\bar{X}}
\end{equation}
and
\begin{equation}
    \frac{d\bar{Z}}{d\bar{t}} = - \bar{Z} + \frac{\bar{r}_{0Z} + \bar{r}_{1Z} \, \mathcal{X}}{1+\mathcal{X} +\mathcal{Y} + \omega \mathcal{X} \mathcal{Y}},
\end{equation}
with $\mathcal{X} = p_\text{act}^X(c_X)\bar{X}/\bar{K}_{XZ}$ and $\mathcal{Y} = p_\text{act}^Y(c_Y)\bar{Y}/\bar{K}_{YZ}$. Again the bar indicates the time in units of $1/\gamma$, concentrations and dissociation rates in units of $K_{XY}$ and rates in units of $\gamma K_{XY}$. 

Intuitively, pulses emerge because of the delay in the repression exerted by $Y$. Following an ON step in $c_X$, $\bar{Y}$ increases exponentially to its final value. In the early phase of the response, $X$ already activates $Z$ strongly, but $Y$ has not yet accumulated enough to exert repression. As a result, $\bar{Z}$ temporarily overshoots its final steady state. This behavior is depicted in Fig.~\ref{fig:ffl_pulse}(D), where the blue curve exhibits a pronounced pulse, in contrast to the monotonic exponential relaxation of the simple regulation output (orange).

The accelerating nature of the incoherent feed-forward loop can also be captured analytically by repeating the calculation leading to Eqn.~\ref{eq:ffl_coh_full_sol} but adapted to this new setting. The modified prefactor
\begin{align}
    \Phi = - (\bar{r}_{0Z} + \bar{r}_{1Z} \mathcal{X}_f)(1 + \omega \mathcal{X}_f)
    \label{eq:incoh_full_sol}
\end{align}
is always negative. As a result, the average time difference $\langle \Delta\bar{t}\rangle$ as defined in Eqn.~\ref{eq:time_delay} is always positive, both for ON and OFF steps, meaning that the incoherent feed-forward loop accelerates the response for both transitions. In Appendix~\ref{sec:def_of_pulse_incoh}, we remark that the definition and interpretation of $\langle \Delta\bar{t}\rangle$ is subtle when $\bar{Z}$ exhibits a pulse. Thus, we only consider $\langle \Delta\bar{t}\rangle$ when a pulse does not exist, and focus on the difference between the peak of the pulse and final steady state of $\bar{Z}$ for the pulses.

To further quantify this acceleration, we again compute the average time difference $\langle \Delta \bar{t} \rangle$ between the feed-forward loop and simple regulation trajectories with Eqn.~\ref{eq:time_delay}. Fig.~\ref{fig:ffl_pulse}(A) shows $\langle \Delta \bar{t} \rangle$ across the phase space defined by $(\bar{K}_{XZ}, \bar{K}_{YZ})$, when the output does not present a strong pulse. The definition of ``strong" is described in Appendix~\ref{sec:def_of_pulse_incoh}.
We observe that acceleration is limited to a maximum value of 1. This upper bound arises because the most accelerated trajectory would consist of an instantaneous jump to the final steady state (green curves in Fig.~\ref{fig:ffl_pulse}(C) and (D)), corresponding to $\langle \Delta  \bar{t}\rangle = 1$ as shown in Appendix~\ref{sec:def_of_pulse_incoh}. An example of a feed-forward trajectory that does not exhibit a strong pulse is shown in Fig.~\ref{fig:ffl_pulse}(C).
The region of highest acceleration lies near the boundary separating the strongly pulsed and not strongly pulsed regimes. Interestingly, acceleration is only substantial in a restricted region of parameter space. For example, we see in Fig.~\ref{fig:ffl_pulse}(A) that high values of $\bar{K}_{YZ}$—corresponding to weak binding of $Y$—lead to negligible acceleration. 

In addition to acceleration, the presence and magnitude of a pulse is another hallmark of the studied network. In Fig.~\ref{fig:ffl_pulse}(B), we map the pulse height, defined as the maximum deviation of $\bar{Z}(t)$ above its steady state value, across the $(\bar{K}_{XZ}, \bar{K}_{YZ})$ space. Pulses are observed only in a portion of this space—specifically, for small enough values of $\bar{K}_{XZ}$, where $X$ binds strongly. The highest pulses occur when both $\bar{K}_{XZ}$ and $\bar{K}_{YZ}$ are small, indicating that strong binding of both regulators enhances the transient overshoot in this case.

\subsection{Continuous signal}
\label{sec:ffl_cont_signal}

\begin{figure*}
    \centering
    \includegraphics[width=\linewidth]{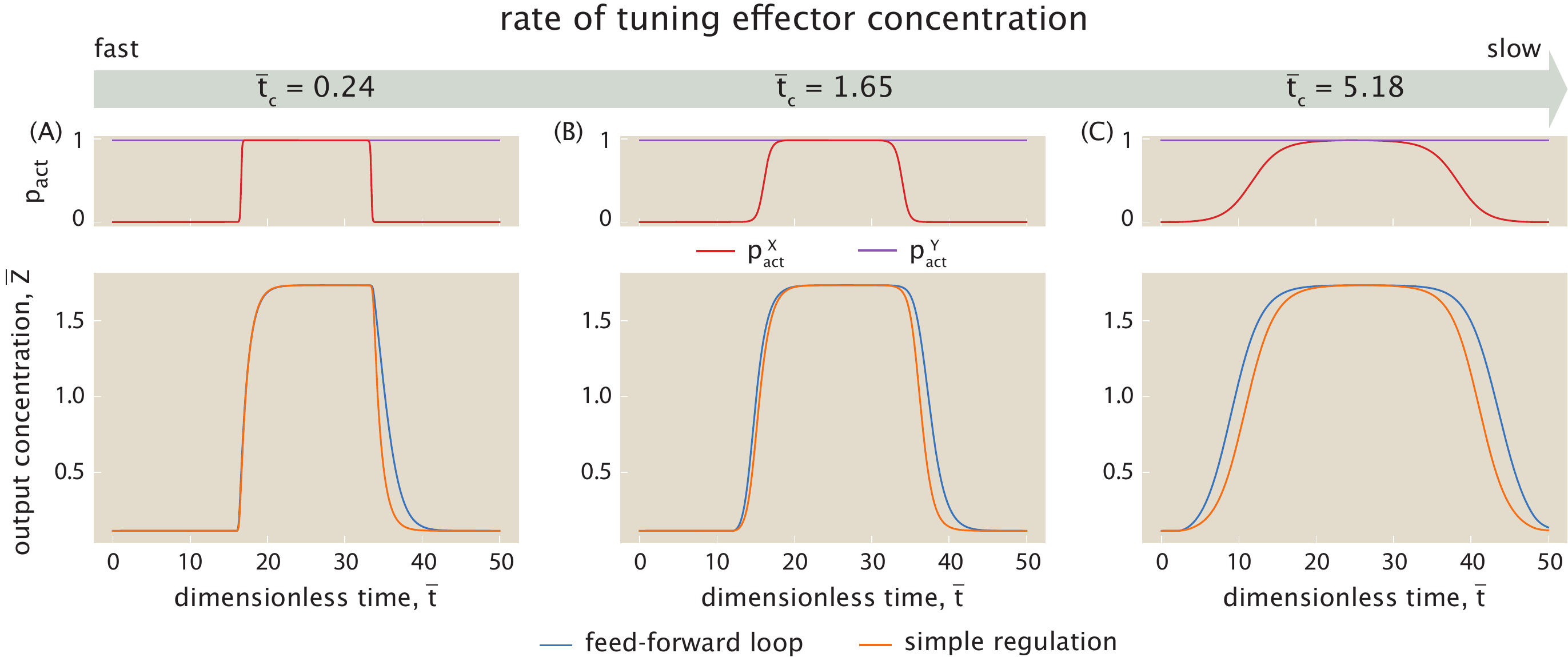}
    \caption{Feed-forward loop response to the rate of continuous tuning of effector concentration. From left to right, the rate of tuning effector concentration slows down while every other parameter is kept constant. Each ON step is a loglinearly increasing function from $c_X^{\min} = 10^{-4}$ M to $c_X^{\min} = 10^{-7}$ M across some time; an OFF step is the reverse. From left to right, the timescale of effector concentration variation are $\bar{t}_c = 0.24$, $\bar{t}_c = 1.65$, $\bar{t}_c = 5.18$. $\bar{t}_c$, as defined in the main text, is the time it takes for $p_\text{act}^X(c_X(\bar{t}))$ to increase from 0.2 to 0.8. $\bar{Y}$ is set to be 1 at all times for simple regulation. 
    Parameters used are the XOR gate parameters in Fig.~\ref{fig:ffl_func_param}.}
    \label{fig:ffl_cont_signal}
\end{figure*}

To conclude our discussion of feed-forward loops, we now consider the system’s response to a continuously tuned effector concentration, rather than an abrupt step change. When the timescale of effector concentration variation, denoted $\bar{t}_c$, is much shorter than the system’s relaxation timescale, the dynamics resemble those observed under step function inputs. Conversely, when $\bar{t}_c$ is much longer than the relaxation time, the system remains quasi-stationary when the input is evolving, effectively tracking steady states values of the output $\bar{Z}$ at each time step.

To analyze this quantitatively, we compare $\bar{t}_c$ to the intrinsic relaxation timescales of both the simple regulation case and the coherent feed-forward loop. In coherent feed-forward loops, we have shown that the output response is consistently delayed relative to simple regulation. However, the precise relaxation timescale of this architecture is not straightforward and depends strongly on the biochemical parameters. As a result, we use the relaxation timescale of the simple regulation system—which is 1 in units of $1/\gamma$—as a reference estimate for the order of magnitude of the feed-forward loop’s relaxation time. We can thus define two limiting regimes: a fast tuning regime where $\bar{t}_c \ll 1$, and a slow tuning regime where $\bar{t}_c \gg 1$.

Before proceeding to the analysis of the delay, we must first clarify the definition of simple regulation in the case of a continuous signal. We adopt the same definition as the step function case, where we fix concentration $\bar{Y} = 1$ here. Further, in our previous analysis, when $c_X(\bar{t})$ is a step function, the choice of $\bar{Y}$ does not affect the dynamics of relaxation of the simple regulation output. However, when $c_X(\bar{t})$ is a continuous function, the choice of $\bar{Y}$ changes these dynamics slightly. Fortunately, the effect of $\bar{Y}$ is small and does not qualitatively change our observations in the rest of the section, allowing us to continue with the comparison between feed-forward loop and simple regulation (for a more detailed discussion, see Appendix~\ref{appendix:ffl_cont_tuning}).

We numerically integrate the dynamical equations under a continuously changing $c_X(\bar{t})$ and illustrate the result in Fig.~\ref{fig:ffl_cont_signal}, where we analyze the response of a coherent feed-forward loop operating as an XOR gate. In this setting, we numerically define the timescale of effector concentration variation $t_c$ as the time it takes for $p_\text{act}^X(c_X)$ to change from 0.2 to 0.8 or vice versa. $t_c$ indicates the time it takes for the switch to be flipped on or off, as regulated by the effector concentration $c_X$.
When the effector concentration $c_X(\bar{t})$ is tuned rapidly as shown in Fig.~\ref{fig:ffl_cont_signal}(A), with $\bar{t}_c\approx 0.24$, we observe a large delay in the output $\bar{Z}(\bar{t})$ on the OFF step and almost no delay on the ON step, consistent with the step function dynamics discussed previously.
In contrast, when $c_X(\bar{t})$ varies slowly, as depicted in Fig.~\ref{fig:ffl_cont_signal}(C), with $\bar{t}_c \approx 5.18$, both the feed-forward loop and the simple regulation trajectories become dominated by their respective steady states. They simply track their steady states dictated by $c_X(\bar{t})$. As a consequence, the responses to ON and OFF steps must be symmetric, as there is exactly one steady state corresponding to a specific $c_X$. Interestingly, the OFF delay is preserved, while for the ON step the feed-forward loop response is accelerated compared to simple regulation. The magnitude of the ON step acceleration is similar to that of the OFF step delay, as required by symmetry. We note that the magnitude of the acceleration/delay depends on the choice of $\bar{Y}$ in simple regulation. 

Nevertheless, for all choices of $\bar{Y}$, the feed-forward loop accelerates the ON step and delays the OFF step when the concentration of effector $c_X$ is slowly tuned.
Between the two limits, the feed-forward loop response smoothly transitions as the tuning rate of $c_X$ decreases. An example in this regime is shown in Fig.~\ref{fig:ffl_cont_signal}(B), with $\bar{t}_c \approx 1.65$, where we begin to observe the ON step acceleration, despite its magnitude being smaller than in Fig.~\ref{fig:ffl_cont_signal}(C). These results demonstrate that the dynamics of feed-forward loops under continuously varying effector concentrations are governed by the timescale of input change: rapid tuning reproduces the asymmetric ON- and OFF-step delays seen with step inputs, whereas slow tuning yields symmetric responses where ON-step acceleration balances OFF-step delay.

\section{\label{sec:discuss}Discussion}

The history of modern molecular biology has been a dazzlingly successful exploration of the way in which genes dictate the function
and dynamics of the cells making up organisms of all kinds.  One of the greatest success stories of that history is the development of our understanding of how genes are connected together in genetic circuits~\cite{Britten1969}, giving rise to an array of stereotyped regulatory motifs such as switches, oscillators, double negative networks and feed-forward networks, to name but a few examples~\cite{Milo2002}. We note that despite all of this progress, there remain gaping holes in our knowledge  of how most genes are regulated.  Even in our best understood organisms such as {\it E. coli}, we lack any knowledge of how more than 60\% of its genes are regulated~\cite{SantosZavaleta2019}.  As a result, we expect that with the advent of the high-throughput era in biology, ever more genetic circuits like those we discussed here will be discovered.  

In addition to our ignorance of the genetic circuits themselves, our understanding of the proteins that mediate those circuits is also very limited. In particular, we often don't know  how effector molecules alter the activity of the transcription factors that control these genes~\cite{Lindsley2006}.  The key point here is that the action of proteins such as transcription factors is often altered through the binding of effector molecules, which induce allosteric conformational changes that in turn change the state of activity of those transcription factors.  However, for many genes, we still remain ignorant of which effector molecules effect those changes.   The central thesis of this paper is that, in fact, most gene circuits have their activity tuned by precisely these kinds of effector molecules and as a result, we need to revisit the theoretical analysis of such circuits to account for the effect of allosteric induction.

In parallel with the impressive progress in molecular biology and the dissection of the rules of regulation, huge progress was made in the study of the behavior of dynamical systems in contexts of all kinds~\cite{Strogatz2015}, and with special reference to genetic circuits themselves~\cite{Cherry2000}.  However,
when theorists have used dynamical systems frameworks to explore the behavior of such circuits, they have largely adopted an approach in which those circuits are tuned in abstract terms using model parameters such as degradation rates $\gamma$, mRNA production rates $r$ and binding constants $K_d$.  As was so importantly discovered in the 1960s, typically these parameters are in fact ``tuned'' in the context of living cells through the action of allosteric transitions of transcription factors between active and inactive conformations as a result of the binding of effector molecules \cite{Gerhart1962, Monod1963,Monod1965,Martins2011,Marzen2013,Changeux2013,Gerhart2014,Phillips2020}.
The history of dynamical systems in these problems largely leaves the all-important effector molecules out of this story, only considering them implicitly.  In this paper we have undertaken a systematic analysis of the role of such effectors in governing the function and dynamics of a variety of fundamental genetic regulatory motifs.

The overarching theme of the work described here is that whereas typical dynamical systems approaches to genetic networks feature the number of transcription factors such as $A(t)$ for activator concentrations and $R(t)$ for repressor concentrations, the variable that the cell actually ``cares about'' is the active number of activators and repressors.  There are a variety of well-defined statistical mechanical approaches that allow us to compute this active fraction by multiplying the total number of transcription factors by the function $p_\text{act}(c)$ as dictated by the Monod-Wyman-Changeux model, for example, and highlighted in Eqn.~\ref{eqn:MWC2site}.
The power of this approach is that now the parameters governing properties such as bistability in genetic circuits will be tuned by experimentally and biologically accessible parameters such as the concentrations of effector molecules.

Throughout the paper, we have shown how
the tuning variable of effector concentration makes it possible for the dynamical systems describing genetic circuits to range across their phase portraits.  We began with perhaps the simplest of such circuits, the auto-activation motif, and showed how tuning effector concentration narrows the range of possible behaviors relative to those found in an unconstrained dynamical system perspective. We also availed ourselves of the opportunity to compare and contrast the conventional Hill function approach to transcription factor-DNA binding and the more mechanistically detailed thermodynamic models that we systematically explore throughout this work.

One interesting conclusion of this comparison between models based on the full set of states and weights demanded by thermodynamic models and the more phenomenological Hill function is that, for certain parameter regimes, each approach will display dramatically different circuit dynamics (i.e., monostability vs. bistability). This insight emphasizes the need to carefully dissect the quantitative parameters underlying the description of gene regulatory architectures in order to justify whether a Hill function description, which is a limiting case of the thermodynamic description, is warranted.

We also used the auto-activation motif as an opportunity for a careful analysis of the temporal relaxation of these genetic circuits to their terminal steady state.  That analysis revealed that for initial conditions that are not ``far'' from the stable fixed points, the relaxation to steady state is exponential with a time constant dictated by
the derivative of the nonlinear protein degradation/production function evaluated at the fixed point.  For initial conditions that start near to the unstable fixed point, the dynamics are richer.

With the description of the induced auto-activation circuit in hand, we turned to  the very important mutual repression motif which is ubiquitous in prokaryotes and eukaryotes alike. Here, again, the capacity to independently tune the effector concentration for each repressor revealed a large flexibility in how cells and synthetic biologists alike can decide to tune the dynamical behavior of this genetic circuit from dictating its steady state behavior to the dynamics of the repressors as they converge to those steady state values.

Finally, we undertook a dissection of the ubiquitous feed-forward loop. Our analysis shows that the dynamic behavior typically associated with feed-forward loops in response to input effector signals is more nuanced and parameter-dependent than previously appreciated \cite{mangan2003coherent}. For the coherent feed-forward loop, we analytically confirm the presence of delay in output response compared to simple regulation. However, we show that both the magnitude and the \textit{sign-sensitivity} of these delays depend on system parameters such as dissociation constants, production rates and cooperativity. This rich range of qualitatively distinct behavior remains true even within the different categories of logic gates that can emerge from special combinations of these parameters. Conversely, incoherent feed-forward loops accelerate output responses compared to simple regulation and can generate transient pulses---again only in certain regions of parameter space.

Within this analysis of feed-forward loops, we demonstrate the crucial roles of effector concentration in our models. The leakiness of $p_\text{act}(c)$, for example, influences key metrics such as delay time. We also highlight how the sigmoidal shape of $p_\text{act}(c)$ enables a continuous change in effector concentration to be translated to a sharp signal when tuning the probability of transcription factors being active. Overall, while the dynamical behaviors of feed-forward loops can be rich, they are not always guaranteed.
This emphasizes how behavior emerges from a delicate interplay of biochemical parameters rather than rigid circuit logic alone, and underscores the need for further experimental and theoretical efforts toward understanding the functions and dynamics of feed-forward loops.

All told, our efforts demonstrate that there is great flexibility inherent in the endogenous signaling modalities adopted by living cells to be contrasted with the way in which model parameters are artificially tuned in many dynamical systems approaches to these same problems.  We are excited for experimental efforts to make a substantial push to solve the huge puzzle of the allosterome, opening the door to more realistic analyses of genetic circuits from a dynamical systems perspective.

\begin{acknowledgments}

We have benefited enormously from conversations with Leonid Mirny, Ned Wingreen, Julie Theriot, Marc Kirschner, Jean Pierre Changeux, Jane Kondev, all of whom have helped us better understand the twin subjects brought together here, namely, genetic circuits and allosteric regulation of the macromolecules of the cell.
We are grateful to the NIH for support through award numbers DP1OD000217 (Director's Pioneer Award) and NIH MIRA 1R35 GM118043-01. 
This research is funded in part by the Gordon and Betty Moore Foundation  GBMF12214 (doi.org/10.37807/GBMF12214) to RP and RJR as part of the grant on the ``Listening to Molecules'' project which has as its central mission to better understand the underlying mechanisms of allosteric regulation.
RP is deeply grateful to the CZI Theory Institute Without Walls both for conversations with its members and support for the work itself.
HGG was supported by NIH R01 Awards R01GM139913 and R01GM152815, by the Koret-UC Berkeley-Tel Aviv University Initiative in Computational Biology and Bioinformatics, by a Winkler Scholar Faculty Award, by the Chan Zuckerberg Initiative Grant CZIF2024-010479, and by the Miller Institute for Basic Research in Science, University of California Berkeley. H.G.G. is also a Chan Zuckerberg Biohub Investigator (Biohub – San Francisco).\\

\end{acknowledgments}

\appendix

\section{Thermodynamic model equivalence to description with polymerase}
\label{Section:CoarseGraining}

In this section, we demonstrate the equivalence of the thermodynamic models used throughout the paper in which we essentially ignored the presence of RNA polymerase (RNAP), to those that explicitly incorporate regulatory interaction with the RNA polymerase. To illustrate the comparison between those that explicitly treat polymerase and those that do not, we begin by examining the auto-activation switch with which the paper opened.

\subsection{Coarse-graining the auto-activation model}\label{appendix:polymerase_autoact}

\begin{figure*}
    \centering
    \includegraphics[width=\linewidth]{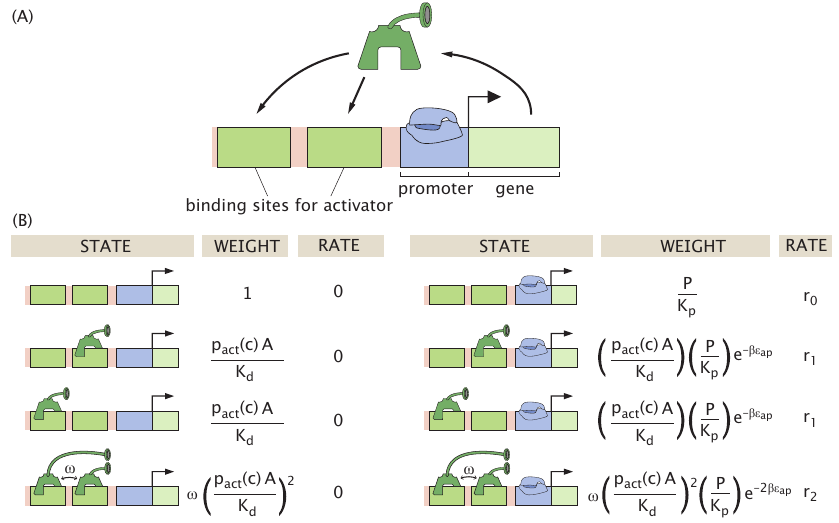}
    \caption{The auto-activation regulatory circuit. (A) A schematic of the circuit operation. Polymerase binding at the promoter (blue) transcribes the gene (encoded in the green region), producing a protein that can activate its own expression at a sufficient concentration. In our model, an activator can bind at two possible sites to enhance gene transcription. (B) The thermodynamic states, weights, and rates for the auto-activation motif including polymerase binding explicitly. The parameter $\omega$ denotes the binding cooperativity between two activators.}
    \label{fig:AutoActivation_P}
\end{figure*}

The statistical mechanical model for auto-activation depicted in Fig.~\ref{fig:AutoActivation} implicitly accounts for interaction between the activator and polymerase. Fig.~\ref{fig:AutoActivation_P} represents the complete accounting of the thermodynamic states, weights and rates, now explicitly accounting for all of the possible polymerase bound states and  denoting the interaction energy between polymerase and activator as $\varepsilon_{ap}$. In light of this complete set of states, weights and rates, We can write the dynamical equation for the number of activators as
\begin{widetext}
\begin{equation}\label{eqn:AutoActivationWithPolymerase}
    \frac{dA}{dt} = -\gamma A + \frac{\frac{P}{K_{P}}[r_{0} + 2r_{1}e^{-\beta\varepsilon_{ap}}\frac{p_\text{act}(c)A}{K_{d}}+r_{2}e^{-2\beta\varepsilon_{ap}}\omega(\frac{p_\text{act}(c)A}{K_{d}})^2]}{Z},
\end{equation}
\end{widetext}
where $P$ is the number of copies of polymerase present and $K_{P}$ is the dissociation constant for $P$. $Z$ is the partition function obtained by summing the statistical weights of all of the states in Fig.~\ref{fig:AutoActivation_P}, which we define as
{{\small
\begin{eqnarray}
    Z&=&\frac{P}{K_{P}}\Bigg[1+ 2e^{-\beta\varepsilon_{ap}}\frac{p_\text{act}(c)A}{K_{d}}+e^{-2\beta\varepsilon_{ap}}\omega\Big(\frac{p_\text{act}(c)A}{K_{d}}\Big)^2\Bigg]\nonumber\\
    &&\hspace{0.5cm} + \Bigg[1 + 2\frac{p_\text{act}(c)A}{K_{d}}+\omega\Big(\frac{p_\text{act}(c)A}{K_{d}}\Big)^2\Bigg]\nonumber
\end{eqnarray}
\begin{eqnarray}
    &=& 1+ \frac{P}{K_P} + 2\frac{p_\text{act}(c)A}{K_{d}} \Big(1+\frac{P}{K_{P}}e^{-\beta\varepsilon_{ap}}\Big)\nonumber\\
    &&\hspace{0.5cm}+\omega\Big(\frac{p_\text{act}(c)A}{K_{d}}\Big)^2\Big(1+e^{-2\beta\varepsilon_{ap}}\frac{P}{K_P}\Big)\nonumber\\
    &=&\Big(1+\frac{P}{K_P}\Big)\Bigg[1 + 2\frac{p_\text{act}(c)A}{K_{d}} \frac{(1+\frac{P}{K_{P}}e^{-\beta\varepsilon_{ap}})}{1+ \frac{P}{K_P}}\nonumber\\
    &&\hspace{0.5cm}+\omega\Big(\frac{p_\text{act}(c)A}{K_{d}}\Big)^2\frac{(1+e^{-2\beta\varepsilon_{ap}}\frac{P}{K_P})}{1+ \frac{P}{K_P}}\Bigg]\nonumber\\
    &\equiv& \Big(1+\frac{P}{K_{P}}\Big)Z_{0}.
\end{eqnarray}}
Note that our goal at this point is to see if by defining the various ``constants'' that appear in 
Eqn.~\ref{eqn:AutoActivationWithPolymerase} we can show that it is equivalent to 
Eqn.~\ref{eqn:AutoActivationNoPolymerase} in which we ignored polymerase altogether.  In particular, we need to find effective versions of the parameters $K_d$, $\omega$, $r_0$, $r_1$ and $r_2$} that have all the polymerase dependence
hidden within them.
To re-express $Z_{0}$ as a sum of states with implicit dependence on polymerase, we note that if we define
\begin{equation}
    K_{d}^{\text{eff}} = \frac{1+\frac{P}{K_{P}}}{1+\frac{P}{K_{P}}e^{-\beta\varepsilon_{ap}}}K_{d},
    \end{equation}
    and 
    \begin{equation}
    \omega^{\text{eff}} = \frac{(1+\frac{P}{K_{P}})(1+\frac{P}{K_{P}}e^{-2\beta\varepsilon_{ap}})}{(1+\frac{P}{K_{P}}e^{-\beta\varepsilon_{ap}})^2}\omega
\end{equation}
then the denominator will have the same form as 
the denominator of Eqn.~\ref{eqn:AutoActivationNoPolymerase}.
Next, we see that if we redefine the rate parameters as
\begin{eqnarray}
    r_{0}^{\text{eff}} &=& \frac{\frac{P}{K_{P}}}{1+\frac{P}{K_{P}}}r_{0},\\
    r_{1}^{\text{eff}} &=& \frac{\frac{P}{K_{P}}e^{-\beta\varepsilon_{ap}}}{1+\frac{P}{K_{P}}e^{-\beta\varepsilon_{ap}}}r_{1},\\
    r_{2}^{\text{eff}} &=& \frac{\frac{P}{K_{P}}e^{-2\beta\varepsilon_{ap}}}{1+\frac{P}{K_{P}}e^{-2\beta\varepsilon_{ap}}}r_{2},
\end{eqnarray}
we recover an equation that is equivalent to the  dynamical equation as described in Eqn.~\ref{eqn:AutoActivationNoPolymerase}. Note that
for convenience, in Eqn.~\ref{eqn:AutoActivationNoPolymerase} we have everywhere dropped the superscript ``eff'' because the notation is way too cumbersome to carry throughout the paper. 
The key point is that we see that the two approaches are formally equivalent.

However, it is important to always bear in mind that the rate parameters, cooperativity, and dissociation constant used in the reduced representation of the paper are thus effective parameters that implicitly depend on the concentration of polymerase present ($P$), the strength of polymerase binding to the DNA ($K_{P}$), and the strength of interaction between activator and RNAP ($\varepsilon_{ap}$).  In a very real sense, this description will lead to a description of the auto-activation switch in which the polymerase serves as a hidden variable.  This analysis is extremely interesting because it  shows that there is a way to rigorously leave explicit treatment of the polymerase out of the problem.

\subsection{Coarse-graining the mutual repression model}\label{appendix:polymerase_mut_rep}

\begin{figure*}
    \centering
    \includegraphics[width=\linewidth]{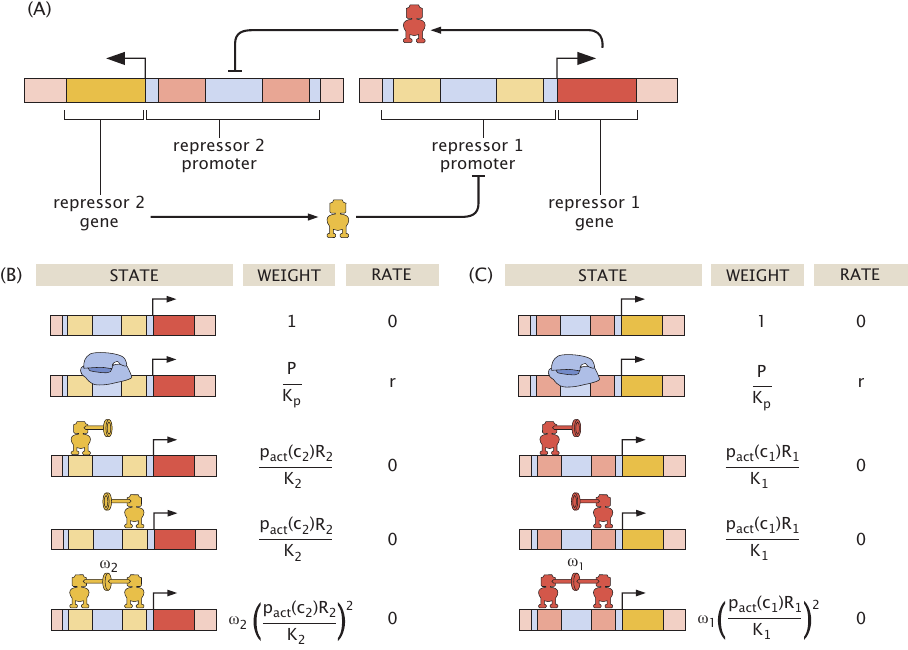}
    \caption{The mutual repression regulatory circuit. (A) Schematic of the operation of the circuit. When the gene for repressor 1 is expressed, the resulting protein downregulates the expression of the gene for repressor 2. Repressor 2, in turn, downregulates the expression of the gene for repressor 1.
     (B) Thermodynamic states, weights, and rates for expression of repressor 1 including the action of the inducer which tunes the number of active repressors. In our model, a repressor can bind non-exclusively at one of two possible sites within the target promoter region to suppress gene transcription. The parameter $\omega_2$ denotes the cooperative strength between two bound repressors $R_{2}$.  (C) Thermodynamic states, weights, and rates for expression of repressor 2 including the action of the inducer which tunes the number of active repressors. The states and weights for the regulation of the promoter responsible for the production of repressor 2 are analogous to those shown for repressor 1. However, the dissociation constant of repressor 1 in this case is given by $K_1$, and the cooperativity term for the interaction of two repressor 1 molecules bound to the DNA is $\omega_1$.}
    \label{fig:MutualRepression_P}
\end{figure*}

In the main body of the paper, just as we did for the auto-activation motif, we treated the mutual repression motif without making explicit reference to
RNA polymerase.  We now consider the full set of states, weights and rates illustrated in
Fig.~\ref{fig:MutualRepression_P}.  The states and weights shown here should be contrasted with those shown in Fig.~\ref{fig:MutRep} where all reference to RNA polymerase is absent.  We now demonstrate the
equivalence of these two descriptions of the mutual repression switch following precisely the same kind of strategy we followed above in the context of the auto-activation switch.

Under this expanded thermodynamic model, the governing equations for the dynamics of $R_1$ and $R_2$ prior to non-dimensionalization can be read off directly from
Fig.~\ref{fig:MutualRepression_P} yielding
\begin{equation}
\frac{dR_1}{dt} =  -\gamma_1 R_1 + \frac{r\frac{P}{K_P}}{1+\frac{P}{K_P} + 2 \frac{p_\text{act}(c_{2})R_2}{K_2}+\omega_2\left( \frac{p_\text{act}(c_{2})R_2}{K_2}\right)^2}
\end{equation}
and
\begin{equation}
\frac{dR_2}{dt} = -\gamma_2 R_2 + \frac{r\frac{P}{K_P}}{1+\frac{P}{K_P} + 2 \frac{p_\text{act}(c_{1})R_1}{K_1}+\omega_1\left( \frac{p_\text{act}(c_{1})R_1}{K_1}\right)^2}.
\end{equation}
These equations can be algebraically manipulated by factoring out $(1+P/K_P)$
from the denominator, resulting in the forms
{\small 
\begin{equation} 
\frac{dR_1}{dt}=-\gamma_1 R_1 + \frac{r\frac{\frac{P}{K_P}}{1+\frac{P}{K_P}}}{1+ 2 \frac{p_\text{act}(c_{2})R_2}{K_2(1+\frac{P}{K_P})}+\omega_2(1+\frac{P}{K_P})\left( \frac{p_\text{act}(c_{2})R_2}{K_2(1+\frac{P}{K_P})}\right)^2},
\end{equation}
\begin{equation}
\frac{dR_2}{dt}=-\gamma_2 R_2 + \frac{r\frac{\frac{P}{K_P}}{1+\frac{P}{K_P}}}{1+ 2 \frac{p_\text{act}(c_{1})R_1}{K_1(1+\frac{P}{K_P})}+\omega_1(1+\frac{P}{K_P})\left( \frac{p_\text{act}(c_{1})R_1}{K_1(1+\frac{P}{K_P})}\right)^2}. 
\end{equation}}
This formulation reveals that the original equations
given in Eqns.~\ref{eq:R1SwtichDynamics}
and~\ref{eq:R2SwtichDynamics} can be recovered through a simple transformation of parameters
in which we once again define effective parameters.  The effective mRNA production rate is given by 
\begin{equation}
r^{\text{eff}}= \dfrac{\frac{P}{K_P}}{1+\frac{P}{K_P}} r,
\end{equation}
the two $K_d$s for transcription factor-DNA binding are given by
\begin{equation}
K_1^{\text{eff}}= (1+\frac{P}{K_P})K_1
\end{equation}
and
\begin{equation}
K_2^{\text{eff}}= (1+\frac{P}{K_P})K_2
\end{equation}
and the two cooperativities are written in effective form as
\begin{equation}
\omega_1^{\text{eff}}= (1+\frac{P}{K_P}) \omega_1
\end{equation}
and
\begin{equation}
\omega_2^{\text{eff}}= (1+\frac{P}{K_P}) \omega_2.
\end{equation}
This demonstrates that polymerase binding can be absorbed into effective parameters, yielding a reduced model equivalent to the one presented in the main text, with renormalized production rate, dissociation constants, and cooperativities.  Once again, the polymerase copy number $P$ and binding strength $K_P$ are hidden variables in the context of the bare model, but the results are exact - this is not an approximation valid only in the limit of weak promoters, for example.  Note also that as in the case of the auto-activation switch, in the main body of the paper we do {\it not} carry around the cumbersome ``eff'' notation, electing instead to simply use the parameters $r$, $K_1$, $K_2$, $\omega_1$ and $\omega_2$ with the convention that those parameters include the hidden variables associated with polymerase.

\section{Minimal bounds on cooperativity and rates for the existence of bistability in auto-activation}

The auto-activation system can exhibit bistability, meaning that it can reach different steady states depending on the initial condition. However, this behavior arises only within a restricted range of parameter values, shown in Fig.~\ref{fig:autoactivation_ccmin_ccmax_sweep} as the red region. To investigate the conditions for which multiple different steady states are possible, we derive analytic bounds in parameter space. Setting $d\bar{A}/d\bar{t} = 0$ and re-expressing in standard polynomial form, $\bar{A}$ must satisfy
\begin{align}
    \omega p_\text{act}^ 2 \bar{A}^3 + p_\text{act}(2-\omega \bar{r}_2p_\text{act})\bar{A}^2+(1-2\bar{r}_1p_\text{act})\bar{A} - \bar{r}_0=0.
    \label{eq:auto_activation_steady_state_polynomial}
\end{align}

If the system exhibits bistability, the corresponding polynomial must have three real roots, as a third-order polynomial cannot have exactly two. Physically, these roots correspond to two stable steady states and one unstable steady state. Further, the presence of only one real root indicates that the system is monostable, as discussed in Appendix~\ref{appendix:auto_activation_more_theorem}.

To identify conditions for bistability, we search for combinations of $\omega$, $\bar{r}{0}$, $\bar{r}{1}$, $\bar{r}_{2}$, and effector concentration $c$ that produce three positive real roots of the polynomial—corresponding to the red region in Fig.~\ref{fig:autoactivation_ccmin_ccmax_sweep}. We may bound this bistable region of parameter space analytically using Descartes' rule of signs, which states that for a single-variable polynomial with real coefficients, the number of positive roots of the polynomial is equal to the number of sign changes between consecutive non-zero coefficients minus an even number. In our case, the polynomial in Eqn.~\ref{eq:auto_activation_steady_state_polynomial} must then have either one or three sign changes. Therefore, three sign changes are necessary for the system to allow bistability.  Evaluating Eqn.~\ref{eq:auto_activation_steady_state_polynomial}, we observe that the coefficient of $\bar{A}^3$ is strictly positive and the constant term is strictly negative. Thus, three (consecutive) coefficient sign changes are only possible if the second term of Eqn.~\ref{eq:auto_activation_steady_state_polynomial} is negative, and the third term of that same equation is positive. Specifically, the condition on the second term implies that
\begin{equation}
    p_\text{act}(c)\Big(2-\omega \bar{r}_2p_\text{act}(c)\Big) < 0 \implies p_\text{act}(c) > \frac{2}{\omega \bar{r}_2},
\end{equation}
while the condition on the second term leads to
\begin{equation}
    1-2\bar{r}_1p_\text{act}(c) > 0 \implies p_\text{act}(c) < \frac{1}{2\bar{r}_1}.
\end{equation}
Thus, these two conditions can be combined to yield
\begin{equation}
    \frac{2}{\omega \bar{r}_2} < p_\text{act}(c) < \frac{1}{2\bar{r}_1}. \label{eq:pactboundrange}
\end{equation}

\subsection{Necessary condition for bistability at some effector concentration $c$}
\label{appendix:autoact_derivation_of_descartes_bound}
Note that if the above condition were to be true at all possible effector concentrations $c$, the system would always be bistable. Rather, we are more specifically interested in the conditions that would allow bistability for at least one value of effector concentration $c_0$. In other words, there exists a concentration such that
\begin{equation}
   \frac{2}{\omega \bar{r}_2} <p_\text{act}(c_0)\label{eq:descartes_min_max1}
\end{equation}
and
\begin{equation}
    \frac{1}{2 \bar{r}_1} >p_\text{act}(c_0).
    \label{eq:descartes_min_max2}
\end{equation}
If the inequality in Eqn.~\ref{eq:descartes_min_max1} holds true, then we also know that
\begin{equation}
    \frac{2}{\omega \bar{r}_2} <p_\text{act}(c_0) \leq \max_{c\in[0,\infty]}[p_\text{act}(c)].
\end{equation}
This result itself then directly requires the existence of some effector concentration for which Eqn.~\ref{eq:descartes_min_max1} is true. We can prove this by considering two possible cases. First, if
\begin{equation}
    \frac{2}{\omega\bar{r}_{2}} < \min_{c\in[0,\infty]}[p_\text{act}(c)] < \max_{c\in[0,\infty]}[p_\text{act}(c)],
\end{equation}
then we know that Eqn.~\ref{eq:descartes_min_max1} holds true for all concentrations $c\geq 0$. Otherwise, if
\begin{equation}
    \min_{c\in[0,\infty]}[p_\text{act}(c)] < \frac{2}{\omega\bar{r}_{2}} < \max_{c\in[0,\infty]}[p_\text{act}(c)],
\end{equation}
then Eqn.~\ref{eq:descartes_min_max1} is true for all non-negative effector concentrations smaller than a threshold concentration
\begin{equation}
    c^* = p^{-1}_\text{act}\Big(\frac{2}{\omega\bar{r}_{2}}\Big),
\end{equation}
(derived from the inverse of Eqn.~\ref{eq:descartes_min_max1}) because $p_\text{act}$ is a continuous and monotonically decreasing function.

Applying similar logic to the inequality in Eqn.~\ref{eq:descartes_min_max2}, we may thus rewrite the necessary conditions for bistability in Eqns.~\ref{eq:descartes_min_max1} and~\ref{eq:descartes_min_max2} as
\begin{eqnarray}
   \frac{2}{\omega \bar{r}_2} &<& \max_{c \in [0, \infty]} \left( p_\text{act}(c) \right) = \frac{1}{1+e^{-\beta \varepsilon}}, \label{eq:pactboundintermed1}\\
   \frac{1}{2 \bar{r}_1} &>& \min_{c \in [0, \infty]} \left( p_\text{act}(c) \right) =\frac{1}{1+e^{-\beta \varepsilon}\bar{K}_c^2}, \label{eq:pactboundintermed2}
\end{eqnarray}
and
\begin{equation}
   \omega \bar{r_2}> 4\bar{r_1},
    \label{eq:pactboundintermed3}
\end{equation}
where we have recalled the saturation (maximum) and leakiness (minimum) of $p_\text{act}(c)$ defined in Eqns.~\ref{eq:pact_limit1} and~\ref{eq:pact_limit2}.
Note that we are in a setting where effector binding stabilizes the inactive form of the activator such that $\bar{K}_c=K_A/K_I>1$.
This fixes the values of saturation and leakiness, which would otherwise be switched if $K_c<1$. After some algebra, we can re-express Eqns.~\ref{eq:pactboundintermed1} and~\ref{eq:pactboundintermed2} such that the necessary conditions for bistability are given by
\begin{eqnarray}
    \frac{\omega \bar{r}_2}{2} &>& 1 + e^{-\beta \varepsilon},\label{eq:bistabconds1}\\
    1 + e^{-\beta \varepsilon}\bar{K}_c^2 &>& 2 \bar{r}_1,\label{eq:bistabconds2}
\end{eqnarray}
and
\begin{equation}
    \omega \bar{r}_2>4\bar{r}_1. \label{eq:bistabconds3}
\end{equation}
Following a similar procedure as in the previous section,  in the next section we derive a necessary condition for bistability that depends on the concentration of effector $c$. For fixed parameter values, this condition defines a bounded range of effector concentrations outside of which the system is guaranteed to be monostable.

\subsection{Necessary condition for bistability for a fixed concentration of effector $c$}
\label{appendix:descart_for_c_auto_act}

We consider the case where the activation probability $p_\text{act}(c)$ is a decreasing function of the effector concentration $c$, as seen in Fig.~\ref{fig:pact}. This monotonicity condition, which requires the derivative of the probability function to be negative for all possible $c$, depends on the parameters of the model, particularly the ratio of dissociation constants $\bar{K}_c$. The derivative of $p_\text{act}(c)$ with respect to $c$ is given from Eqn.~\ref{eqn:MWC2site} by
\begin{equation}
\frac{dp_\text{act}}{dc} = -\frac{2 (1 + c/K_A) e^{\beta \varepsilon} (-1 + \bar{K}_c)(1 + c/K_I)}{\left( (1 + c/K_A)^2 e^{\beta \varepsilon} + (1 + c/K_I)^2 \right)^2},
\end{equation}
which is negative for all $c > 0$ if and only if $\bar{K}_c > 1$. 

Recalling the previously-derived necessary condition for bistability,
\begin{equation}
\frac{2}{\omega \bar{r}_2} < p_\text{act}(c) < \frac{1}{2\bar{r}_1},
\end{equation}
we now investigate what constraint this condition imposes on the effector concentration $c$, assuming the parameters of the system are fixed. First, the inequality
\begin{equation}
    p_\text{act}(c) < \frac{1}{2\bar{r}_1} \label{eq:B2bound1}
\end{equation}
can be re-expressed equivalently using the explicit expression of $p_{\text{act}}(c)$ in Eqn.~\ref{eqn:MWC2site} as
\begin{align}
\displaystyle g(c) =\; &\left(\frac{c}{K_A}\right)^2\frac{1}{1+e^{-\beta \varepsilon}}\left[\frac{1}{2\bar{r}_1}-\frac{1}{1+e^{-\beta \varepsilon}\bar{K}_c^2}\right]\nonumber \\
&+ 2\frac{c}{K_A}\frac{\frac{1}{2\bar{r}_1}(1+e^{-\beta \varepsilon}\bar{K}_c)-1}{(1+e^{-\beta \varepsilon})(1+e^{-\beta \varepsilon}\bar{K}_c^2)} \nonumber \\
&+ \left[\frac{1}{2\bar{r}_1}-\frac{1}{1+e^{-\beta \varepsilon}}\right]\frac{1}{1+e^{-\beta \varepsilon}\bar{K}_c^2} > 0.
\end{align}
Similarly, the condition 
\begin{equation}
    p_\text{act}(c) > \frac{2}{\omega \bar{r}_2} \label{eq:B2bound1}
\end{equation}
is equivalent to requiring that
\begin{align}
h(c) =\; &\left(\frac{c}{K_A}\right)^2\frac{1}{1+e^{-\beta \varepsilon}}\left[\frac{2}{\omega \bar{r}_2}-\frac{1}{1+e^{-\beta \varepsilon}\bar{K}_c^2}\right] \\
&+ 2\frac{c}{K_A}\frac{\frac{2}{\omega \bar{r}_2}(1+e^{-\beta \varepsilon}\bar{K}_c)-1}{(1+e^{-\beta \varepsilon})(1+e^{-\beta \varepsilon}\bar{K}_c^2)} \nonumber \\
&+ \left[\frac{2}{\omega \bar{r}_2}-\frac{1}{1+e^{-\beta \varepsilon}}\right]\frac{1}{1+e^{-\beta \varepsilon}\bar{K}_c^2} < 0. \nonumber
\end{align}

We can now apply Descartes’ Rule of Signs to the polynomials $g(c)$ and $h(c)$ to determine when the inequalities are satisfied. Since we are working under the assumption that $\bar{K}_c > 1$, this means that,
\begin{equation}
p_\text{act}^{\text{max}} = \frac{1}{1+e^{-\beta \varepsilon}} > \frac{1}{1+e^{-\beta \varepsilon} \bar{K}_c} > \frac{1}{1+e^{-\beta \varepsilon} \bar{K}_c^2} = p_\text{act}^{\text{min}}.
\end{equation}
For the polynomial $g(c)$, three cases then arise. First, if
\begin{equation}
    \frac{1}{2\bar{r}_1} > \frac{1}{1+e^{-\beta \varepsilon}},
\end{equation}
then all coefficients of $g(c)$ are positive and $g(c) > 0$ for all $c \geq 0$, so Eqn.~\ref{eq:B2bound1} is always satisfied. Second, if
\begin{equation}
    \frac{1}{2\bar{r}_1} > \frac{1}{1+e^{-\beta \varepsilon} \bar{K}_c^2},
\end{equation}
then all coefficients are negative, and the condition is never satisfied for any $c$. Finally, if the intermediate case
\begin{equation}
\frac{1}{1 + e^{-\beta \varepsilon} \bar{K}_c} > \frac{1}{2\bar{r}_1} > \frac{1}{1 + e^{-\beta \varepsilon} \bar{K}_c^2}
\end{equation}
holds, then the coefficient of the term proportional to $c^2$ in $g(c)$ is positive, while those of the remaining terms proportional to $c^1$ and $c^0$ are negative. This results in exactly one sign change, so by Descartes’ Rule of Signs, the polynomial $g(c)$ has exactly one positive root. This defines the minimal concentration for bistability, denoted $c_{\text{bistab}}^{\min}(\bar{r}_1)$, and given by
{\footnotesize
\begin{equation}
c_{\text{bistab}}^{\min}(\bar{r}_1) = K_A
\frac{e^{-\beta \varepsilon} \bar{K}_c + 2\bar{r}_1 - 1 + \sqrt{e^{-\beta \varepsilon} (1 + \bar{K}_c)^2 (2\bar{r}_1 - 1)}}{e^{-\beta \varepsilon} \bar{K}_c^2 - 2\bar{r}_1 + 1}.
\end{equation}}
Under these conditions, the inequality $g(c) > 0$ holds for all $c > c_{\text{bistab}}^{\min}(\bar{r}_1)$.

We now turn to the polynomial $h(c)$. If
\begin{equation}
    \frac{2}{\omega \bar{r}_2} > \frac{1}{1 + e^{-\beta \varepsilon}},
\end{equation}
then all coefficients are positive and the polynomial is strictly positive for all $c$, meaning that the condition in Eqn.~\ref{eq:B2bound1} is never satisfied. Conversely, if 
\begin{equation}
    \frac{2}{\omega \bar{r}_2} < \frac{1}{1 + e^{-\beta \varepsilon} \bar{K}_c^2},
\end{equation}
then all coefficients are negative and the condition is always satisfied. Lastly, in the intermediate case
\begin{equation}
\frac{1}{1 + e^{-\beta \varepsilon} \bar{K}_c} > \frac{2}{\omega \bar{r}_2} > \frac{1}{1 + e^{-\beta \varepsilon} \bar{K}_c^2},
\end{equation}
Descartes’ Rule of Signs again implies exactly one positive root of $h(c)$, corresponding to the upper bound of the bistable region. This upper concentration threshold is denoted $c_{\text{bistab}}^{\max}(\omega \bar{r}_2)$ and given by
{\footnotesize
\begin{equation}
c_{\text{bistab}}^{\max}(\omega \bar{r}_2) = K_A
\frac{e^{-\beta \varepsilon} \bar{K}_c + \frac{\omega \bar{r}_2}{2} - 1 + \sqrt{e^{-\beta \varepsilon} (1 + \bar{K}_c)^2 \left( \frac{\omega \bar{r}_2}{2} - 1 \right)}}{e^{-\beta \varepsilon} \bar{K}_c^2 - \frac{\omega \bar{r}_2}{2} + 1}.
\end{equation}}
Under these conditions, the inequality $h(c) < 0$ holds for all $c < c_{\text{bistab}}^{\max}(\omega \bar{r}_2)$.

Summing up the case-by-case analysis, we derive an effector concentration-dependent necessary condition for bistability. The full set of conditions allowing for bistability in different ranges of effector concentrations is given by
\begin{equation}
    \begin{cases}
    c_{\text{bistab}}^{\max} > c > c_{\text{bistab}}^{\min} & \text{if } \frac{1}{1 + e^{-\beta \varepsilon} \bar{K}_c^2} < \frac{2}{\omega \bar{r}_2} < \frac{1}{2\bar{r}_1} < \frac{1}{1 + e^{-\beta \varepsilon}}, \\
    c_{\text{bistab}}^{\max} > c & \text{if } \frac{1}{1 + e^{-\beta \varepsilon} \bar{K}_c^2} < \frac{2}{\omega \bar{r}_2} < \frac{1}{1 + e^{-\beta \varepsilon}} < \frac{1}{2\bar{r}_1}, \\
    c > c_{\text{bistab}}^{\min} & \text{if } \frac{2}{\omega \bar{r}_2} < \frac{1}{1 + e^{-\beta \varepsilon} \bar{K}_c^2} < \frac{1}{2\bar{r}_1} < \frac{1}{1 + e^{-\beta \varepsilon}}, \\
    c \geq 0 & \text{if } \frac{2}{\omega \bar{r}_2} < \frac{1}{1 + e^{-\beta \varepsilon} \bar{K}_c^2} < \frac{1}{1 + e^{-\beta \varepsilon}} < \frac{1}{2\bar{r}_1}, \\
    \text{no bistability} & \text{if } \frac{2}{\omega \bar{r}_2} > \frac{1}{1 + e^{-\beta \varepsilon}} \text{ or } \frac{1}{1 + e^{-\beta \varepsilon} \bar{K}_c^2} > \frac{1}{2\bar{r}_1}.
    \end{cases}
    \label{eq:auto_act_all_decartes}
\end{equation}

As noted, the parameter $\bar{r}_0$ does not enter into these Descartes-based bounds and thus does not influence the existence of bistability in this analysis. From these expressions, we recover the necessary conditions for bistability, stated in the previous section as
\begin{equation}
    \frac{2}{\omega \bar{r}_2} < \frac{1}{1 + e^{-\beta \varepsilon}},
\end{equation}
\begin{equation}
    \frac{1}{1 + e^{-\beta \varepsilon} \bar{K}_c^2} < \frac{1}{2\bar{r}_1},
\end{equation}
and
\begin{equation}
    \frac{2}{\omega r_2} < \frac{1}{2\bar{r}_1}.
\end{equation}
and re-expressed in Eqns.~\ref{eq:bistabconds_final1} -~\ref{eq:bistabconds_final3}.
\begin{figure*}[t]
    \centering
    \includegraphics[width=\linewidth]{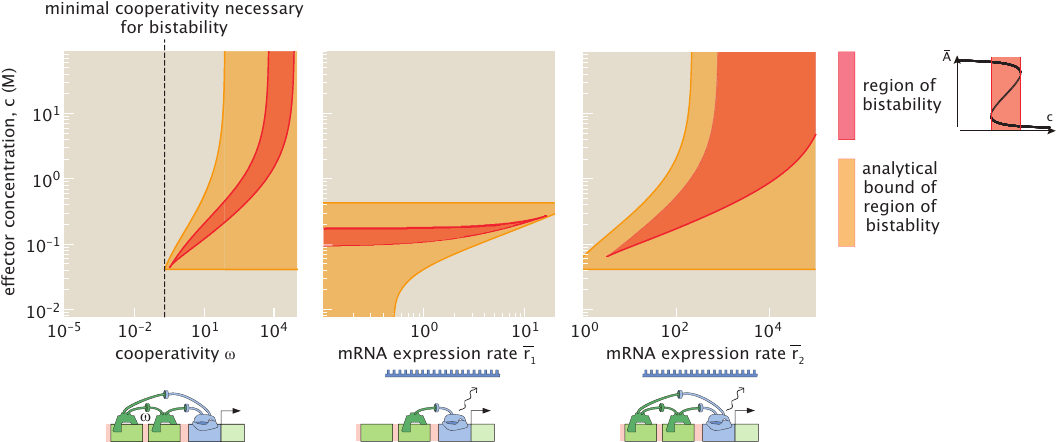}
    \caption{Minimal and maximal values of effector concentration between which the system is bistable. Given baseline parameter values $\omega =7.5$, $\bar{r}_0=0.1$, $\bar{r}_1=1$, $\bar{r}_2=20$, each figure varies a different parameter, keeping all others fixed. %The concentration of effector is sampled for $c/K_A \in [10^{-4},10^2]$. 
    For each panel, the shaded region in red is the region of effector concentration for which there is bistability. The shaded region in orange denotes the analytically-bounded region of bistability. 
    The dotted line in the first figure is an analytical lower bound for the minimal cooperativity required for bistability. Note that the analytic approach discussed here, summed up in Eqn.~\ref{eq:auto_act_all_decartes}, invokes a necessary but not sufficient condition for bistability, and thus always encompasses a larger region of parameter space than the system's observed region in red. }
    \label{fig:autoactivation_ccmin_ccmax_sweep_SI}
\end{figure*}
These conditions can equivalently be rewritten as
\begin{equation}
    \omega \bar{r}_2 > \max\left(2(1 + e^{-\beta \varepsilon}),\ 4\bar{r}_1\right)
\end{equation}
and
\begin{equation}
    \bar{r}_1 < \min\left(\frac{1 + e^{-\beta \varepsilon} \bar{K}_c^2}{2},\ \frac{\omega \bar{r}_2}{4}\right).
\end{equation}

From Eqn.~\ref{eq:auto_act_all_decartes}, we identify necessary conditions under which the system displays bistability for all effector concentrations above a minimal threshold. This corresponds to being in either the third or fourth case of Eqn.~\ref{eq:auto_act_all_decartes}. These cases are captured by the inequality
\begin{equation}
\omega \bar{r}_2 > 2(1 + e^{-\beta \varepsilon} \bar{K}_c^2),
\end{equation}
which implies that, for a sufficiently large product $\omega \bar{r}_2$, the system permits bistability across a semi-infinite range of effector concentrations.

To complement the analytical results summarized in Eqn.~\ref{eq:auto_act_all_decartes}, we compare the derived necessary conditions for bistability with numerically computed bistability regions across different parameters. As shown in Figure~\ref{fig:autoactivation_ccmin_ccmax_sweep_SI}, the analytically predicted bounds—represented in orange—are in close agreement with the numerically determined region of bistability—shown in red—near the onset of bistability. For larger values of the cooperativity parameter $\omega$ or the activation rate $\bar{r}_2$, as well as for smaller values of the intermediate rate $\bar{r}_1$, the analytical bounds significantly overestimate the true bistable region. This discrepancy arises because the analytical bounds are necessary but not sufficient conditions, and therefore do not capture the full behavior of the system. Nevertheless, these bounds offer a valuable predictor of the minimal and maximal effector concentrations that can support bistability under a given set of parameters.

\section{Fixed point structure of the auto-activation system as a gradient flow}
\label{appendix:auto_activation_more_theorem}
The auto-activation dynamical system defined in Eqn.~\ref{eq:auto_activation_master_equation} is a dynamical system that derives from a gradient. Indeed, we can write this equation as
\begin{equation}
    \frac{d\bar{A}}{d\bar{t}}=-\frac{dV}{d\bar{A}},
\end{equation}
with
\begin{equation}
   -\frac{dV}{d\bar{A}}= \frac{P(\bar{A})}{1 +  2 p_\text{act}(c)\bar{A} + \omega (p_\text{act}(c)\bar{A})^2},
\label{eq:appendix_aut_act_definition_of_potentiel}
\end{equation}
and 
\begin{align}
P(\bar{A})&=r_0 + (r_1 2 p_\text{act}(c)-1)\bar{A} \\
&+ (r_2\omega (p_\text{act}(c))^2-2 p_\text{act}(c))\bar{A}^2-\omega (p_\text{act}(c))^2\bar{A}^3 \nonumber\\
&=\omega p^2_{act}(c)(\bar{A}-\bar{A}_1)(\bar{A}-\bar{A}_2)(\bar{A}-\bar{A}_3),\nonumber
\end{align}
 where $(\bar{A}_1,\bar{A}_2,\bar{A}_3) \in \mathbb{R}
$, if there is bistability.

Given Eqn.~\ref{eq:appendix_aut_act_definition_of_potentiel}, our goal now is to determine the landscape $V(A)$ itself.  To that end, we  need to integrate Eqn.~\ref{eq:appendix_aut_act_definition_of_potentiel}.  We invoke the strategy of separation of variables, resulting in
\begin{equation}
   -dV = d\bar{A} \cdot \frac{1 + 2 p_\text{act}(c)\bar{A} + \omega (p_\text{act}(c)\bar{A})^2}{\omega p_\text{act}^2(c)(\bar{A}-\bar{A}_1)(\bar{A}-\bar{A}_2)(\bar{A}-\bar{A}_3)}.
\end{equation}
To make progress with this integral, 
we express the right-hand side using partial fraction decomposition. This yields
\begin{equation}
    \frac{1 + 2 p_\text{act}(c)\bar{A} + \omega (p_\text{act}(c)\bar{A})^2}{\omega p_\text{act}^2(c)\prod_{i=1}^3 (\bar{A} - \bar{A}_i)} = \sum_{i=1}^3 \frac{C_i}{\bar{A} - \bar{A}_i}.
\end{equation}
We find the coefficients $C_1$, $C_2$, and $C_3$ by multiplying through by the common denominator and evaluating at $\bar{A} = \bar{A}_i$, for $i \in \{1, 2, 3\}$. The resulting expressions are
\begin{equation}
\begin{cases}
    C_1 = \displaystyle \frac{\frac{1}{p_\text{act}^2(c)\omega} +  \frac{2}{p_\text{act}(c)\omega}\bar{A}_1 + \bar{A}_1^2}{(\bar{A}_1 - \bar{A}_2)(\bar{A}_1 - \bar{A}_3)}, \\[8pt]
    C_2 = \displaystyle \frac{\frac{1}{p_\text{act}^2(c)\omega} +  \frac{2}{p_\text{act}(c)\omega}\bar{A}_2 + \bar{A}_2^2}{(\bar{A}_2 - \bar{A}_1)(\bar{A}_2 - \bar{A}_3)} ,\\[8pt]
    C_3 = \displaystyle \frac{\frac{1}{p_\text{act}^2(c)\omega} +  \frac{2}{p_\text{act}(c)\omega}\bar{A}_3 + \bar{A}_3^2}{(\bar{A}_3 - \bar{A}_1)(\bar{A}_3 - \bar{A}_2)}.
\end{cases}
\end{equation}
We can then write the potential function $V(\bar{A})$ as
\begin{equation}
    V(\bar{A}) = C_1 \ln|\bar{A} - \bar{A}_1| + C_2 \ln|\bar{A} - \bar{A}_2| + C_3 \ln|\bar{A} - \bar{A}_3|.
\end{equation}

Since the auto-activation system derives from a gradient, we can apply classical results from one-dimensional gradient dynamics: namely, that the number of stable steady states is equal to the number of unstable steady states plus one ~\cite{Strogatz2015}.
Indeed let's assume that
\begin{equation}
    \frac{dV(\bar{A})}{d\bar{A}}|_{\bar{A}=\bar{A}^i}=0
\end{equation}
at finitely many points $i \in [1,n]$ and 
\begin{equation}
    \frac{d^2V(\bar{A})}{d\bar{A}^2}|_{\bar{A}=\bar{A}^i}\neq0
\end{equation}
at those points (the stable points are not degenerate). We take $\bar{A}_1<...<\bar{A}_n$. With two minima in $V$, the function must then reach a local maximum between the two to transition between these minima. We therefore see that the local minima and maxima of $V$ must alternate. A last key point is why that the first and last extrema of $V$ must be minima. If the first extremum of $V$ were a maximum—corresponding to an unstable steady state—a small perturbation toward smaller $\bar{A}$ would drive the system toward the boundary of the domain, where no minimum of $V$ exists and no steady state is defined. This would render the system ill-posed. A similar reasoning can be applied to understand why the last steady state also has to be a minimum. So we can apply this to our system. Intuitively, imagining our dynamical system as a one-dimensional energy landscape, two stable steady state ``valleys'' must be connected by an unstable steady state ``hill.'' Therefore, bistability implies that our system has three steady states, two stable and one unstable. 

\section{Auto-activation : No bistability at high cooperativity and rate $\bar{r}_2$.}
\label{appendix:auto_act_no_bistab_high_coop}

\begin{figure*}[t]    
\centering
    \includegraphics[width=\linewidth]{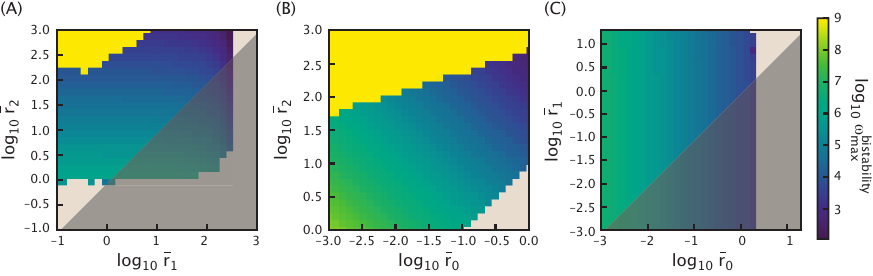}
\caption{Parameter space exploration of the maximal cooperativity $\omega_{\max}^{\text{bistable}}$ above which the system becomes monostable for all effector concentrations.
The cooperativity $\omega$ is sampled over the interval $\omega \in [10^{1}, 10^9]$. Effector concentrations are effectively scanned from 0 to $\infty$ by sweeping $p_\text{act}$ between its biologically constrained bounds: the leakiness level $p_\text{act}^{\text{min}} = \frac{1}{1 + e^{-\beta \varepsilon} \bar{K}_c^2}$ and the saturation level $p_\text{act}^{\text{max}} = \frac{1}{1 + e^{-\beta \varepsilon}}$, with fixed parameters $\beta \varepsilon = 4.5$ and $\bar{K}_c = 2.6 \times 10^2$. Three two-dimensional parameter sweeps are performed. In panel (A), $(\bar{r}_1, \bar{r}_2)$ are varied in $[\bar{r}_0, 10^5] \times [\bar{r}_0, 10^5]$ with $\bar{r}_0 = 0.1$ held constant. In panel (B), $(\bar{r}_0, \bar{r}_2)$ are varied in $[10^{-5}, \bar{r}_1] \times [\bar{r}_1, 10^5]$ with $\bar{r}_1 = 1$ fixed. In panel (C), $(\bar{r}_0, \bar{r}_1)$ are varied in $[10^{-5}, \bar{r}_2] \times [10^{-5}, \bar{r}_2]$ with $\bar{r}_2 = 20$ fixed. Regions shaded in gray correspond to parameter combinations that violate the auto-activation condition $\bar{r}_0 \leq \bar{r}_1 \leq \bar{r}_2$, and for which the system no longer functions as an auto-activating unit. In regions where no maximal cooperativity values for bistability are reported, the system remains monostable across the entire range of cooperativity values sampled.}
\label{fig:autoact_max_coop_for_bifurcation}
\end{figure*}

As shown in Fig.~\ref{fig:autoactivation_ccmin_ccmax_sweep}, for sufficiently large values of $\omega$ and $\bar{r}_2$, the system does not exhibit bistability for any effector concentration $c$. In this section, we support this observation using bi-dimensional numerical parameter sweeps and provide analytical arguments explaining its origin.

In Fig.~\ref{fig:autoact_max_coop_for_bifurcation}, we report the maximal cooperativity $\omega$ above which the system is monostable for all values of effector concentration. The rate parameters are varied two at a time while keeping the third fixed. For each triplet $(\bar{r}_0, \bar{r}_1, \bar{r}_2)$, we sample all effector concentrations by sweeping over values of $p_\text{act}$ between leakiness and saturation. We then determine the maximal value of $\omega$ for which the system is bistable for at least one value of $c$. These parameter sweeps reveal a finite—but potentially large—upper bound on cooperativity beyond which bistability is lost. The yellow regions in Fig.~\ref{fig:autoact_max_coop_for_bifurcation}(A–B) indicate that no upper bound was found within our sampled cooperativity range ($10^1$ to $10^9$); this absence does not imply the bound does not exist, but rather reflects the limits of our numerical exploration, which did not extend beyond $10^9$ due to sampling choices and diminishing biophysical relevance. However, since a finite bound exists in other parts of parameter space, we hypothesize that such a bound also exists in these regions. In the next section, we confirm this analytically.
Interestingly, the appearance of yellow regions in Fig.~\ref{fig:autoact_max_coop_for_bifurcation}(A–B)—where no numerical upper bound on cooperativity is observed—correlates with increasing values of $\bar{r}_2$, consistent with the fact that raising $\bar{r}_2$ initially promotes bistability. While these bounds appear only at very high cooperativities (typically $\omega > 10^2$), and may exceed biologically plausible values, they nonetheless depend on the system’s rate parameters and could be lower in other settings.

We explain this behavior analytically. As discussed in Appendix~\ref{appendix:descart_for_c_auto_act}, bistability can be assessed by examining the number of non-negative roots of the steady-state polynomial Eqn.~\ref{eq:auto_activation_steady_state_polynomial}. To be bistable, the system has to admit more than one steady state, which corresponds to the polynomial having three real non-negative roots. A necessary condition for this, is that the polynomial has three real roots, regardless of their sign. While this condition does not guarantee bistability—since some of the roots may be negative—it is nonetheless governed directly by the sign of the polynomial's discriminant. For a general cubic polynomial 
\begin{equation}
    Q(x) = ax^3 + bx^2 + cx + d,
\end{equation}
the discriminant is given by 
\begin{equation}
   \Delta = b^2c^2 - 4ac^3 - 4b^3d - 27a^2d^2 + 18abcd.
\end{equation}
If $\Delta > 0$, the polynomial has three distinct real roots; if $\Delta < 0$, it has only one real root; and if $\Delta = 0$, it has at least one repeated root.

For the auto-activation system, letting $p_\text{act}(c) = p$, the discriminant of the steady-state polynomial can be written as
{\small
\begin{align}
&\Delta =\  p^2 \Big[ 
 4 + 32p\, \bar{r}_0 + 16p\, \bar{r}_1 (-1 + p\, \bar{r}_1) \\
& + 4 \omega (-1 + 2p\, \bar{r}_1)\left(1 + 9p\, \bar{r}_0 + 4p\, \bar{r}_1 (-1 + p\, \bar{r}_1)\right) \nonumber\\
& - 4 \omega p \left(1 + 4p (3 \bar{r}_0 + \bar{r}_1 (-1 + p\, \bar{r}_1))\right) \bar{r}_2 \nonumber \\
& - 4 \omega^3 p^4 \bar{r}_0\, \bar{r}_2^3 \nonumber \\
& + \omega^2 p^2 \left[ -27\, \bar{r}_0^2 + \left(1 - 2p\, \bar{r}_1\right)^2 \bar{r}_2^2 
+ 6 \bar{r}_0\, \bar{r}_2 \left(3 - 6p\, \bar{r}_1 + 4p\, \bar{r}_2\right) \right] \Big].\nonumber
\end{align}
}

We now study the asymptotic behavior of this discriminant in the limits of infinite $\omega$, $\bar{r}_2$, and $\bar{r}_0$. We do not consider the limit of infinite $\bar{r}_1$, since, according to the bounds in Eqns.~\ref{eq:bistabconds_final1} -~\ref{eq:bistabconds_final3}, a necessary condition for bistability is that $\bar{r}_1$ remains below a threshold set by the other system parameters. Therefore, in this limit, the system is necessarily monostable.

In each of the asymptotic limits, we derive the leading-order term of the discriminant and infer the discriminant diverges negatively.
Respectively for $\omega$,
\begin{equation}
\begin{cases}
\Delta \underset{\omega \to \infty}{\sim} -4p^6 \bar{r}_0 \bar{r}_2^3 \omega^3,\\
\displaystyle\lim_{\omega \to \infty} \Delta = -\infty,
\label{eq:discriminant_large_omega}
\end{cases}
\end{equation}
for $\bar{r_2}$,
\begin{equation}
\begin{cases}
\Delta \underset{\bar{r}_2 \to \infty}{\sim} -4p^6 \bar{r}_0 \bar{r}_2^3 \omega^3,\\
\displaystyle\lim_{\bar{r}_2 \to \infty} \Delta = -\infty,
\label{eq:discriminant_large_r2}
\end{cases}
\end{equation}
and for $\bar{r}_0$,

\begin{equation}
\begin{cases}
\Delta \underset{\bar{r}_0 \to \infty}{\sim} -27 \omega^2 p^4 \bar{r}_0^2,\\
\displaystyle\lim_{\bar{r}_0 \to \infty} \Delta = -\infty.
\label{eq:discriminant_large_r0}
\end{cases}
\end{equation}
These asymptotic results indicate that bistability becomes impossible in the limit of arbitrarily large $\omega$, $\bar{r}_2$, or $\bar{r}_0$.

Since a negative discriminant implies that the polynomial admits only one real root, the system is necessarily monostable in these asymptotic regimes. This analytically supports the existence of upper bounds on $\omega$ and $\bar{r}_2$ observed in Fig.~\ref{fig:autoact_max_coop_for_bifurcation} and Fig.~\ref{fig:autoactivation_ccmin_ccmax_sweep}(A,D), consistent with the scaling behaviors shown in Eqns.\ref{eq:discriminant_large_omega} and \ref{eq:discriminant_large_r2}. Moreover, the symmetric structure of these leading-order terms highlights the seemingly interchangeable roles of $\omega$ and $\bar{r}_2$ in promoting bistability. The loss of bistability for large $\bar{r}_0$, as seen in Fig.~\ref{fig:autoactivation_ccmin_ccmax_sweep}(B), is similarly explained by the negative divergence of the discriminant in Eqn.~\ref{eq:discriminant_large_r0}.

\section{Conditions for activation in auto-activation circuit}
\label{appendix:condition_r1_r2_autoact}

We define the range of parameters on which we will focus in the setting of the study of auto-activation. In this framework, the production term in Eqn.~\ref{eq:auto_activation_master_equation}, which we will refer to as $y(\bar{A})$, must be monotonically increasing. In other words, we want $dy/d\bar{A} \geq 0$ for all $\bar{A} \geq 0$. To simplify the computation, let $x = p_\text{act}(c)\bar{A}$. We then have 
\begin{equation}
    \frac{dy}{dx} = \frac{1}{p_\text{act}(c)}\frac{dy}{d\bar{A}},
\end{equation}
and the condition thus becomes $dy/dx \geq 0$ for all $x \geq 0$. 

Writing down the expression of $y(x)$,
\begin{equation}
    y(x)=\frac{\bar{r}_0+2\bar{r}_1x+\omega \bar{r}_2 x^2}{1+2x+\omega x^2},
\end{equation}
we compute the derivative
\begin{equation}
\frac{dy}{dx}=\frac{2 \left(\bar{r}_1 - \bar{r}_0 + x\omega (\bar{r}_2 - \bar{r}_0) + x^2 \omega (\bar{r}_2 - \bar{r}_1)\right)}{\left(1 + x \left(2 + x\omega\right)\right)^2}.
\end{equation}
For this expression to be non-negative for all $x$, it must be non-negative for $x=0$ and for $x\to \infty$. This then requires that $\bar{r}_2 \geq \bar{r}_1$ and $\bar{r}_1 \geq \bar{r}_0$, further implying that $\bar{r}_2 \geq \bar{r}_0$. This is enough to assert that $y(x)$ is an increasing function of $x$ as its derivative is always non-negative for all $x \geq 0$.

\section{Comparison between the Hill and thermodynamic models for auto-activation}
\label{appendix:hill_v_thermo_sup}

\begin{figure*}
    \centering
    \includegraphics[width=0.6\linewidth]{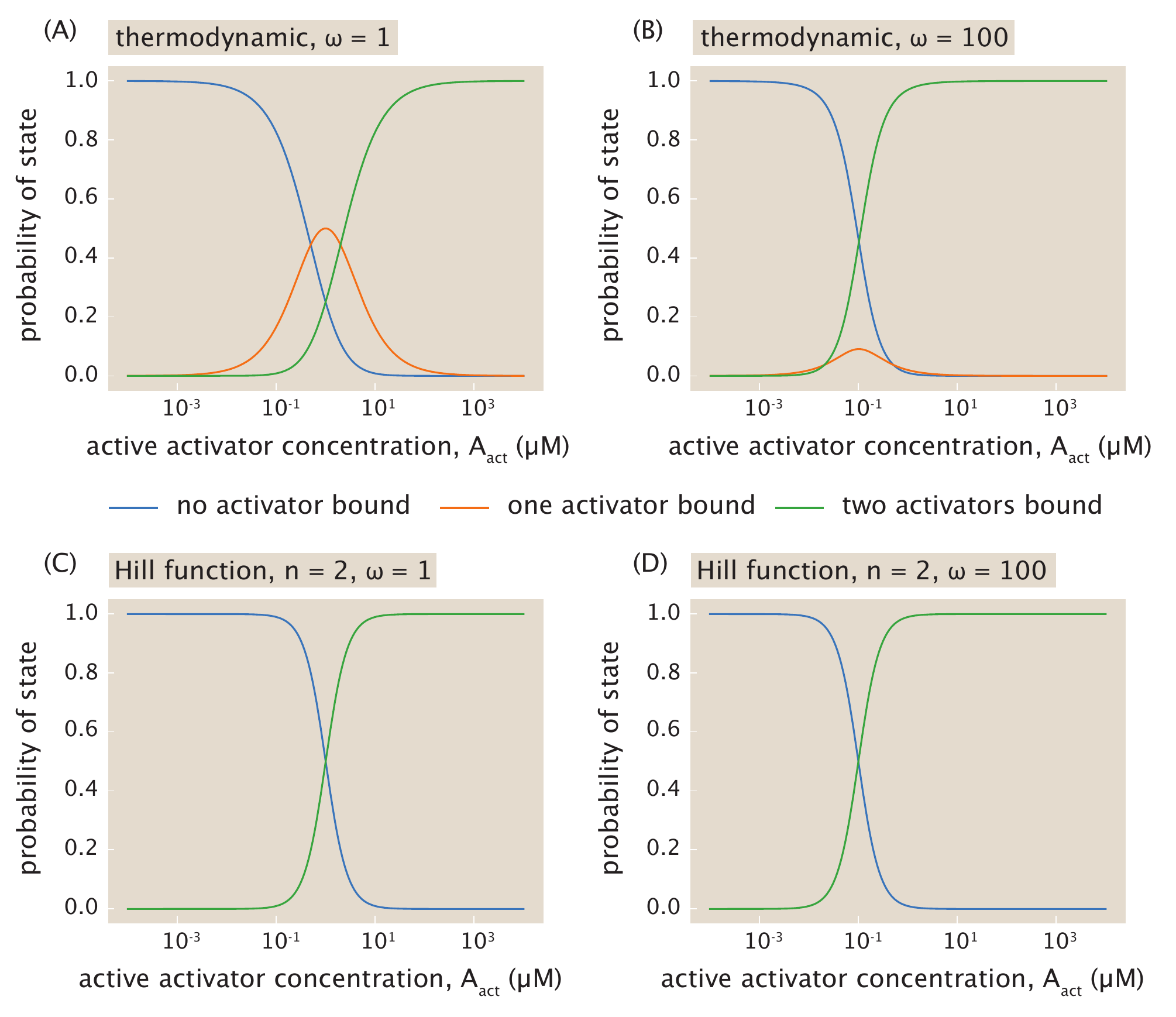}
    \caption{Comparing probabilities of activator binding states as a function of active activator concentration $A_{\text{act}}$ in the thermodynamic and Hill model. $K_d = 1\mu$M across all panels. (A) Thermodynamic model with cooperativity $\omega = 1$. (B) Thermodynamic model with $\omega = 100$. (C) Hill function model with $\omega=1$. (D) Hill function model with $\omega=100$. The blue curve corresponds to the state with no activator bound. The orange curve corresponds to the state with one activator bound. The green curve corresponds to the state with two activators bound. Notably, there is no orange curve in (C) as the Hill function approximates away the state with one activator bound.}
    \label{fig:hill_v_thermo_sup2}
\end{figure*}

In this section, we provide support for some of the claims made in Section~\ref{sec:hill_v_thermo} when comparing the use of thermodynamic and Hill function approaches to model auto-activation.

First, we demonstrate how a Hill function emerges in the high cooperativity limit of the thermodynamic model, a result known previously in the literature \cite{Sherman2012}. To reiterate, the dynamical equation for protein production through auto-activation using a thermodynamical model is
\begin{align}
    \begin{split}
    \frac{dA}{dt} = -\gamma A &+ \frac{r_0 + r_1 (2p_\text{act}(c)\frac{A}{K_d}) + r_2 \omega (p_\text{act}(c)\frac{A}{K_d})^2}{1+2p_\text{act}(c)\frac{A}{K_d} + \omega (p_\text{act}(c)\frac{A}{K_d})^2}.
\end{split}
\end{align}
Letting $K_d^\text{eff} =K_d/\sqrt{\omega}$, we can then re-express the previous equation as
\begin{align}
    \frac{dA}{dt} = -\gamma A + \frac{r_0 + r_1 2p_\text{act}(c) \frac{A}{K_d^\text{eff}\sqrt{\omega}} + r_2 (p_\text{act}(c) \frac{A}{K_d^\text{eff}})^2}{1 + 2p_\text{act}(c) \frac{A}{K_d^\text{eff}\sqrt{\omega}} + (p_\text{act}(c) \frac{A}{K_d^\text{eff}})^2}.
\end{align}
In the limit as $\omega \to \infty$ with finite $K_d^{\text{eff}}$, the single activator-bound state vanishes and the dynamics simplify to 
\begin{align}
    \frac{dA}{dt} = -\gamma A &+ \frac{r_0+r_2(p_\text{act}(c)\frac{A}{K_d^\text{eff}})^2}{1 + (p_\text{act}(c)\frac{A}{K_d^\text{eff}})^2}.
\end{align}
The production term thus takes the form of a Hill function with $n=2$, as in Eqn.~\ref{eq:autoactdim}.

We next compare the probabilities of different state occupancies as derived from the thermodynamic and Hill function models. Fig.~\ref{fig:hill_v_thermo_sup2} plots the probabilities of no TF bound, one TF bound, and two TFs bound as functions of the active activator concentration $A_{\text{act}} = p_{\text{act}}(c) A$. In the thermodynamic model the probability of no activator bound is
\begin{align}
    \frac{1}{1+2\frac{A_{\text{act}}}{K_d} + \omega \left(\frac{A_{\text{act}}}{K_d}\right)^2}.
\end{align}
In the high cooperativity regime, the singly-bound state has negligible weight, and the probabilities of zero or two activators being bound, as plotted in Fig.~\ref{fig:hill_v_thermo_sup2}(B), closely match those given by the Hill function in Fig.~\ref{fig:hill_v_thermo_sup2}(D). The correspondence is weaker in the low cooperativity regime but still visible. The probabilities of zero or two activators being bound, plotted in Fig.~\ref{fig:hill_v_thermo_sup2}(A) and (C) for both models, share a sigmoidal shape and have similar EC50. Where the probability curves intersect at $A_{\text{act}} = 1\,\mu$M, however, the Hill model assigns higher probabilities than the thermodynamic model because the singly bound state contributes strongly in the latter setting.

Overall, these differences appear moderate, and it is difficult to anticipate from these results alone the substantive differences between the two models in the small $\omega$ regime that we present in the main text. This highlights the importance of comparing the two models not only through their production curves but also through their bifurcation curves.

\section{Oligomerization and DNA Looping in the auto-activation regulation unit}\label{app:oligo}

In the main text, transcription factors (TFs) are treated as single binding units that occupy one site at a time, with cooperative interactions between TFs bound at two neighboring sites on the DNA. This description, however, omits two important possibilities: (i) TFs may oligomerize before binding DNA (for example as dimers, like the transcription factor PhoP in \textit{E. coli} \cite{perron2005dimerization}, or as tetramers, like the transcription factor ComK \cite{hamoen1998competence} in \textit{B. subtilis}); and (ii) they may bridge distant sites to form DNA loops~\cite{morelli2009dna,anderson2008dna,vilar2011control, vilarSaiz2013}. Here we consider how these features would change the behavior of an auto-activating transcription factor, with similar analysis possible for mutual repression and other regulatory motifs.

To isolate the effects of oligomerization, the following discussion does not include inducers, though in general they could interact differently with monomers and oligomers. To extend the model, Eqn. \ref{eqn:AutoActivationNoPolymerase} in Section~\ref{subsec:selfact}, we now allow the transcription factor to dimerize, and each protein copy to exist as a free monomer or as part of a dimer. We track the total concentration of monomer subunits as
\begin{equation}
A_{\text{tot}} = A + 2A_2,
\end{equation}
where $A_2$ denotes dimers. Protein synthesis increases $A_{\text{tot}}$, and assuming rapid dimerization the newly produced monomers are instantly redistributed between monomers and dimers according to equilibrium. We assume that dilution and degradation act at the same effective rate $\gamma$ on both monomers and dimers, which is reasonable for stable TFs in fast-growing \emph{E.~coli} where dilution dominates, as a few candidates of auto-activating feedback loops are present in this organism \cite{groisman2001pleiotropic, schleif2010arac}.

Assuming the reaction $2A\leftrightarrow A_2$ is much faster than synthesis or degradation, the dimer concentration is given by
\begin{equation}
A_2 = \frac{A^2}{K_{\text{dim}}}.
\end{equation}
Mass balance then gives
\begin{equation}
A_{\text{tot}} = A + \frac{2A^2}{K_{\text{dim}}},
\end{equation}
which, solving for $A$, yields an explicit definition for the free monomer concentration as a function of $A_{\text{tot}}$,
\begin{equation}
A(A_{\text{tot}})=\frac{K_{\text{dim}}}{4}\!\left(\sqrt{1+\frac{8A_{\text{tot}}}{K_{\text{dim}}}}-1\right),
\end{equation}
and, given Eqn.~\ref{eq:A2}, the corresponding free dimer concentration as a function of $A_{\text{tot}}$,
\begin{equation}
A_2(A_{\text{tot}})=\frac{A(A_{\text{tot}})^2}{K_{\text{dim}}}.\label{eq:A2}
\end{equation}

\begin{figure}
    \centering
    \includegraphics[width=\linewidth]{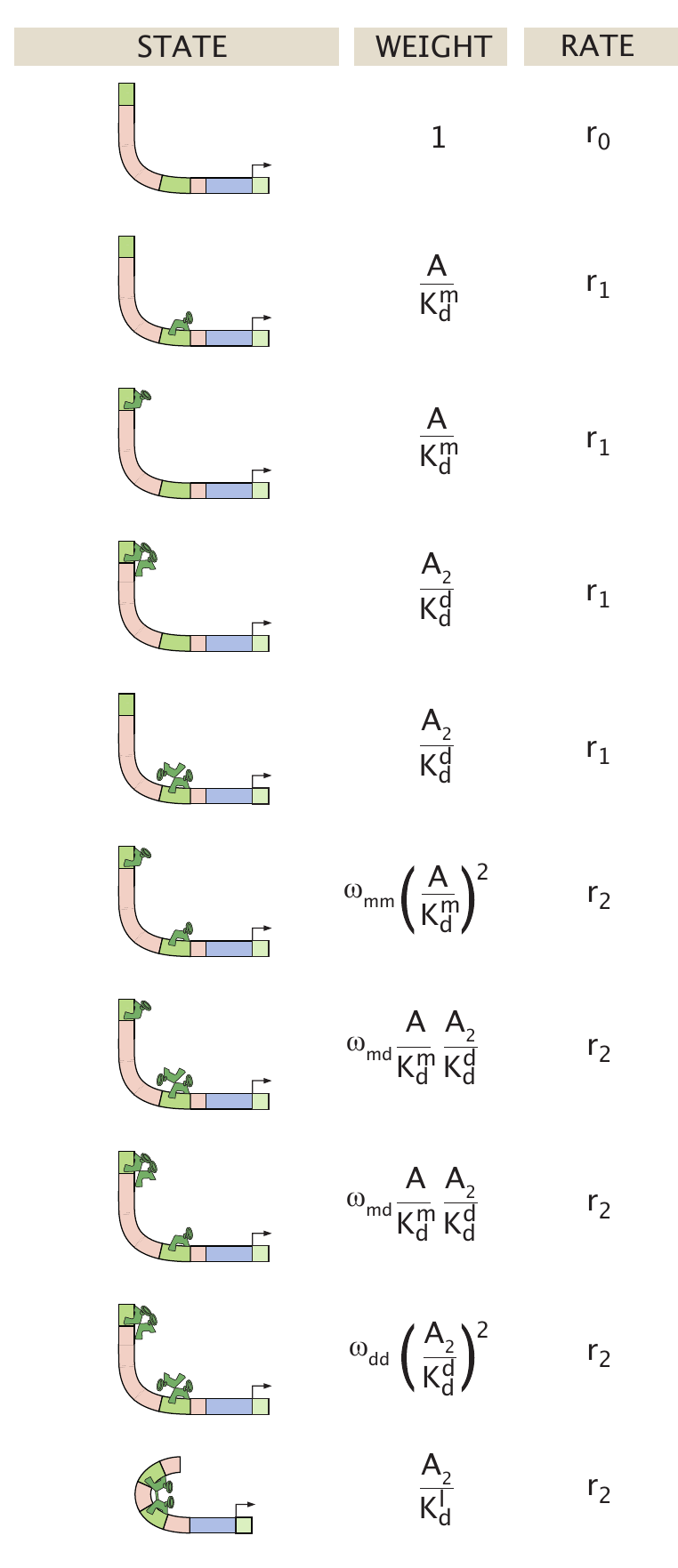}
    \caption{Promoter states for a dimerizing auto-activating transcription factor.
    Each row shows a possible promoter occupancy state when two binding sites are available. States differ by the number and type of bound transcription factor (monomer or dimer) and by whether the two dimers form a DNA loop. Each state is associated with a statistical weight determined by its binding configuration and a transcription rate ($r_0$, $r_1$, or $r_2$) that depends on the number of bound molecules.}
    \label{fig:auto_act_dimerSI}
\end{figure}

The transcription factor can bind DNA either as a monomer, with dissociation constant $K_d^m$, or as a dimer, with dissociation constant $K_d^d$. We now consider a promoter with two identical binding sites. The binding of two proteins can be cooperative, described by cooperativity parameters $\omega_{mm}$ (monomer–monomer), $\omega_{md}$ (monomer–dimer), and $\omega_{dd}$ (dimer–dimer). When both sites are occupied by dimers, they can also bridge the two sites to form a DNA loop, with a dissociation constant $K^l_d$ describing the dimer bound to DNA in its looped state. Each promoter state is associated with a transcription rate $r_0$, $r_1$, or $r_2$, corresponding to zero, one, or two bound proteins, respectively. Given the statistical weights of all promoter states as presented in Fig.~\ref{fig:auto_act_dimerSI}, the mean transcription rate is then
\begin{align}
F(A,A_2)& =
\frac{ r_0 }{Z}
+ r_1\frac{ 2\frac{A}{K^m_d} + 2\frac{A_2}{K^d_d} }{Z} \nonumber\\[4pt]
&\quad
+ r_2\frac{
      \omega_{mm}\!\left(\frac{A}{K^m_d}\right)^2
    + 2\omega_{md}\frac{A}{K^m_d}\frac{A_2}{K^d_d}
    + \omega_{dd}\!\left(\frac{A_2}{K^d_d}\right)^2}{Z}
\nonumber\\[4pt]
&\quad    
    + r_2\frac{\frac{A_2}{K^l_d}
    }{Z},
\end{align}
with a partition function
\begin{equation}
\begin{aligned}
Z =\; & 1 
+ 2\frac{A}{K^m_d} 
+ 2\frac{A_2}{K^d_d}
+ \omega_{mm}\!\left(\frac{A}{K^m_d}\right)^2
+ 2\omega_{md}\frac{A}{K^m_d}\frac{A_2}{K^d_d} \\[4pt]
&\quad
+ \omega_{dd}\!\left(\frac{A_2}{K^d_d}\right)^2
+ \frac{A_2}{K^l_d},
\end{aligned}
\end{equation}
and the corresponding dynamics of the system become
\begin{equation}
\frac{dA_{\text{tot}}}{dt} = -\gamma A_{\text{tot}} + F\!\big(A(A_{\text{tot}}),A_2(A_{\text{tot}})\big).\label{eq:oligoeq}
\end{equation}
Eqn.~\ref{eq:oligoeq} shows that once oligomerization is included, the system must be described in terms of $A_{\text{tot}}$ rather than the free monomer alone. Importantly, the dynamics become more nonlinear, allowing new behaviors to emerge.

We now consider an alternative setting in which several oligomeric forms coexist in solution, but only one can bind to DNA. As an example, suppose that monomers and dimers are found in solution with equilibrium constant $K_\text{dim}$, but only monomers can bind to the gene promoter site. We then have
\begin{equation}
\frac{dA_{\text{tot}}}{dt}=-\gamma A_\text{tot} + \frac{r_0+2r_1\frac{A}{K^m_d}+r_2\omega_{mm} (\frac{A}{K^m_d})^2}{1+2\frac{A}{K^m_d}+\omega_{mm} (\frac{A}{K^m_d})^2}.
\end{equation}
Substituting $A_{\text{tot}} = A + 2A^2/K_\text{dim}$ and using the chain rule, we can ultimately arrive at an ordinary differential equation with additional terms beyond those of the simple monomer case,
\begin{align}
\begin{split}
    \frac{dA}{dt} = &-\gamma \frac{K_\text{dim}A + 2A^2}{K_\text{dim} + 4A} \\
    &+ \frac{K_\text{dim}}{K_\text{dim} + 4A}\frac{r_0+2r_1\frac{A}{K^m_d}+r_2\omega_{mm} (\frac{A}{K^m_d})^2}{1+2\frac{A}{K^m_d}+\omega_{mm} (\frac{A}{K^m_d})^2}.
\end{split}
\end{align}
The presence of multiple oligomeric states could therefore alter the number and stability of steady states, although in practice usually only one oligomeric form is dominant.

We next consider looping when only the dimeric form of the transcription factor is present. If a dimer both binds and loops DNA, looping can be seen as adding binding configurations that strengthen effective binding and cooperativity. The production term of a looping dimer $A_2$ can then be written as
\begin{equation}
\frac{r_0+r_1(2+\frac{K^d_d}{K^l_d})p_{\text{act}}\frac{A_2}{K^d_d} + r_2\omega_{dd}\left(\frac{p_{\text{act}}A_2}{K^d_d}\right)^2}{1+(2+\frac{K^d_d}{K^l_d})p_{\text{act}}\frac{A_2}{K^d_d} + \omega_{dd}\left(\frac{p_{\text{act}}A_2}{K^d_d}\right)^2},
\end{equation}
which is algebraically equivalent to the standard model using renormalized parameters
\begin{equation}
\tilde{K}^d_d = \frac{2K^d_d}{2+\frac{K^d_d}{K^l_d}}
\end{equation}
and 
\begin{equation}
\tilde{\omega_{dd}} = \frac{4\omega_{dd}}{(2+\frac{K^d_d}{K^l_d})^2}.
\end{equation}
Looping therefore does not change the functional form of the model but simply rescales $K^d_d$ and $\omega_{dd}$.

Finally, when dimers form from monomers with equilibrium constant $K_{\text{dim}}$, and there is no looping and all cooperativity terms are equal ($\omega_{mm}=\omega_{md}=\omega_{dd}=\omega$), the production term can be reduced exactly to the monomer-only functional form as
\begin{equation}
f_{\text{mono}}(x) = \frac{ r_0 + 2r_1x + r_2\,\omega x^2}
                          { 1 + 2x + \omega x^2},
\end{equation}
where
\begin{equation}
x = \frac{A}{K^m_d} + \frac{A_2}{K^d_d}.
\end{equation}
The dynamical equation can then be written as
\begin{equation}
\frac{dA_{\text{tot}}}{dt} = -\gamma\,A_{\text{tot}} + f_{\text{mono}}\!\big(x(A_{\text{tot}})\big).
\end{equation}
At steady state this means that
\begin{equation}
\gamma(x)x=f_{\text{mono}}(x),
\end{equation}
with
\begin{equation}
\gamma(x)=\frac{2 K^d_d \left(K^d_d - K^m_d \left(K_{dim} + \sqrt{K^d_d \left(\frac{K^d_d}{(K^m_d)^2} + 4x\right)}\right)\right)}
{K^d_d \left(1 - 2 K_{dim}\right) - K^m_d \sqrt{K^d_d \left(\frac{K^d_d}{(K^m_d)^2} + 4x\right)}}.
\end{equation}
 Thus, the production curve retains the same shape as in the monomeric model, but the effective degradation rate becomes concentration-dependent.

The various settings discussed in this appendix demonstrate how incorporating oligomerization alters the equations of the dynamical system. Specifically, accounting for oligomerization introduces additional molecular states, and makes the effective production term a function of the total protein pool rather than only the active monomer. This transforms the dynamics from a simple one-variable system into a more nonlinear one. By contrast, looping does not change the mathematical structure of the model but can be absorbed into effective parameters that strengthen binding and cooperativity. When only a single oligomeric form binds DNA, the production term can still be written in the same functional form as the monomer-only model, but the degradation term becomes effectively concentration-dependent.

\section{Auto-activation : Bistability is possible for non cooperative systems ($\omega=1$).}
\label{appendix:auto_act_bistab_low_coop}
\subsection{Definition of the effective Hill coefficient}
\label{appendix:def_of_hill}
We first recall how to compute the Hill coefficient of an activating Hill function with constitutive expression, denoted $g(x)$, defined by
\begin{equation}
    g(x) = \frac{\bar{r}_0 + \bar{r}_2 x^n}{1 + x^n}.
\end{equation}
Its log-derivative is
\begin{equation}
    \frac{d \ln g}{d \ln x} = n \cdot \frac{x^n (\bar{r}_2 - \bar{r}_0)}{(1 + x^n)(\bar{r}_0 + \bar{r}_2 x^n)}.
\end{equation}
Let us define $x^*$ such that
\begin{equation}
    g(x^*) = \frac{\bar{r}_2 + \bar{r}_0}{2}.
\end{equation}
This holds for $x^* = 1$. Evaluating the derivative at this value then gives
\begin{equation}
    \left. \frac{d \ln g}{d \ln x} \right|_{x = x^*} = \frac{1}{2}n \cdot \frac{\bar{r}_2 - \bar{r}_0}{\bar{r}_2 + \bar{r}_0},
\end{equation}
and solving for $n$, we obtain
\begin{equation}
    n = \left. \frac{d \ln g}{d \ln x} \right|_{x = x^*} \cdot 2 \frac{\bar{r}_2 + \bar{r}_0}{\bar{r}_0 - \bar{r}_2}.
\end{equation}

We now define a similar expression for a thermodynamic model $w(x)$ given by
\begin{equation}
    w(x) = \frac{\bar{r}_0 + 2\bar{r}_1 p_{\text{act}} x + \omega \bar{r}_2 p_{\text{act}}^2 x^2}{1 + 2p_{\text{act}} x + \omega p_{\text{act}}^2 x^2}.
\end{equation}
We observe that $w(x)$ can be written as $\hat{w}(p_{\text{act}}x)$ with
\begin{equation}
    \hat{w}(\hat{x}) = \frac{\bar{r}_0 + 2\bar{r}_1 \hat{x} + \omega \bar{r}_2 \hat{x}^2}{1 + 2h \hat{x} + \omega \hat{x}^2},
\end{equation}
where $h=1$ in our case. Since $d \ln w/d \ln x = d \ln \hat{w}/d \ln \hat{x}$ when $h=1$, the effective Hill coefficient does not depend on $p_{\text{act}}$. The derivative of $\hat{w}$ is
\begin{equation}
    \frac{d \ln \hat{w}}{d \ln \hat{x}} = \frac{2(1 + h \hat{x})}{1 + \hat{x}(2h + \omega \hat{x})} - \frac{2(\bar{r}_0 + \bar{r}_1 \hat{x})}{\bar{r}_0 + \hat{x}(2\bar{r}_1 + \omega \bar{r}_2 \hat{x})}.
\end{equation}
Letting $\hat{x}^*$ be defined by $\hat{w}(\hat{x}^*) = \frac{\bar{r}_2 + \bar{r}_0}{2}$ yields
\begin{equation}
    \hat{x}^* = \frac{-h \bar{r}_0 + 2 \bar{r}_1 - h \bar{r}_2}{(\bar{r}_0 - \bar{r}_2) \omega} + \sqrt{S},
\end{equation}
where
\begin{equation}
S = \frac{
\splitfrac{\textstyle h^2 \bar{r}_0^2 - 4 h \bar{r}_0 \bar{r}_1 + 4 \bar{r}_1^2 + 2 h^2 \bar{r}_0 \bar{r}_2}
          {\textstyle - 4 h \bar{r}_1 \bar{r}_2 + h^2 \bar{r}_2^2 + \bar{r}_0^2 \omega - 2 \bar{r}_0 \bar{r}_2 \omega + \bar{r}_2^2 \omega}
}{
(\bar{r}_0 - \bar{r}_2)^2 \omega^2}.
\end{equation}
This then gives the derivative at $\hat{x}^*$ as
\begin{equation}
\frac{d \ln \hat{w}}{d \ln \hat{x}} \bigg|_{\hat{x} = \hat{x}^*} = \frac{\bar{r}_2 - \bar{r}_0}{\bar{r}_2 + \bar{r}_0} \cdot \frac{(h(\bar{r}_0 + \bar{r}_2) - 2\bar{r}_1) u + t}{(2h \bar{r}_2 - 2\bar{r}_1) u + t},
\label{eq:dlnwhatdlnx}
\end{equation}
with
\begin{equation}
\begin{cases}
\alpha = \sqrt{\frac{(-2\bar{r}_1 + h(\bar{r}_0 + \bar{r}_2))^2}{(\bar{r}_0 - \bar{r}_2)^2} + 4\omega}, \\
u = -2\bar{r}_1 + h(\bar{r}_0 + \bar{r}_2) + (-\bar{r}_0 + \bar{r}_2) \alpha, \\
t = 4(\bar{r}_0 - \bar{r}_2)^2 \omega.
\end{cases}
\end{equation}

We defined $\hat{w}$ with an extra parameter $h$ to allow a mapping between the thermodynamic model and the Hill function. For the Hill case, we set $h = 0$, $\bar{r}_1 = 0$, and $\omega = 1$. In our thermodynamic model, $h = 1$ and the expression of Eqn.~\ref{eq:dlnwhatdlnx} becomes
\begin{equation}
\frac{d \ln \hat{w}}{d \ln \hat{x}} \bigg|_{\hat{x} = \hat{x}^*} = \frac{\bar{r}_2 - \bar{r}_0}{\bar{r}_2 + \bar{r}_0} \cdot \frac{(\bar{r}_0 + \bar{r}_2 - 2\bar{r}_1) u + t}{(2\bar{r}_2 - 2\bar{r}_1) u + t},
\label{eq:dlnwdlnxheq1}
\end{equation}
with
\begin{equation}
\begin{cases}
\alpha = \sqrt{\frac{(-2\bar{r}_1 + \bar{r}_0 + \bar{r}_2)^2}{(\bar{r}_0 - \bar{r}_2)^2} + 4\omega}, \\
u = -2\bar{r}_1 + \bar{r}_0 + \bar{r}_2 + (-\bar{r}_0 + \bar{r}_2) \alpha, \\
t = 4(\bar{r}_0 - \bar{r}_2)^2 \omega.
\label{eq:alpha_u_t_heq1}
\end{cases}
\end{equation}

To ensure consistency with the Hill model case, we therefore define the effective Hill coefficient as
\begin{equation}
    n_{\text{eff}} = \left. \frac{d \ln w}{d \ln x} \right|_{x = x^*} \cdot 2 \frac{\bar{r}_2 + \bar{r}_0}{\bar{r}_2 - \bar{r}_0}\label{appendix:neff}
\end{equation}
for
\begin{equation}
    w(x) = \frac{\bar{r}_0 + 2\bar{r}_1 p_{\text{act}} x + \omega \bar{r}_2 p_{\text{act}} x^2}{1 + 2 p_{\text{act}} x + \omega p_{\text{act}} x^2}
\end{equation}
and $x^*$ defined such that
\begin{equation}
    w(x^*) = \frac{\bar{r}_2 + \bar{r}_0}{2}.
    \label{eq:w_half_value}
\end{equation}

We note that for the auto activation system, in which $\bar{r}_0\leq\bar{r}_1\leq\bar{r}_2$, having an effective Hill coefficient larger than unity ($n_{eff}>1$) is equivalent to 
\begin{equation}
   \omega>\frac{(\bar{r}_1-\bar{r}_0)(\bar{r}_2-\bar{r}_1)}{(\bar{r}_2-\bar{r}_0)^2}. 
\end{equation}
Therefore, depending on the parameters of the system, the effective Hill coefficient can be larger or smaller than one, despite the activator having two binding sites.

\subsection{Numerical sweep for minimal cooperativity above which there is bistability.}
\begin{figure*}[t]
    \centering
    \includegraphics[width=0.8\linewidth]{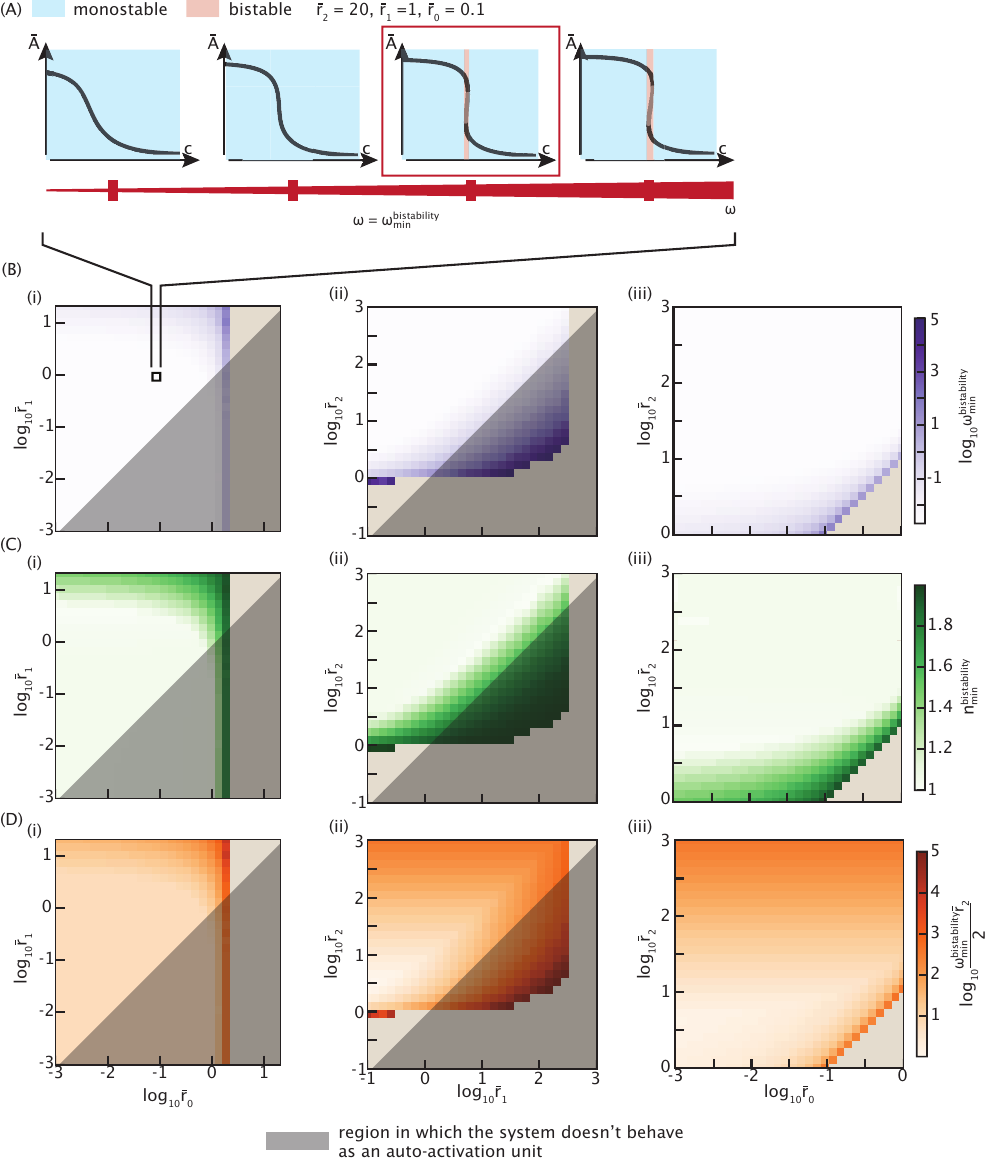}

\caption{
Parameter space exploration tracking the minimal cooperativity required for bistability over a range of effector concentrations. The cooperativity $\omega$ is sampled over the interval $\omega \in [1, 10^5]$. The condition $\bar{r}_0 \leq \bar{r}_1 \leq \bar{r}_2$ is imposed to ensure auto-activation behavior; regions where this is not satisfied are shaded in gray. (A) Illustration of the method used to determine the minimal cooperativity required for bistability. (B) Minimal cooperativity values identified using the approach in (A). (C) Corresponding effective Hill coefficient. (D) Effective cooperativity estimated from the a necessary condition for bistability $\omega\bar{r}_2/2>1$. In panels (B–D), subpanels (i)–(iii) show different slices of parameter space: (i) $(\bar{r}_0, \bar{r}_1) \in [10^{-5}, \bar{r}_2]^2$ with $\bar{r}_2 = 20$ fixed; (ii) $(\bar{r}_1, \bar{r}_2) \in [\bar{r}_0, 10^5]^2$ with $\bar{r}_0 = 0.1$ fixed; (iii) $(\bar{r}_0, \bar{r}_2) \in [10^{-5}, \bar{r}_1] \times [\bar{r}_1, 10^5]$ with $\bar{r}_1 = 1$ fixed. In regions where no minimal cooperativity values for bistability are reported, the system remains monostable across the entire range of cooperativity values sampled.}
    \label{fig:autoact_min_coop_for_bifurcation}
\end{figure*}
To better understand how the parameters of the auto-activation system constrain the emergence of bistability, we explore the minimal cooperativity $\omega_{\min}^{\text{bistable}}$ required to observe bistability across a broad range of rate parameters. Specifically, we perform numerical parameter sweeps over $(\bar{r}_0, \bar{r}_1, \bar{r}_2)$, systematically enforcing the auto-activation condition $\bar{r}_0 \leq \bar{r}_1 \leq \bar{r}_2$. Regions where this condition is violated are shaded in gray. For each valid triplet, we determine the minimal value of $\omega$ for which bistability occurs over a finite range of effector concentrations.

Fig.~\ref{fig:autoact_min_coop_for_bifurcation}(A) illustrates the numerical method used to identify this minimum cooperativity. The resulting values are displayed in Fig.~\ref{fig:autoact_min_coop_for_bifurcation}(B), where we observe that the required cooperativity varies significantly across parameter space. Notably, bistability can be achieved even in the case where $\omega \leq1$. Particularly where $\bar{r}_2$ is sufficiently large relative to $\bar{r}_1$ and $\bar{r}_0$. This includes cases where $\omega < 1$, which corresponds to anti-cooperative behavior—i.e., where the binding of the first activator decreases the likelihood of a second one binding and cases where $\omega=1$, which corresponds to no cooperativity.

While cooperativity in the strict thermodynamic sense may not be required, the system still exhibits an effective nonlinearity sufficient to support bistability. To assess this, we compute the effective Hill coefficient of the production term, derived with a thermodynamical model, shown in Fig.~\ref{fig:autoact_min_coop_for_bifurcation}(C). When the system is bistable, the effectively Hill coefficient always exceeds 1, consistent with theoretical expectations ~\cite{griffith1968mathematics}. Furthermore, in Fig.~\ref{fig:autoact_min_coop_for_bifurcation}(D), we evaluate an effective cooperativity based on the inequality $\omega \bar{r}_2/2 > 1$, which serves as a necessary (though not sufficient) condition for bistability. The consistency of this bound with the numerically determined $\omega_{\min}^{\text{bistable}}$ highlights its predictive value.

Together, these analyses reveal that bistability is not strictly dependent on cooperative binding in the classical sense, but rather emerges from the combined effects of system parameters—particularly the balance between production rates. This underscores the importance of kinetic tuning in biological systems and the potential for bistable behavior even in regimes of weak or anti-cooperative interactions.

\section{Relaxation timescale to equilibrium for the auto-activation system}
\label{appendix:auto_act_timescale}
We examine the relaxation timescales to steady state in the auto-activation system as a function of the initial concentration of activator $A$, denoted $\bar{A}_0$. To define the timescale, we employ two different methods. The first method, referred to as the \textit{threshold approach}, involves measuring the time it takes for the system to evolve from the initial condition to a fixed fraction of its steady state. We track the time-dependent trajectory $\bar{A}(\bar{t})$ and define the relaxation timescale as the time $\bar{t}^*$ such that
\begin{equation}
    \frac{\bar{A}(\bar{t}^*) - \bar{A}_i}{\bar{A}_f - \bar{A}_i} = \epsilon,
\end{equation}
where $\epsilon$ is the chosen threshold. In practice, since the system is simulated numerically over $N$ discrete time points, the relaxation time is computed as the earliest sampled time $\bar{t}_i$ for which the normalized deviation exceeds $\epsilon$,
\begin{equation}
\bar{t}^* = \min_{j \in [1, N]} \left\{ \bar{t}_j \,\middle|\, \frac{\bar{A}(\bar{t}_j) - \bar{A}_i}{\bar{A}_f - \bar{A}_i} > \epsilon \right\}.
\end{equation}

\begin{figure}
    \centering
    \includegraphics[width=\linewidth]{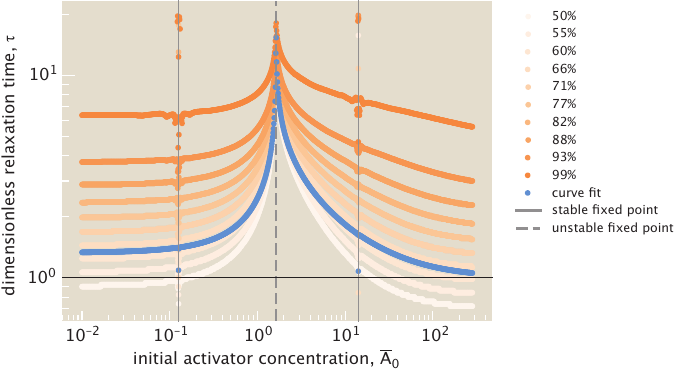}
\caption{
Relaxation timescales as a function of the initial concentration of gene $A$ ($\bar{A}_0$). The parameters of the system are fixed at the following values: $\omega = 7.5$, $\bar{r}_0 = 0.1$, $\bar{r}_1 = 1$, $\bar{r}_2 = 20$, and $c=2 \cdot 10^{-5}\text{M}$. Timescales are computed using two approaches: a threshold-based method (orange curves) and exponential curve fitting (blue curve). Each orange curve corresponds to a different threshold value $\epsilon$ indicated as percentages in the legend. The threshold timescale $\bar{t}^*$ is the time required for $\bar{A}(\bar{t})$ to reach a fraction $\epsilon$ of the total change from initial condition $\bar{A}_0$ to steady state $\bar{A}_\infty$. The blue curve represents the relaxation time obtained from exponential fits to $\bar{A}(\bar{t})$ trajectories. Vertical lines mark positions of stable (solid lines) and unstable (dashed line) fixed points. The horizontal black line indicates the unity timescale as a reference for comparison.}
\label{fig:autoactivation_dynamics_thresh_sampling}
\end{figure}

We report the relaxation timescale obtained using various values of $\epsilon$ in Fig.~\ref{fig:autoactivation_dynamics_thresh_sampling}. All resulting curves exhibit similar behavior, indicating that the precise value of the threshold does not significantly affect the overall system dynamics.

We compare these threshold-based relaxation timescales with those obtained by fitting an exponential function to the trajectory $\bar{A}(\bar{t})$. If $\bar{A}(\bar{t})$ followed a purely exponential decay---as it does near stable fixed points---then the timescale from the exponential fit would match the threshold-based timescale for $\epsilon \approx 0.63$. As shown in Fig.~\ref{fig:autoactivation_dynamics_thresh_sampling}, the curve corresponding to the exponential fit closely matches the threshold-based curve with $\epsilon = 0.63$, even when the initial condition is far from the stable fixed points and the system is not strictly exponential, as illustrated in Fig.~\ref{fig:autoactivation_dynamics}(A). Nevertheless, near the stable fixed points, we observe that the threshold method reports a longer timescale. This is because the initial condition is very close to that of the stable fixed point, and the system takes longer to cross the relative threshold. Therefore, we rely on the timescale obtained from the exponential fitting, as shown in Fig.~\ref{fig:autoactivation_dynamics}(B).

\section{Bistability regimes in the mutual repression circuit}
\label{appendix:mutrep_bistab}

\subsection{A necessary condition for bistability}
\label{appendix:mutrep_bistab_bounds}
In the mutual repression system at steady state, we set the time derivatives to zero, leading to the equations
\begin{equation}
   \bar{R}_1 = \dfrac{\bar{r}}{1 + 2p_2 \bar{R}_2 + \omega_2 (p_2 \bar{R}_2)^2} 
   \label{eq:R1ss}
\end{equation}
and 
\begin{equation}
     \bar{R}_2 = \dfrac{\bar{r}}{1 + 2p_1 \dfrac{\bar{R}_1}{\bar{K}} + \omega_1 \left( p_1 \dfrac{\bar{R}_1}{\bar{K}} \right)^2},
     \label{eq:R2ss}
\end{equation}
where for convenience we define $p_1 = p_\text{act}(c_1) $ and $p_2 = p_\text{act}(c_2) $. Substituting the expression for $\bar{R}_1$ of Eqn.~\ref{eq:R1ss} into Eqn.~\ref{eq:R2ss} for $\bar{R}_2$, and rewriting in standard polynomial form, $M(\bar{R}_2) = 0$, we obtain
{\small
\begin{align}
    M(\bar{R}_2)=&p_2^4\omega_2^2\bar{R}_2^5 
    + p_2^3\omega_2(4-\bar{r}\omega_2p_2)\bar{R}_2^4 \nonumber\\
    &+ 2p_2^2(2+\omega_2(1+\bar{r}(\frac{p_1}{\bar{K}}-2p_2)))\bar{R}_2^3 \nonumber\\
    &+ 4p_2(1+\bar{r}(\frac{p_1}{\bar{K}}-p_2-\frac{\omega_2p_2}{2}))\bar{R}_2^2 \nonumber\\
    &+(1+\frac{2p_1\bar{r}}{\bar{K}}+\frac{\omega_1p_1^2\bar{r}^2}{\bar{K}^2}-4p_2r)\bar{R}_2-\bar{r}.
    \label{eq:poly_mut_rep}
\end{align} }

To assess whether the system is monostable, we examine the number of non-negative roots of the polynomial $M(\bar{R}_2) $. If the second derivative of $M(\bar{R}_2) $ does not change sign, then the polynomial can have at most two real roots. In particular, if the polynomial is convex for all non-negative values of $\bar{R}_2 $, i.e., $M''(\bar{R}_2) > 0 $, then the system cannot be bistable.
The second derivative of $M(\bar{R}_2) $ is given by
\begin{align}
M''(\bar{R}_2) &= 20 p_2^4 \omega_2^2 \bar{R}_2^3 + 12 p_2^3 \omega_2 (4 - \bar{r} \omega_2 p_2) \bar{R}_2^2 \nonumber\\
&\quad + 12 p_2^2 \left(2 + \omega_2 \left(1 + \bar{r} \left( \frac{p_1}{\bar{K}} - 2p_2 \right) \right) \right) \bar{R}_2 \nonumber \\
&\quad + 8p_2 \left(1 + \bar{r} \left( \frac{p_1}{\bar{K}} - p_2 - \frac{\omega_2 p_2}{2} \right) \right).
\end{align}
To guarantee that $M''(\bar{R}_2) > 0$ for all non-negative values of $\bar{R}_2$, we require that all coefficients in the polynomial expression of $M''(\bar{R}_2)$ remain strictly positive. 
This condition translates into three distinct inequalities. First, the positivity of the quadratic term, requires that
\begin{equation}
4 - \bar{r} \omega_2 p_2 > 0.
\end{equation} 
Next, positivity of the linear term imposes the constraint 
\begin{equation}
2 + \omega_2 \left(1 + \bar{r} \left( \frac{p_1}{\bar{K}} - 2p_2 \right) \right) > 0.
\end{equation}
Finally, the positivity of the constant term yields 
\begin{equation}
1 + \bar{r} \left( \frac{p_1}{\bar{K}} - p_2 - \frac{\omega_2 p_2}{2} \right) > 0. 
\end{equation}
These conditions can be equivalently rewritten in terms of upper bounds on $p_2 $ and combinations of $p_1 $ and $p_2 $, yielding 
\begin{align}
p_2 &< \frac{4}{\bar{r} \omega_2}, \\
2p_2 - \frac{p_1}{\bar{K}} &< \frac{1}{\bar{r}} \left( \frac{2}{\omega_2} + 1 \right),\\
\left(1 + \frac{\omega_2}{2} \right) p_2 - \frac{p_1}{\bar{K}} &< \frac{1}{\bar{r}}.
\end{align}
To ensure that the system remains monostable for all values of effector concentrations $c_1 $ and $c_2 $, we require that these inequalities hold for the maximum possible values for the different functions of $c_1$ and $c_2$. Thus, we obtain the sufficient conditions
\begin{align}
p_{\text{max}} &< \frac{4}{\bar{r} \omega_2}, \\
2p_{\text{max}} - \frac{p_{\text{min}}}{\bar{K}} &< \frac{1}{\bar{r}} \left( \frac{2}{\omega_2} + 1 \right), \\
\left(1 + \frac{\omega_2}{2} \right) p_{\text{max}} - \frac{p_{\text{min}}}{\bar{K}} &< \frac{1}{\bar{r}},
\end{align}
Finally, in the special case where $\bar{K} = 1$, a sufficient conditions under which the system remains monostable for all effector concentrations simplify, using Mathematica, to
\begin{equation}
    \bar{r}<\frac{1}{p_{\text{max}}-p_{\text{min}}+\omega_2p_{\text{max}}/2}.
\end{equation}

Taking the contrapositive, we obtain a necessary condition for the system to exhibit bistability at some value of the effector concentration
\begin{equation}
    \bar{r}>\frac{1}{p_{\text{max}}-p_{\text{min}}+\omega_2p_{\text{max}}/2}
    \label{eq:boundnecessaryomega2}
\end{equation}
The bound stated in Eqn.~\ref{eq:boundnecessaryomega2} depends on both $\omega_2$ and $\bar{r}$, again, similarly to the auto-activation system, acting together to determine wether bistability can be accessed or not. 

For $\bar{K}=1$ the two cooperativities play a symmetric role.  Therefore necessary conditions for bistability are
\begin{equation}
\bar{r}>\frac{1}{p_{\text{max}}-p_{\text{min}}+\omega_2p_{\text{max}}/2}    \label{eq:necessary_condition_2} 
\end{equation}
and
\begin{equation}
\bar{r}>\frac{1}{p_{\text{max}}-p_{\text{min}}+\omega_1p_{\text{max}}/2}
\label{eq:necessary_condition_1}.
\end{equation}
From Eqns.~\ref{eq:necessary_condition_1} and~\ref{eq:necessary_condition_2} a necessary condition for bistability for $K=1$ is that
{\small
\begin{equation}
    \bar{r}>\max(\frac{1}{p_{\text{max}}-p_{\text{min}}+\omega_2p_{\text{max}}/2},\frac{1}{p_{\text{max}}-p_{\text{min}}+\omega_1p_{\text{max}}/2})\label{eq:necessary_condition_max_w1w2}.
\end{equation}}
which simplifies to
\begin{equation}
    \bar{r}>\frac{1}{p_{\text{max}}-p_{\text{min}}+\min(\omega_2,\omega_1)p_{\text{max}}/2}\label{eq:necessary_condition_max_w1w2final}.
\end{equation}

\subsection{Effective Hill coefficient of the production terms}

In the case of auto-activation, analyzing the effective Hill coefficient provided insight into how bistability can arise even in non-cooperative systems—$\omega>1$ is not a necessary condition for bistability, but from our numerical sweeps displayed in Fig.~\ref{fig:autoact_min_coop_for_bifurcation}, $n_{\text{eff}}>1$ is. Motivated by this, we now examine whether a similar criterion might help explain the restriction of bistability to specific zones of parameter space in mutual repression circuits. 

We defined (Eqn.~\ref{appendix:neff}, Eqn.~\ref{eq:w_half_value}) and derived an analytical formula (Eqn.~\ref{eq:alpha_u_t_heq1}, Eqn.~\ref{eq:dlnwdlnxheq1}) for the effective Hill coefficient for a general production term
\begin{equation}
    w(x) = \frac{\bar{r}_0 + 2\bar{r}_1 p_{\text{act}} x + \omega \bar{r}_2 p_{\text{act}} x^2}{1 + 2 p_{\text{act}} x + \omega p_{\text{act}} x^2},
    \label{eq:general_production_term}
\end{equation}
in Appendix~\ref{appendix:def_of_hill}.
In the mutual repression system, the production terms of interest are the production of $R_1$ driven by promoter 1 and regulated by $R_2$
\begin{equation}
    f_1(\bar{R}_2) = \frac{\bar{r}}{1 + 2(p_\text{act}(c_2)\bar{R}_2) + \omega_2 (p_\text{act}(c_2)\bar{R}_2)^2}
\end{equation}
and the production of $R_2$ driven by promoter 2 and regulated by $R_1$ 
\begin{equation}
    f_2(\bar{R}_1) = \frac{\bar{r}}{1 + 2(p_\text{act}(c_1)\frac{\bar{R}_1}{K}) + \omega_1 (p_\text{act}(c_1)\frac{\bar{R}_1}{K})^2}.
\end{equation}
With $\bar{r}_0 = \bar{r}$, $\bar{r}_1 = 0$, and $\bar{r}_2 = 0$, $p_{\text{act}}\equiv p_{\text{act}}(c_2)$ for $f_1(\bar{R}_2)$ and $p_{\text{act}}\equiv p_{\text{act}}(c_1)/K$ for $f_2(\bar{R}_1)$; we see that those two production terms fall in to the more general from $w(x)$ of Eqn.~\ref{eq:general_production_term}. We can therefore apply the reasoning and algebra derived in Appendix~\ref{appendix:def_of_hill}. The corresponding effective Hill coefficients are given by
\begin{equation}
    n_1 = 2 - \frac{\sqrt{1 + 4\omega_2} - 1}{2\omega_2},
\end{equation}
for the production term $f_1(\bar{R}_2)$ and
\begin{equation}
    n_2 = 2 - \frac{\sqrt{1 + 4\omega_1} - 1}{2\omega_1},
\end{equation}
for the production term $f_2(\bar{R}_1)$. We see that the two expressions of the effective Hill coefficients have the same functional form. Therefore showing that $n_1>1$ for all $\omega_2>0$, implies that $n_2>1$ for all $\omega_1>0$. We therefore establish that $n_1>1$ for all $\omega_2>0$, and $n_2>1$ for all $\omega_1>0$ then follows. This amounts to showing
\begin{equation}
  2\omega_2 \;>\; \sqrt{1+4\omega_2}-1.
\end{equation}
Adding \(1\) to each side and squaring both sides gives
\begin{equation}
  \bigl(2\omega_2 + 1\bigr)^2 \;>\; 1 + 4\omega_2.
\end{equation}
Expanding the square and canceling identical terms leaves
\begin{equation}
  4\omega_2^{\,2} \;>\; 0,
\end{equation}
which is true for every $\omega_2>0$.

We therefore showed that the effective Hill coefficients of the production terms are always greater than one in this system. As a result, while this observation is consistent with conventional expectations ~\cite{griffith1968mathematics}, it provides little discriminatory power for identifying regions of bistability, since the condition is satisfied across parameter space.

\subsection{Effect of cooperativity on the bistability region}
\label{appendix:mutrep_colorbar_coop}
\begin{figure*}
    \centering    \includegraphics[width=0.7\linewidth]{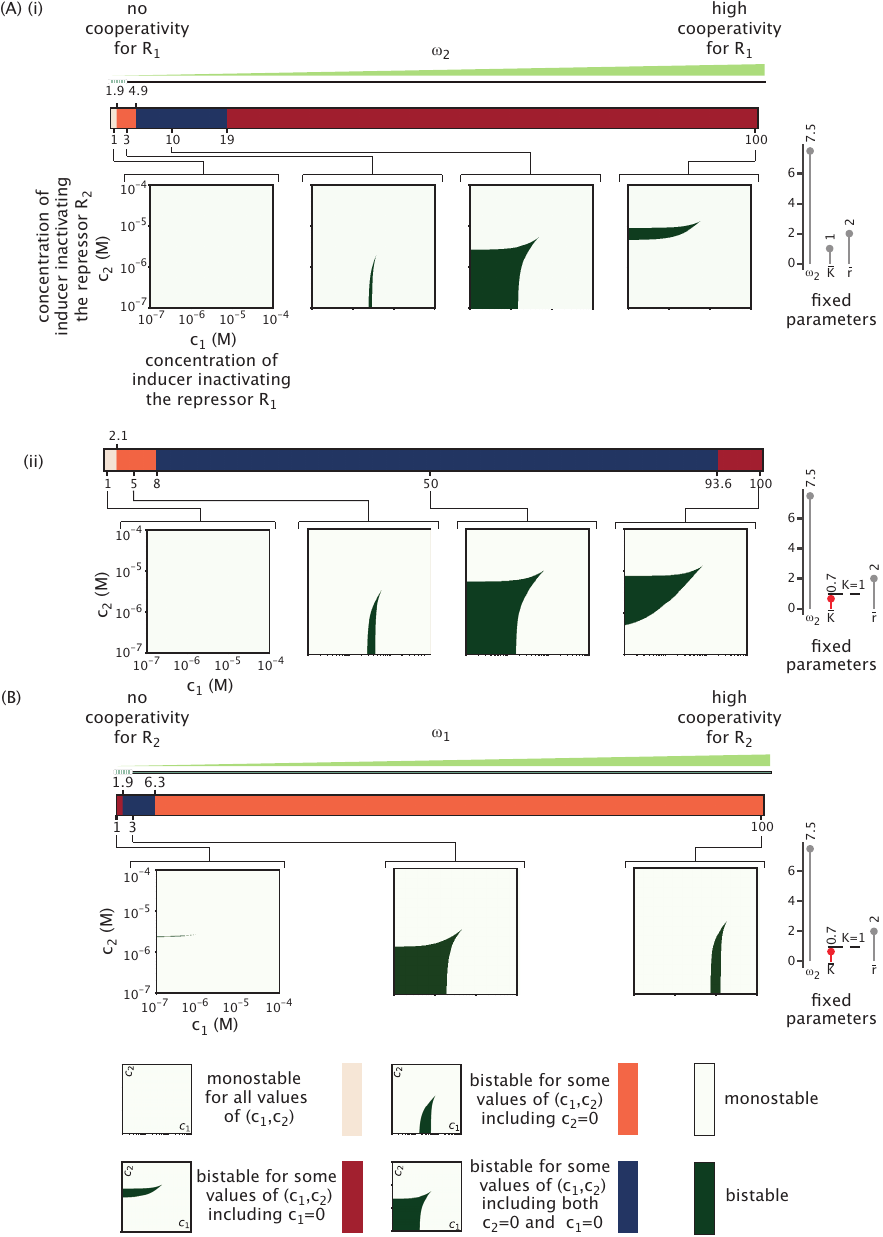}
    \caption{Bistability regimes in mutual repression as a function of cooperativity. Colored regions denote distinct bistable phase space geometries, defined by whether bistability occurs at very small $c_1$ ($c_1 = 10^{-7}\text{ M}$), very small $c_2$ ($c_2 = 10^{-7}\text{ M}$), both, or neither. (A) Evolution of the geometry of the bistability phase space, sweeping on inducer concentrations $(c_1,c_2)$, for fixed $\omega_1 = 7.5$ and $\bar{r} = 2$ and respectively $\bar{K}=1$ and $\bar{K}=0.7$ for (i) and (ii), when the parameter $\omega_2$ is varied. (B) Evolution of the geometry of the bistability phase space, sweeping on inducer concentrations $(c_1,c_2)$, for fixed $\omega_2 = 7.5$, $\bar{r} = 2$  and $\bar{K}=0.7$, when the parameter $\omega_1$ is varied.}
    \label{fig:color_bar_phase_space_SI}
\end{figure*}
Fig.~\ref{fig:color_bar_phase_space_SI}(A)(i) shows how the geometry of the bistable region evolves as the cooperativity of repressor $R_2$ ($\omega_2$) is varied, while $\omega_1 = 7.5$, $\bar{K} = 1$, and $\bar{r} = 2$ are held fixed. This corresponds to a symmetric case where both repressors bind their respective promoters with equal affinity. At low $\omega_2$, the system is monostable for all inducer concentrations, which is consistent with the known requirement for a minimal degree of nonlinearity to enable bistability. As $\omega_2$ increases beyond a threshold, bistability emerges, but initially in a constrained region where the tunability is mostly limited by $c_1$. When $\omega_2$ lies approximately between 2 and 5, a nonzero concentration of $c_1$ is required to inactivate a portion of the repressors $R_1$, thereby reducing their ability to bind DNA efficiently. In this regime, we are still in a setting where $\omega_1 > \omega_2$, meaning that $R_1$ binds more strongly to the DNA than $R_2$—as they have equal binding constants, the difference in binding arises solely from the cooperativity parameters. To support two distinct expression states—one with high $R_1$ and low $R_2$, and another with the reverse—the binding strength of $R_1$ must be reduced. This enables a more balanced competition between the two repressors, making it possible for both stable states to coexist. When $\omega_2$ becomes large (e.g., $\omega_2 \gtrsim 20$), the situation reverses: the bistable region in the $(c_1, c_2)$ phase space becomes constrained along the $c_2$ axis, as higher concentrations of inducer are required to counteract the strong DNA binding of $R_2$. 

For intermediate values of cooperativity, approximately between 5 and 20, the $(c_1, c_2)$ phase space is less constrained. In this regime, the concentrations of $c_1$ and $c_2$ need to be small enough to maintain repression by $R_1$ and $R_2$. If either inducer concentration becomes too high, the system is not repressed anymore and only has a unique steady state with high concentrations of both repressors.

While increasing $\omega_2$ initially expands the bistable region and enhances its robustness, we find that beyond a certain threshold, further increases in cooperativity begin to shrink the bistable domain. This reflects a general principle also observed for the parameter $\bar{K}$: pushing the system too far in one direction strongly constrains bistability. In the case of $\omega_2$, overly strong cooperativity amplifies the binding of $R_2$ and therefore the repression of $R_1$, so the system commits to one state, thereby reducing the range of inducer concentrations for which multiple steady states coexist.

Fig.~\ref{fig:color_bar_phase_space_SI}(A)(ii) explores the impact of tuning $\omega_2$ in an asymmetric setting, where $\bar{K} = 0.7$, $\omega_1 = 7.5$, and $r = 2$ are fixed. In this case, repressor $R_1$ binds more tightly to its promoter than $R_2$ does, breaking the symmetry observed in Fig.~\ref{fig:color_bar_phase_space_SI}(A)(i). At low values of $\omega_2$, the system is monostable, consistent with insufficient nonlinearity to support multiple steady states. As $\omega_1$ increases, bistability appears, but the geometry of the bistable region is notably skewed. Compared to Fig.~\ref{fig:color_bar_phase_space_SI}(A)(i), it is interesting to note that the bistability region at higher cooperativity is larger than in the symmetric case for the range of inducer concentrations considered. Finally in Fig.~\ref{fig:color_bar_phase_space_SI}(B), we tune the cooperativity $\omega_1$ instead of $\omega_2$ as was done in the previous panel. Therefore the roles of $c_1$ and $c_2$ are mirrored.

\section{Separatrix for mutual repression}
\label{appendix:sep_mut_rep}

We recall the differential equations governing the mutual repression system,
\begin{align}
\label{eq:mutRep1}
    \frac{d\bar{R}_{1}}{d\bar{t}} &= - \bar{R}_{1} + \bar{r}\frac{1}{1 + 2p_\text{act}(c_2)\bar{R}_{2} + \omega_2 \left[p_\text{act}(c_2)\bar{R}_{2} \right]^2} \nonumber\\
    &= F(\bar{R}_1,\bar{R}_2),
\end{align}
\begin{align}
    \frac{d\bar{R}_{2}}{d\bar{t}} &= - \bar{R}_{2} + \bar{r}\frac{1}{1 + 2p_\text{act}(c_1)\frac{\bar{R}_{1}}{\bar{K}} + \omega_1 \left[p_\text{act}(c_1)\frac{\bar{R}_{1}}{\bar{K}}\right]^2} \nonumber\\
    &= G(\bar{R}_1,\bar{R}_2).\label{eq:mutRep2}
\end{align}

The separatrix is defined as the curve $\bar{R}_2(\bar{R}_1)$ that satisfies the differential equation
\begin{equation}
    \frac{d\bar{R}_2}{d\bar{R}_1} = \frac{G(\bar{R}_1,\bar{R}_2)}{F(\bar{R}_1,\bar{R}_2)},
\end{equation}
which tracks the trajectory along which the system transitions between the basins of attraction of the two stable steady states.

\section{Coherent feed-forward loop response to a step function signal}
\label{appendix: step_func_signal}

\subsection{Analytical solution for the output $\bar{Z}(\bar{t})$}
We now rewrite the dynamical equations of the coherent feed-forward loop.  In particular, we introduce a simplifying notation for the activation terms for gene products  $Y$ and $Z$.  The reason such a definition is useful is that these terms are independent of $Y$ and $Z$ themselves and depend only upon $X$ itself and the concentration of inducer.  To that end, we write the dynamical equations for $Y$ and $Z$ as
\begin{align}
    \frac{d\bar{Y}}{d\bar{t}} &= - \bar{Y} + f_Y(t)\label{eq:yfllapp}\\
    \frac{d\bar{Z}}{d\bar{t}} &= - \bar{Z} + f_Z(t),
    \label{eq:zfllapp}
\end{align}
with the simplifying notation
\begin{equation}
    f_Y(\bar{t})= \frac{\bar{r}_{0Y} + \bar{r}_{1Y} \, p_\text{act}^X(c_X(\bar{t}))\bar{X}}{1+p_\text{act}^X(c_X(\bar{t}))\bar{X}},
\end{equation}
and
\begin{equation}
        f_Z(\bar{t})= \frac{\bar{r}_{0Z} + \bar{r}_{1Z} (\mathcal{X}(\bar{t}) + \mathcal{Y}(\bar{t})) + \omega \bar{r}_{2Z} \mathcal{X}(\bar{t})\mathcal{Y}(\bar{t})}{1+\mathcal{X}(\bar{t}) +\mathcal{Y}(\bar{t}) + \omega \mathcal{X}(\bar{t}) \mathcal{Y}(\bar{t})}.
\end{equation}
We recall that the bar indicates quantities where time is measured in units of $1/\gamma$, and where concentration and dissociation constants are measured in units of $K_{XY}$. The rates are then in units of $\gamma K_{XY}$.
The notations $\mathcal{X}$ and $\mathcal{Y}$ and are defined as
$\displaystyle \mathcal{X}(\bar{t})=p_\text{act}^X(c_X(\bar{t}))\bar{X}/\bar{K}_{XZ}$ and $\displaystyle \mathcal{Y}(\bar{t})=p_\text{act}^Y(c_Y)\bar{Y}(\bar{t})/\bar{K}_{YZ}$.  We study the response of the coherent feed-forward loop to a step function in effector concentration acting on $X$, namely, 
\begin{equation}
    c_X (\bar{t})= \begin{cases}
    \begin{aligned}
        c_{X}^{i} \quad & \text{if} \quad \bar{t}\leq 0,\\
        c_{X}^{f}\quad & \text{if} \quad \bar{t} > 0. \label{eq:mathcalXdef}
    \end{aligned}
    \end{cases}
\end{equation}

The step in the active concentration of $X$ and the rescaled concentration $\mathcal{X}$ are themselves subject to a step and can be written as
\begin{equation}
    \mathcal{X} (\bar{t})= \begin{cases}
    \begin{aligned}
        \mathcal{X}_{\rm i} = \frac{p_\text{act}^X(c_{X}^{i})\bar{X}}{\bar{K}_{XZ}} \quad & \text{if} \quad \bar{t}\leq 0,\\
        \mathcal{X}_{\rm f} = \frac{p_\text{act}^X(c_{X}^{f})\bar{X}}{\bar{K}_{XZ}} \quad & \text{if} \quad \bar{t} > 0. \label{eq:mathcalXdef}
    \end{aligned}
    \end{cases}
\end{equation}
The concentration of effector acting on $Y$, $c_Y(\bar{t})$ is taken to be constant $c_Y(\bar{t})=c_Y^0$.
Our goal here is to solve for the feed-forward dynamics analytically and obtain insights into the system on the basis of such a solution. Such a solution is possible because Eqns.~\ref{eq:coh_ffl_nondim1} and~\ref{eq:coh_ffl_nondim2} have a simple form, the time derivative of a variable equals the negative of itself plus a function of time,
\begin{equation}
    \frac{dG(t)}{dt} = -G(t) + f(t).
    \label{eq:general_diff_eq}
\end{equation}
Such equations can be solved in their most general form as 
\begin{equation}
     G(t) = e^{-t} \Big(G(0) + \int_0^t e^{t'} f(t') dt' \Big).
    \label{eq:general_diff_eq}
\end{equation}
In the cases of interest here, $G(t)$ is either $\bar{Y}(\bar{t})$ or $\bar{Z}(\bar{t})$, and $f(t)$ correspondingly is either $f_Y(\bar{t})$ or $f_Z(\bar{t})$, the activation terms for $Y$ and $Z$, respectively, in Eqns.~\ref{eq:zfllapp} and~\ref{eq:yfllapp}. 

We first solve Eqn.~\ref{eq:yfllapp}, because its dynamics is not coupled to $\bar{Z}(\bar{t})$. Notice that for $\bar{t} > 0$, $f_Y(\bar{t})$ is constant. Referring to Eqn.~\ref{eq:general_diff_eq}, we see that $\bar{Y}$ evolves from initial to final state purely exponentially according to the time evolution
\begin{align}
    \bar{Y}(\bar{t}) = -\Delta \bar{Y} e^{-\bar{t}} + \bar{Y}_f.
\end{align}
Here $\Delta \bar{Y} = \bar{Y}_f - \bar{Y}_i$ is the difference between the final and initial concentration of $\bar{Y}$.
We write the explicit expression of those initial and final steady states hereafter,
\begin{align}
    \bar{Y}_{i} &= \frac{\bar{r}_{0Y} + \bar{r}_{1Y} \, p_\text{act}^X(c_X^{i})\bar{X}}{1+p_\text{act}^X(c_X^{i})\bar{X}} \\
    \bar{Y}_{f} &= \frac{\bar{r}_{0Y} + \bar{r}_{1Y} \, p_\text{act}^X(c_X^{f})\bar{X}}{1+p_\text{act}^X(c_X^{f})\bar{X}}.
\end{align}

As $p_\text{act}^Y(c_Y(\bar{t}))=p_\text{act}^Y(c_Y^0)$ is constant, $\bar{Y} \to \mathcal{Y}$ is a proportional mapping. Therefore, like $\bar{Y}(t)$, $\mathcal{Y}(\bar{t})$ also evolves exponentially in time by the similar form 
\begin{align}
    \mathcal{Y}(\bar{t}) = -\Delta \mathcal{Y} e^{-\bar{t}} + \mathcal{Y}_f,
\end{align}
with $\displaystyle \mathcal{Y}_{f/i}=p_\text{act}^Y(c_Y^0)\bar{Y}_{f/i}/\bar{K}_{YZ}$ and $\Delta \mathcal{Y}=\mathcal{Y}_f-\mathcal{Y}_i$.
Given this expression for $\mathcal{Y}(\bar{t})$, we now know the full time dependence of $f_Z(\bar{t})$ in Eqn.~\ref{eq:zfllapp}. Next, we can solve for the dynamics of $\bar{Z}$ by substituting $f_Z(\bar{t})$ into the general solution Eqn.~\ref{eq:general_diff_eq}. By evaluating the integral, we find that
\begin{align}
\begin{split}
    \bar{Z}(\bar{t}) &= \bar{Z}_ie^{-\bar{t}} + \Big( e^{-\bar{t}}\int_0^{\bar{t}} e^{t'} f_Z(t') dt' \Big) \\[5pt]
    &= \bar{Z}_ie^{-\bar{t}} + \Big(\bar{Z}_f (1 - e^{-\bar{t}}) + \Theta(\bar{t}) \Big) \\[5pt]
    &= \bar{Z}_{\rm simple}(\bar{t}) + \Theta(\bar{t}),
    \label{eq:coh_full_sol}
\end{split}
\end{align}
with
\begin{align}
\begin{split}
    \Theta(\bar{t}) &= -\frac{\Phi \Delta \mathcal{Y}}{S^2} e^{-\bar{t}} \log \left( \frac{Se^{\bar{t}} - \Delta \mathcal{Y} (1 + \omega \mathcal{X}_f)}{S - \Delta \mathcal{Y} (1 + \omega \mathcal{X}_f)} \right), \\[5pt]
    \Phi &=  \omega\mathcal{X}_f^2(\bar{r}_{2Z} - \bar{r}_{1Z}) + \omega \mathcal{X}_f (\bar{r}_{2Z} - \bar{r}_{0Z}) \\
     & \hspace{2cm} + (\bar{r}_{1Z} - \bar{r}_{0Z}), \\[5pt]
    S &= 1 + \mathcal{X}_f + \mathcal{Y}_f + \omega \mathcal{X} _f\mathcal{Y}_f,
    \label{eq:coh_full_sol2}
\end{split}
\end{align}
as shown in the main text. The solution of $\bar{Z}(\bar{t})$ cleanly splits into two parts. The first two terms describe the exponential behavior one expects from simple regulation 
\begin{equation}
\bar{Z}_{\rm simple} (\bar{t})= \bar{Z}_ie^{-\bar{t}} + \bar{Z}_f (1 - e^{-\bar{t}})
\label{eq:zsimple}
\end{equation}
with $\bar{Z}_i$ and $\bar{Z}_f$, respectively the initial and final steady state concentration of the output $\bar{Z}$. Their explicit expression is given by
\begin{align}
    \bar{Z}_{i}&=\frac{\bar{r}_{0Z} + \bar{r}_{1Z} (\mathcal{X}_{i} + \mathcal{Y}_{i}) + \omega \bar{r}_{2Z} \mathcal{X}_{i}\mathcal{Y}_{i}}{1+\mathcal{X}_{i} +\mathcal{Y}_{i} + \omega \mathcal{X}_{i} \mathcal{Y}_{i}} \\
    \bar{Z}_{f}&=\frac{\bar{r}_{0Z} + \bar{r}_{1Z} (\mathcal{X}_{f} + \mathcal{Y}_{f}) + \omega \bar{r}_{2Z} \mathcal{X}_{f}\mathcal{Y}_{f}}{1+\mathcal{X}_{f} +\mathcal{Y}_{f} + \omega \mathcal{X}_{f} \mathcal{Y}_{f}}.
\end{align}
We see that $\Theta(\bar{t})$ accounts for the difference between the feed-forward trajectory $\bar{Z}$ and the simple regulation trajectory $\bar{Z}_{\rm simple}$. As a sanity check, we see $\Theta(\bar{t}) = 0$ when $\bar{t} = 0$ and $\bar{t} \to \infty$, confirming that the feed-forward loop and the simple regulation trajectory have the same initial and final state, as expected.

\subsection{Derivation and sign of the average delay $\langle \Delta \bar{t} \rangle$}

\label{appendix: ffl_derive_delay}

From the analytical expression of $\bar{Z}(\bar{t})$, we can then also derive the average time delay from the offset $\Theta(\bar{t})$
\begin{equation}
    \langle \Delta\bar{t} \rangle = \:\frac{1}{\bar{Z}_{f}-\bar{Z}_{i}}\int_{0}^\infty\Theta(\bar{t})d\bar{t}.
    \label{eq:time_delay_si}
\end{equation}

With Eqn.~\ref{eq:coh_full_sol2}, we can analytically evaluate the integral and obtain the following expression for the average delay
\begin{align}
    \langle\Delta \bar{t} \rangle
    = \frac{\Phi(\bar{Z}_{f} - \bar{Z}_{i}) ^{-1}}{S(1+\omega \mathcal{X}_{f})} \log \Big( \frac{1 + \mathcal{X}_{f} + \mathcal{Y}_{i}+\omega \mathcal{X}_{f} \mathcal{Y}_{i}}{S}\Big).
    \label{eq:area_between_curves_si}
\end{align}
As a reminder, $\langle\Delta \bar{t} \rangle$ signifies the average time difference between the feed-forward loop response and the simple regulation response. The sign of $\langle\Delta \bar{t} \rangle$ indicates whether the feed-forward loop delays ($\langle\Delta \bar{t} \rangle < 0$) or accelerates ($\langle\Delta \bar{t} \rangle > 0$). From Eqn.~\ref{eq:area_between_curves_si}, we can analytically determine whether the feed-forward loop delays or accelerates by treating the contribution from each component. To begin with, we have $S > 0$ and $1 + \omega \mathcal{X}_f > 0$ as concentrations are strictly non-negative. For the coherent feed-forward loop, we have $\Phi \geq 0$ as $\bar{r}_{2Z} \geq \bar{r}_{1Z} \geq \bar{r}_{0Z}$ since both $X$ and $Y$ activate $Z$. The term $(\bar{Z}_f - \bar{Z}_i)$ depends on the direction of the step. For an ON step, $(\bar{Z}_f - \bar{Z}_i) > 0$ and for an OFF step, $(\bar{Z}_f - \bar{Z}_i)<0$. Finally, the logarithm also depends on the direction of the step. For an ON step, we have 
\begin{align}
    &1 + \mathcal{X}_{f} + \mathcal{Y}_{i}+\omega \mathcal{X}_{f} \mathcal{Y}_{i} \leq 1 + \mathcal{X}_{f} + \mathcal{Y}_{f}+\omega \mathcal{X}_{f} \mathcal{Y}_{f} = S \\
    &\implies \log \Big( \frac{1 + \mathcal{X}_{f} + \mathcal{Y}_{i}+\omega \mathcal{X}_{f} \mathcal{Y}_{i}}{S}\Big) \leq 0,
\end{align}
since $\mathcal{X}_f \geq \mathcal{X}_i$ and $\mathcal{Y}_f \geq \mathcal{Y}_i$. Combined with the effect of other terms, we find $\langle\Delta \bar{t} \rangle \leq 0$ for an ON step. Similarly, for the OFF step, we have
\begin{align}
    &1 + \mathcal{X}_{f} + \mathcal{Y}_{i}+\omega \mathcal{X}_{f} \mathcal{Y}_{i} \geq 1 + \mathcal{X}_{f} + \mathcal{Y}_{f}+\omega \mathcal{X}_{f} \mathcal{Y}_{f} = S \\
    &\implies  \log \Big( \frac{1 + \mathcal{X}_{f} + \mathcal{Y}_{i}+\omega \mathcal{X}_{f} \mathcal{Y}_{i}}{S}\Big) \geq 0,
\end{align}
and we thus find that $\langle\Delta \bar{t} \rangle \leq 0$ for the OFF step as well. 

We have shown that the average time difference $\langle\Delta \bar{t} \rangle$ is negative for both the ON and OFF steps. To complete the analysis, we will further demonstrate that the time difference $\Delta\bar{t}(\bar{Z})$, defined in Section~\ref{sec:ffl_charac_delay}, has the same sign for any concentration $\bar{Z}$. This amounts to saying that the feed-forward trajectory and the simple regulation trajectory never cross each other.
We can show this by proving that $\Theta(\bar{t})$ has the same sign for any $\bar{t} > 0$. In Eqn.~\ref{eq:coh_full_sol2}, we observe that the time dependence in $\Theta(\bar{t})$ appears in $e^{-\bar{t}}$ and in $Se^{\bar{t}}$ inside the logarithm. $e^{-\bar{t}} > 0$ for any $\bar{t}$, thus only the logarithm term might change its sign as time evolves. However, we observe that
\begin{eqnarray}
    &S - \Delta \mathcal{Y} (1 + \omega \mathcal{X})  
    = 1 + \mathcal{X}_f +\mathcal{Y}_i \nonumber \\
    &+\omega\mathcal{X}_f\mathcal{Y}_i > 0.
\end{eqnarray}
In other words, $S > \Delta \mathcal{Y} (1 + \omega \mathcal{X})$. Since $e^{\bar{t}} > 1$ for any $\bar{t} > 0$, it is always true that $Se^{\bar{t}} > \Delta \mathcal{Y} (1 + \omega \mathcal{X})$. Therefore, the sign of the logarithm does not change with time. Thus, we have proven that $\Delta \bar{t} (\bar{Z})$ has the same sign as $\langle\Delta \bar{t} \rangle$ for any concentration $\bar{Z}$.

\subsection{Logic gate analysis in coherent feed-forward loop}
\label{appendix:ffl_logic_gates}

In this appendix section, we will prove that for the XOR gate in the coherent feed-forward loop, the OFF step delay is always greater than the ON step delay. 
In the XOR limit, $\omega = 0$. Eqn.~\ref{eq:area_between_curves} simplifies to
\begin{align}
    \langle \Delta\bar{t} \rangle &= \frac{1}{(\bar{Z}_{f} - \bar{Z}_{i})}\frac{\bar{r}_{1Z} - \bar{r}_{0Z}}{1 + \mathcal{X}_f + \mathcal{Y}_f} \log \Big( \frac{1 + \mathcal{X}_f + \mathcal{Y}_i}{1 + \mathcal{X}_f + \mathcal{Y}_f}\Big).
    \label{eq:or_gate}
\end{align}
The time delay $\langle \Delta\bar{t} \rangle$ is different for ON and OFF steps because $\mathcal{X}$ and $\mathcal{Y}$ values at $\bar{t} = 0$ and $\bar{t} = \infty$ are different for ON and OFF steps. Specifically, for ON step, $\mathcal{X}_f = \mathcal{X}_{\max}$ and $\mathcal{Y}_f = \mathcal{Y}_{\max}$; while for OFF step $\mathcal{X}_f = \mathcal{X}_{\min}$ and $\mathcal{Y}_f = \mathcal{Y}_{\min}$. Let's now compare the time delay $\langle \Delta\bar{t} \rangle$ for ON and OFF steps by taking their ratio
\begin{align}
\begin{split}
    \left| \frac{\langle \Delta\bar{t} \rangle_\text{OFF}}{\langle \Delta\bar{t} \rangle_\text{ON}} \right| &= \left| \frac{1 + \mathcal{X}_{\max} + \mathcal{Y}_{\max}}{1 + \mathcal{X}_{\min} + \mathcal{Y}_{\min}} \frac{\log(\frac{1 + \mathcal{X}_{\min} + \mathcal{Y}_{\max}}{1 + \mathcal{X}_{\min} + \mathcal{Y}_{\min}})}{\log (\frac{1 + \mathcal{X}_{\max} + \mathcal{Y}_{\min}}{1 + \mathcal{X}_{\max} + \mathcal{Y}_{\max}})} \right| \\[5pt]
    &= \frac{1 + \mathcal{X}_{\max} + \mathcal{Y}_{\max}}{1 + \mathcal{X}_{\min} + \mathcal{Y}_{\min}} \frac{\log(\frac{1 + \mathcal{X}_{\min} + \mathcal{Y}_{\max}}{1 + \mathcal{X}_{\min} + \mathcal{Y}_{\min}})}{\log (\frac{1 + \mathcal{X}_{\max} + \mathcal{Y}_{\max}}{1 + \mathcal{X}_{\max} + \mathcal{Y}_{\min}})}. \\[5pt]
\end{split}
\end{align}
Note that $1 + \mathcal{X}_{\max} + \mathcal{Y}_{\max} \geq 1 + \mathcal{X}_{\min} + \mathcal{Y}_{\min}$, and that
\begin{align}
    \frac{1 + \mathcal{X}_{\min} + \mathcal{Y}_{\max}}{1 + \mathcal{X}_{\min} + \mathcal{Y}_{\min}} \geq \frac{1 + \mathcal{X}_{\max} + \mathcal{Y}_{\max}}{1 + \mathcal{X}_{\max} + \mathcal{Y}_{\min}},
\end{align}
because for any fraction $a/b$ where $a \geq b > 0$, $a/b \geq (a+c)/(b+c)$ for any $c \geq 0$. Here, $a= 1 + \mathcal{X}_{\min}+\mathcal{Y}_{\max}$, $b = 1 +\mathcal{X}_{\min}+\mathcal{Y}_{\min}$, and $c = \mathcal{X}_{\max} - \mathcal{X}_{\min}$. As logarithm is an increasing function, this means
\begin{align}
    \log\frac{1 + \mathcal{X}_{\min} + \mathcal{Y}_{\max}}{1 + \mathcal{X}_{\min} + \mathcal{Y}_{\min}} \geq \log\frac{1 + \mathcal{X}_{\max} + \mathcal{Y}_{\max}}{1 + \mathcal{X}_{\max} + \mathcal{Y}_{\min}}.
\end{align}
Therefore, the ratio $\left| \langle \Delta\bar{t} \rangle_\text{OFF}/\langle \Delta\bar{t} \rangle_\text{ON} \right| \geq 1$.
We have demonstrated that for the XOR gate, the OFF step delay is larger than the ON step delay. We cannot say much analytically about the magnitude of the ratio of delay. As we see in Fig.~\ref{fig:ffl_func_param}(A), this ratio can range anywhere from close to 1 to a large number.

From here, we might be tempted to argue that for the AND gate, the ratio always satisfies $\left| \langle \Delta\bar{t} \rangle_\text{OFF}/\langle \Delta\bar{t} \rangle_\text{ON} \right| \leq 1$. However, Fig.~\ref{fig:ffl_func_param}(B)(iii) provides a counter example. The statement for AND gate is thus false.

\subsection{Analytic solutions of other feed-forward loop networks.}
\label{appendix:ffl_other_networks}

\begin{figure*}
    \centering
    \includegraphics[width=\linewidth]{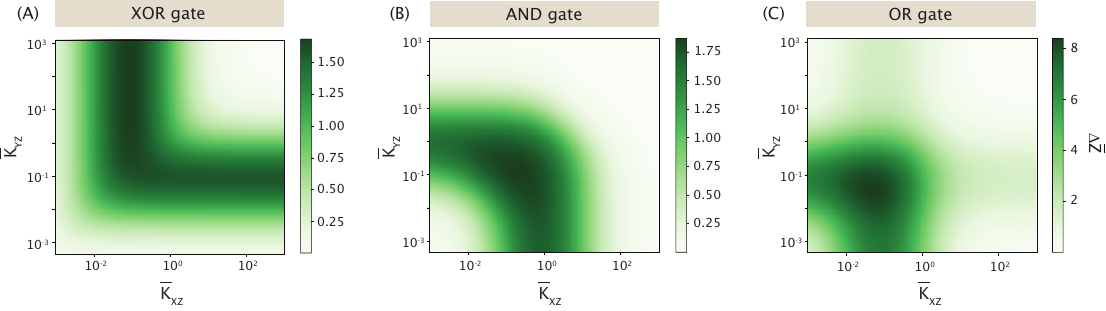}
    \caption{Systematic sweep to find region of large $\Delta \bar{Z} = \bar{Z}_{\max} - \bar{Z}_{\min}$. For every logic gate, we sweep across $(\bar{K}_{XZ}, \bar{K}_{YZ}) \in [10^{-3}, 10^3] \times  [10^{-3}, 10^3]$ and inspect the region of large relative $\Delta\bar{Z}$. (A) XOR gate. (B) AND gate. (C) OR gate. The XOR gate parameters are $\bar{r}_{0Y} = \bar{r}_{0Z} = 0$, $\bar{r}_{1Y} = \bar{r}_{1Z} = 2$, and $\omega = 0$. The AND gate parameters are $\bar{r}_{0Y} = \bar{r}_{0Z} = \bar{r}_{1Z} = 0$, $\bar{r}_{1Y} = \bar{r}_{2Z} = 2$, and $\omega = 10$. The OR gate parameters are $\bar{r}_{0Y} = \bar{r}_{0Z} = 0$, $\bar{r}_{1Y} = \bar{r}_{1Z} = 2$, $\bar{r}_{2Z} = 10$, and $\omega = 1$. These are the same as in Fig.~\ref{fig:ffl_func_param}.}
    \label{fig:ffl_delZ}
\end{figure*}

In any feed-forward loop, $X$ regulates $Y$ and $Z$, and $Y$ regulates $Z$. As there are three regulatory interactions, there exist $2^3 = 8$ different regulatory logics. For the dynamical equation of $\bar{Y}$, if $X$ activates $Y$, then
\begin{align}
    \frac{d\bar{Y}}{d\bar{t}} = - \bar{Y} + \frac{\bar{r}_{0Y} + \bar{r}_{1Y} p_\text{act}^X(c_X)\bar{X}}{1+p_\text{act}^X(c_X)\bar{X}}.
\end{align}
Otherwise, if $X$ represses $Y$, then
\begin{align}
    \frac{d\bar{Y}}{d\bar{t}} = - \bar{Y} + \frac{\bar{r}_{0Y}}{1+p_\text{act}^X(c_X)\bar{X}}.
\end{align}
There exist 4 different possibilities for how $X$ and $Y$ regulate $Z$. In Section~\ref{sec:FFL}, we showed the case where $X$ and $Y$ both activate $Z$, and the case where $X$ activates but $Y$ represses $Z$. If $X$ represses and $Y$ activates $Z$, we have
\begin{align}
    \frac{d\bar{Z}}{d\bar{t}} &= - \bar{Z} + \frac{\bar{r}_{0Z} + \bar{r}_{1Z} \mathcal{Y}}{1+\mathcal{X} +\mathcal{Y} + \omega \mathcal{X} \mathcal{Y}}.
\end{align}
If both $X$ and $Y$ repress $Z$, then
\begin{align}
    \frac{d\bar{Z}}{d\bar{t}} &= - \bar{Z} + \frac{\bar{r}_{0Z}}{1+\mathcal{X} +\mathcal{Y} + \omega \mathcal{X} \mathcal{Y}}.
\end{align}
Using the procedures described in the main text, we can similarly find analytic solutions describing all the other regulatory logics of
the feed-forward architectures.  
 Note that for a step function signal in $c_X$ and $c_Y$,
\begin{align}
    \mathcal{Y}(\bar{t}) = -\Delta \mathcal{Y} e^{-\bar{t}} + \mathcal{Y}_f.
\end{align}
Thus, we can compute the analytic expression of $\bar{Z}$ without worrying about how $Y$ is regulated. We find that the solutions all have the same form, except with a different $\Phi$ for each architecture of logic. Specifically, we have
\begin{align}
    \Phi_{XY} &= \omega\mathcal{X}^2(\bar{r}_{2Z} - \bar{r}_{1Z}) + \omega \mathcal{X} (\bar{r}_{2Z} - \bar{r}_{0Z}) \\
     & \hspace{2cm} + (\bar{r}_{1Z} - \bar{r}_{0Z}) \nonumber \\[5pt]
\Phi_X &= - (\bar{r}_{1Z}\mathcal{X} + \bar{r}_{0Z})(1 + \omega \mathcal{X}) \\[5pt]
\Phi_Y &= \bar{r}_{1Z}(1 + \mathcal{X}) - \bar{r}_{0Z}(1 + \omega \mathcal{X}) \\[5pt]
\Phi_0 &= -\bar{r}_{0Z}(1 + \omega \mathcal{X}),
\end{align}
where the subscript on $\Phi$ indicates which TF activates $Z$. $\mathcal{X} = \mathcal{X}_f$ is the value of $\mathcal{X}$ after the step function jump. Interestingly, we can write these expressions as
\begin{align}
    \Phi_{XY} &= \omega\mathcal{X}(1+\mathcal{X})\bar{r}_{2Z}+\Phi_X + \Phi_Y -\Phi_0\\[5pt]
\Phi_X &= - \bar{r}_{1Z}\mathcal{X}(1 + \omega \mathcal{X}) + \Phi_0 \\[5pt]
\Phi_Y &= \bar{r}_{1Z}(1 + \mathcal{X}) + \Phi_0 \\[5pt]
\Phi_0 &= -\bar{r}_{0Z}(1 + \omega \mathcal{X}).
\end{align}
A way to interpret this is that $\Phi$ gets a new term associated with a weight of a state when that state changes from no expression to expression.

\section{Functionality condition comparison}
\label{appendix:ffl_func_region}

In Section~\ref{sec:FFLlogicsweep}, we mentioned that besides the time delay, another helpful criterion for a functional feed-forward loop is the existence of a large difference between the maximum and minimum steady state values of $Z$. Let's define $\Delta \bar{Z} = |\bar{Z}_f - \bar{Z}_i|$. We want this to be big so that the dynamical change between low and high concentrations is meaningful. The steady state concentrations $\bar{Z}_f$ and $\bar{Z}_i$ have expressions given by
\begin{align}
    \bar{Z}_i &= \frac{\bar{r}_{0Z} + \bar{r}_{1Z} (\mathcal{X}_i + \mathcal{Y}_i) + \omega \bar{r}_{2Z} \mathcal{X}_i\mathcal{Y}_i}{1+\mathcal{X}_i +\mathcal{Y}_i + \omega \mathcal{X}_i \mathcal{Y}_i} \\
    \bar{Z}_f &= \frac{\bar{r}_{0Z} + \bar{r}_{1Z} (\mathcal{X}_f + \mathcal{Y}_f) + \omega \bar{r}_{2Z} \mathcal{X}_f\mathcal{Y}_f}{1+\mathcal{X}_f +\mathcal{Y}_f + \omega \mathcal{X}_f \mathcal{Y}_f}.
\end{align}
We observe that both $\bar{Z}_i$ and $\bar{Z}_f$ scale linearly with the rate parameters $\bar{r}_{iZ}$ for $i = 0, 1, 2$. For this reason, in the theoretical setting, a large $\Delta \bar{Z}$ can always be obtained by tuning the rate parameters high. Therefore, we did not include the discussion of this criterion in the main text. Nevertheless, it might be worthwhile to demonstrate the dependence of $\Delta \bar{Z}$ on the dissociation constants $\bar{K}_{XZ}$ and $\bar{K}_{YZ}$. We perform similar sweeps as Fig.~\ref{fig:ffl_func_param}, except we plot $\Delta \bar{Z}$ for each choice of $(\bar{K}_{XZ}, \bar{K}_{YZ})$. The result is shown in Fig.~\ref{fig:ffl_delZ}. We employ the same logic gate parameters as in Fig.~\ref{fig:ffl_func_param}. We observe that the region of large relative $\Delta\bar{Z}$ tends to have an L-shape. The XOR and AND gate have regions of large $\Delta\bar{Z}$ that extend in opposite directions. While XOR gate tends to prefer weak binding, the AND gate benefits from strong binding. The OR gate shape is a superposition of the XOR and AND gate, which is perhaps not surprising since XOR and AND gates are in a sense the limit cases of OR gate. Note that the difference between the magnitude of $\Delta\bar{Z}$ across logic gates is artificial. The absolute magnitude of OR gate $\Delta\bar{Z}$ is large only because $\bar{r}_{2Z} = 10$. We verify the previous claim that $\Delta\bar{Z}$ scales linearly with production rates.

\begin{figure*}
    \centering
    \includegraphics[width=\linewidth]{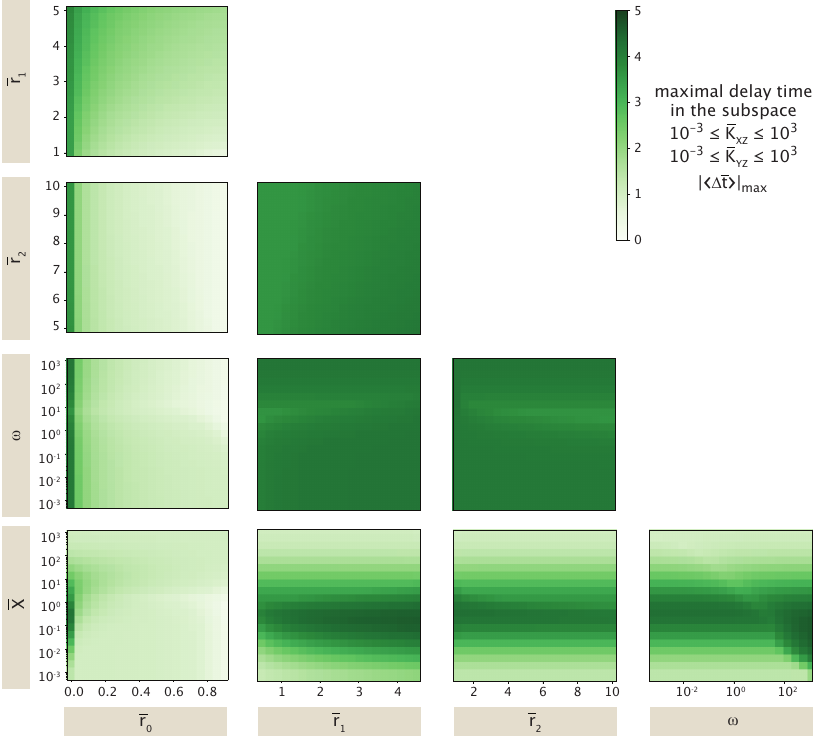}
    \caption{Systematically sweeping across all tunable parameters in coherent feed-forward loops to find the largest $\langle \Delta \bar{t} \rangle$ for the step in $p_\text{act}^X(c_X)$ used throughout the feed-forward loop section, where $c_X^{max} = 10^{-4}$ and $c_X^{min} = 10^{-7}$. We pick a default set of parameters: $\bar{r}_{0Y}  = \bar{r}_{0Z} = \bar{r}_{0} = 0$, $\bar{r}_{1Y}  = \bar{r}_{1Z} = \bar{r}_1 = 1$, $\bar{r}_{2Z} = 5$, $\omega = 5$, $\bar{X} = 1$. Note that we set $\bar{r}_{0Y}  = \bar{r}_{0Z}$ and $\bar{r}_{1Y}  = \bar{r}_{1Z}$ to decrease the number of degrees of freedom in the parameter space without losing too much information.
    For every colormap, we select a pair of parameters, and perform sweeps on a grid of values, while the other parameters remain the default value. For each value pair, they are used to search in the space $(\bar{K}_{XZ}, \bar{K}_{YZ}) \in [10^{-3}, 10^3] \times  [10^{-3}, 10^3]$, and the maximum $\langle \Delta \bar{t} \rangle$ is recorded.}
    \label{fig:ffl_func_param_sweep}
\end{figure*}

Regarding the average time delay $\langle \Delta \bar{t} \rangle$, we demonstrate computational evidence for the existence of an upper bound on $\langle \Delta \bar{t} \rangle$, given a step in $p_\text{act}^X(c_X)$ (fix the high and low values of $p_\text{act}^X$). The full set of tunable parameters of the system is $\bar{r}_{iY}, \bar{r}_{jZ}, \omega, \bar{X}, \bar{K}_{XZ}, \bar{K}_{YZ}$, where $i \in \{0, 1\}$ and $j \in \{0, 1, 2 \}$. They span a semi-infinite 9-dimensional parameter space. To sweep across this entire space is computationally prohibitive. As a result, we instead take a few 2-dimensional slices to illustrate the existence of the upper bound on $\langle \Delta \bar{t} \rangle$. Specifically, we pair-wise tune 5 different parameters: $\bar{r}_0 = \bar{r}_{0Y} = \bar{r}_{0Z}$, $\bar{r}_1 = \bar{r}_{1Y} = \bar{r}_{1Z}$, $\bar{r}_{2Z}$, $\omega$, and $\bar{X}$. For each parameter combination, we search in $K$-subspace (as in Fig.~\ref{fig:ffl_func_param}, $(\bar{K}_{XZ}, \bar{K}_{YZ}) \in [10^{-3}, 10^3] \times  [10^{-3}, 10^3]$) and find the combination of $(\bar{K}_{XZ}, \bar{K}_{YZ})$ that generates the maximum $\langle \Delta \bar{t} \rangle$ and record the value of the largest time delay as $\langle \Delta \bar{t} \rangle_{\max}$. In Fig.~\ref{fig:ffl_func_param_sweep}, we plot $\langle \Delta \bar{t} \rangle_{\max}$ as a function of 10 different pair-wise parameters. We find that all parameter combinations yield $\langle \Delta \bar{t} \rangle < 5$. $\langle \Delta \bar{t} \rangle$ remains finite when any parameter (when possible) is tuned towards $\infty$.  

\begin{figure*}
    \centering
    \includegraphics[width=\linewidth]{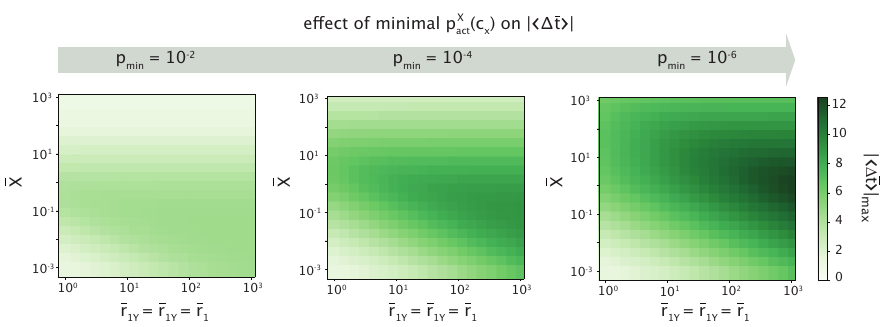}
    \caption{The effect of leakiness of $p_\text{act}^X(c_X)$ on $\langle \Delta \bar{t} \rangle$. We examine the dependence $\langle \Delta \bar{t} \rangle$ on minimal $p_\text{act}^X(c_X)$ in a step, $p_{\min}$, in the XOR gate setting. Parameters used are $\bar{r}_{0Y} = \bar{r}_{0Z} = 0$, $\omega = 0$. The value of $\bar{r}_{1Y}  = \bar{r}_{1Z} = \bar{r}_1$ and $\bar{X}$ are tuned to find the largest $\langle \Delta \bar{t} \rangle$ in that parameter subspace. The detailed sweep process is identical to that in Fig.~\ref{fig:ffl_func_param_sweep}.}
    \label{fig:ffl_pact_sweep}
\end{figure*}

Finally, we address the dependence of $\langle \Delta \bar{t} \rangle$ on the leakiness and saturation of $p_\text{act}^X(c_X)$. Here, we denote the minimal $p_\text{act}^X(c_X)$ in a step as $p_{\min}$ and the maximal $p_\text{act}^X(c_X)$ as $p_{\max}$. After many experimentation with the numerics, we find that the saturation $p_{\max}$ plays a smaller role in determining $\langle \Delta \bar{t} \rangle_{\max}$. The limit $p_{\max} = 1$ yields a similar $\langle \Delta \bar{t} \rangle_{\max}$ as $p_{\max} = 0.95$, and decreasing $p_{\max}$ only makes $\langle \Delta \bar{t} \rangle_{\max}$ smaller. Contrary to $p_{\max}$, $p_{\min}$ plays a much bigger role. We demonstrate this dependency in Fig.~\ref{fig:ffl_pact_sweep}. Here, we examine the OFF step delay in the XOR coherent feed-forward loop, as from previous sweeps in Fig.~\ref{fig:ffl_func_param} and Fig.~\ref{fig:ffl_func_param_sweep}, we see that this tends to be the setting that generates the largest amount of delay. For a given $p_{\min}$, we sweep across $\bar{r}_{1Y} = \bar{r}_{1Z} = \bar{r}_1$, $\bar{X}$. For each combination of $\bar{r}_1$ and $\bar{X}$, we again perform another sweep in $(\bar{K}_{XZ}, \bar{K}_{YZ}) \in [10^{-3}, 10^3] \times  [10^{-3}, 10^3]$ to find largest $\langle \Delta \bar{t} \rangle$. We see from Fig.~\ref{fig:ffl_pact_sweep} that as $p_{\min}$ decreases, the maximal $\langle \Delta \bar{t} \rangle_{\max}$ increases. We note, however, that $\langle \Delta \bar{t} \rangle_{\max}$ still converges computationally for finite $p_{\min}$. Due to the constraint of the explicit effector function $p_\text{act}^X(c_X)$, for any biological parameter $p_{\min}$ is finite. For this reason, we demonstrate the maximum delay corresponding to a biologically sensible set of parameters, and investigate its significance in the main text.

\section{Technical details in incoherent feed-forward loop}
\label{sec:def_of_pulse_incoh}

In this appendix, we will return to some technical details regarding pulses in the incoherent feed-forward loop, as presented in Section~\ref{sec:pulse}. To begin with, we discuss the quantity $\langle \Delta\bar{t}\rangle$ in incoherent feed-forward loops. In the coherent feed-forward loop, the definition of $\langle \Delta\bar{t}\rangle$ in Eqn.~\ref{eq:time_delay_physical} and Eqn.~\ref{eq:time_delay} are equivalent, allowing us to interpret it as the average time delay across concentrations. Here, when $\bar{Z}$ does not exhibit a pulse, there is no difference from the coherent case.
However, Eqn.~\ref{eq:time_delay_physical} becomes ill-defined when $\bar{Z}$ exhibits a pulse. This is because the feed-forward loop response can reach $\bar{Z}$ in a way that the simple regulation curve cannot. A value can still be computed for $\langle \Delta\bar{t}\rangle$ using Eqn.~\ref{eq:time_delay}, but it contains information about the relative size of the pulse rather than acceleration. Because it has unit of time, we regard the direct magnitude of the pulse $\bar{Z}_{\max} - \bar{Z}_f$ for an increasing $\bar{Z}$ response in the main text.

Next, let's derive the maximal average time acceleration $\langle \Delta\bar{t}\rangle_{\max}$ when a pulse does not exist. The green curve shown in Fig.~\ref{fig:ffl_pulse}(C) and (D) is the trajectory that has the largest acceleration; denote this trajectory to be $\bar{Z}_{\text{step}}(\bar{t})$. Mathematically, $\bar{Z}_{\text{step}}(\bar{t}) = \bar{Z}_f$ for $\bar{t}>0$. The simple regulation curve is again
\begin{align}
    \bar{Z}_{\text{simple}}(\bar{t}) = \bar{Z}_ie^{-\bar{t}} + \bar{Z}_f(1-e^{-\bar{t}}).
\end{align}
From Eqn.~\ref{eq:time_delay}, we can then compute the maximal average time acceleration as
\begin{align}
\begin{split}
    \langle \Delta\bar{t}\rangle_{\max} &= \frac{1}{\bar{Z}_f - \bar{Z}_i} \int_0^\infty d\bar{t}\left(\bar{Z}_{\text{step}}(\bar{t})  - \bar{Z}_{\text{simple}}(\bar{t}) \right) \\
    &= \frac{1}{\bar{Z}_f - \bar{Z}_i} \int_0^\infty d\bar{t}\left(\bar{Z}_f  - \bar{Z}_ie^{-\bar{t}} - \bar{Z}_f(1-e^{-\bar{t}}) \right) \\
    &= \frac{1}{\bar{Z}_f - \bar{Z}_i}  \int_0^\infty d\bar{t}(\bar{Z}_f - \bar{Z}_i)e^{-\bar{t}} \\
    & = \frac{\bar{Z}_f - \bar{Z}_i}{\bar{Z}_f - \bar{Z}_i} = 1.
\end{split}
\end{align}

Finally, we discuss the definition of strong pulse used in Fig.~\ref{fig:ffl_pulse}. Numerically, we characterize a pulse to be strong when the maximum concentration that the transient $\bar{Z}(\bar{t})$ reaches, $\bar{Z}_{\max}$, satisfies $(\bar{Z}_{\max} - \bar{Z}_f)/(\bar{Z}_f - \bar{Z}_i) > 0.05$ for an increasing $\bar{Z}$ response. The threshold of 0.05 is an arbitrary choice. Due to this threshold, there exists a region of trajectories that are technically pulses but are not strong enough to be considered. The trajectory in Fig.~\ref{fig:ffl_pulse}(C) in fact belongs to this region. It exhibits a pulse strictly speaking. Nevertheless, the magnitude of the pulse is vanishingly small. Functionally, it acts in the same manner as the trajectories that do not possess a pulse. 

\section{Effect of $\bar{Y}$ in continuous tuning}
\label{appendix:ffl_cont_tuning}

\begin{figure*}
    \centering
    \includegraphics[width=\linewidth]{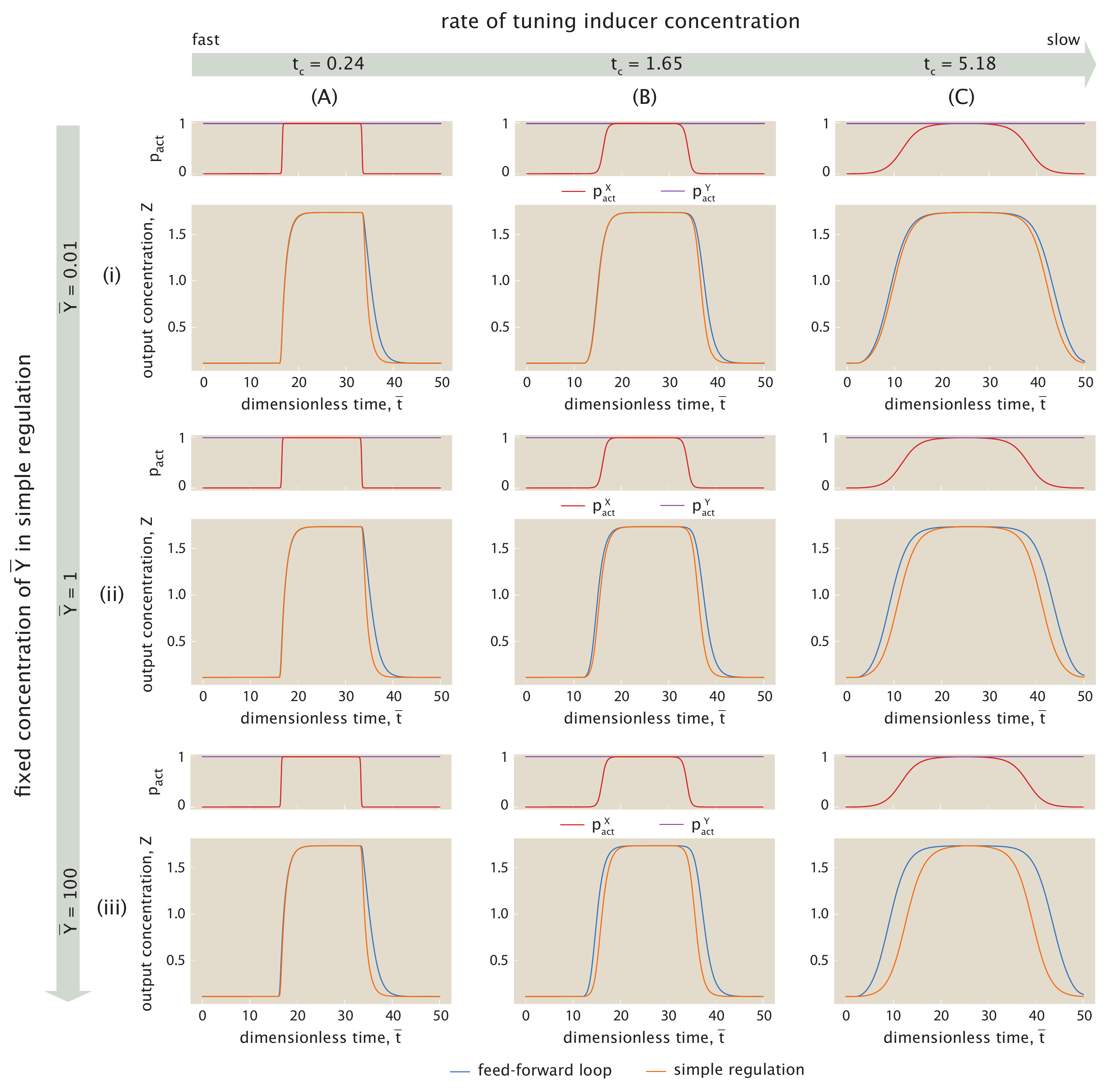}
    \caption{Demonstrating the how the arbitrary choice of $\bar{Y}$ affects the simple regulation dynamics. We replicate Fig.~\ref{fig:ffl_cont_signal} with different choices of $\bar{Y}$. From top to bottom, $\bar{Y}$ is set to be 0.01 in (i), 1 in (ii), which matches the setting in Fig.~\ref{fig:ffl_cont_signal}, and 100 in (iii). From left to right, the rate of concentration is tuned the same way as in Fig.~\ref{fig:ffl_cont_signal}. System parameters are also identical to those in Fig.~\ref{fig:ffl_cont_signal}.}
    \label{fig:ffl_effect_of_Y}
\end{figure*}

In Section~\ref{sec:ffl_cont_signal}, we discussed the scenario where the signal $c_X(\bar{t})$ is no longer a step function, but a continuous function in time. We mentioned that the comparison with simple regulation in this case is subtle, as the arbitrary choice of value $\bar{Y}$ affects the shape of simple regulation response. Here, we expand on Fig.~\ref{fig:ffl_cont_signal} where we repeat the numerical integration for three distinct values of $\bar{Y}$, as shown in Fig.~\ref{fig:ffl_effect_of_Y}. For the fast tuning case, different choices of $\bar{Y}$ have no effect on the simple regulation trajectory. This is expected since in the limit case of a step function signal in $c_X$, the value of $\bar{Y}$ strictly has no effect on the shape of the trajectory. As the rate of tuning $c_X$ slows down, the effect of $\bar{Y}$ becomes more and more pronounced. Specifically, as $\bar{Y}$ increases, the rescaled simple regulation curve ``shrinks". As a result, the magnitude of apparent delay/acceleration between the feed-forward loop and simple regulation response increases. Due to the artificial nature of the choice of $\bar{Y}$ in simple regulation, we cannot make a claim about the magnitude of delay/acceleration in the slow tuning limit. Nevertheless, our qualitative results stand. The choice of $\bar{Y}$ does not change the type of response to a step. For example, even though $\bar{Y}$ affects the magnitude of delay on the OFF step, the feed-forward loop in this case always delays on the OFF step.\\

\section{Code availability}\label{app:code}
All Jupyter notebooks used to generate graphs in figures throughout this paper are available \footnote{Supporting Python code is available on GitHub at \href{https://github.com/RPGroup-PBoC/2025_inducers}{github.com/RPGroup-PBoC/2025\_inducers}}.

\bibliography{main}% Produces the bibliography via BibTeX.

@article{mangan2003structure,
  title={Structure and function of the feed-forward loop network motif},
  author={Mangan, Shmoolik and Alon, Uri},
  journal={Proc. Nat. Acad. Sci. USA},
  volume={100},
  number={21},
  pages={11980--11985},
  year={2003},
}

@article{Jacob1961,
   author = {F. Jacob and J. Monod},
   title = {{Genetic regulatory mechanisms in the synthesis of proteins}},
   journal = {J. Mol. Biol.},
   volume = {3},
   pages = {318-56},
   year = {1961}
}

@article{Muller1996,
   author = {J. M\"{u}ller and S. Oehler and B. M\"{u}ller-Hill},
   title = {{Repression of {\it lac} promoter as a function of distance, phase and quality of an auxiliary {\it lac} operator}},
   journal = {J. Mol. Biol.},
   volume = {257},
   number = {1},
   pages = {21-9},
   year = {1996}
}

@article{Oehler1994,
   author = {Oehler, S. and Amouyal, M. and Kolkhof, P. and von Wilcken-Bergmann, B. and M{\"u}ller-Hill, B.},
   title = {{Quality and position of the three {\it lac} operators of \emph{E. coli} define efficiency of repression}},
   journal = {EMBO J.},
   volume = {13},
   number = {14},
   pages = {3348-55},
   year = {1994}
}

@article{Britten1969,
   author = {Britten, R. J. and Davidson, E. H.},
   title = {{Gene regulation for higher cells: a theory}},
   journal = {Science},
   volume = {165},
   number = {891},
   pages = {349-57},
   year = {1969}
}

@article{Englesberg1965,
   author = {Englesberg, E. and Irr, J. and Power, J. and Lee, N.},
   title = {Positive control of enzyme synthesis by gene {{\it C}} in the {{\it L-arabinose}} system},
   journal = {J. Bacteriol.},
   volume = {90},
   number = {4},
   pages = {946-57},
   year = {1965}
}

@article{Gardner2000,
   author = {Gardner, T. S. and Cantor, C. R. and Collins, J. J.},
   title = {Construction of a genetic toggle switch in {{ \it Escherichia coli}}},
   journal = {Nature},
   volume = {403},
   number = {6767},
   pages = {339-42},
   year = {2000}
}

@article{Elowitz2000,
   author = {Elowitz, M. B. and Leibler, S.},
   title = {{A synthetic oscillatory network of transcriptional regulators}},
   journal = {Nature},
   volume = {403},
   number = {6767},
   pages = {335-8},
   year = {2000}
}

@article{Hill1910,
 author = {A. V. Hill},
   title = {{The possible effects of the aggregations of the molecules of haemoglobin on its dissociation curves}},
   journal = {The Journal of Physiology},
   volume = {40},
   number = {suppl},
   pages = {i-vii},
   year = {1910}
}

@article{Hill1913,
   author = {A. V. Hill},
   title = {{The combinations of haemoglobin with oxygen and with carbon monoxide. I}},
   journal = {Biochem. J.},
   volume = {7},
   number = {5},
   pages = {471-480},
   year = {1913}
}

@book{Ptashne2004,
   author = {M. Ptashne},
   title = {{A Genetic Switch: Phage Lambda Revisited}},
   publisher = {Cold Spring Harbor Laboratory Press},
   address = {Cold Spring Harbor, N.Y.},
   edition = {3},
   year = {2004}
}

@book{Ptashne2002,
   author = {M. Ptashne and A. Gann},
   title = {{Genes and Signals}},
   publisher = {Cold Spring Harbor Laboratory Press},
   address = {New York},
   year = {2002}
}

@book{Phillips2020,
   author = {Phillips, Rob},
   title = {The Molecular Switch: Signaling and Allostery},
   publisher = {Princeton University Press},
   year = {2020}
}

@article{Marzen2013,
   author = {Marzen, S. and Garcia, H. G. and Phillips, R.},
   title = {{Statistical mechanics of Monod-Wyman-Changeux (MWC) models}},
   journal = {J. Mol. Biol.},
   volume = {425},
   number = {9},
   pages = {1433-60},
   year = {2013}
}

@article{Martins2011,
   author = {Martins, B. M. and Swain, P. S.},
   title = {{Trade-offs and constraints in allosteric sensing}},
   journal = {PLoS Comput. Biol.},
   volume = {7},
   number = {11},
   pages = {e1002261},
   year = {2011}
}

@article{Monod1965,
   author = {Monod, J. and Wyman, J. and Changeux, J. P.},
   title = {{On the nature of allosteric transitions: a plausible model}},
   journal = {J. Mol. Biol.},
   volume = {12},
   pages = {88-118},
   year = {1965}
}

@article{Gerhart1962,
   author = {Gerhart, J. C. and Pardee, A. B.},
   title = {{Enzymology of control by feedback inhibition}},
   journal = {J. Biol. Chem.},
   volume = {237},
   number = {3},
   pages = {891-+},
   year = {1962}
}

@article{Gerhart2014,
   author = {Gerhart, J.},
   title = {{From feedback inhibition to allostery: the enduring example of aspartate transcarbamoylase}},
   journal = {FEBS J.},
   volume = {281},
   number = {2},
   pages = {612-620},
   year = {2014}
}

@book{Strogatz2015,
   author = {Strogatz, Steven H.},
   title = {{Nonlinear Dynamics and Chaos : With Applications to Physics, Biology, Chemistry, and Engineering}},
   publisher = {Westview Press, a member of the Perseus Books Group},
   address = {Boulder, CO},
   edition = {2},
   year = {2015}
}

@article{Milo2002,
   author = {Milo, R. and Shen-Orr, S. and Itzkovitz, S. and Kashtan, N. and Chklovskii, D. and Alon, U.},
   title = {{Network motifs: simple building blocks of complex networks}},
   journal = {Science},
   volume = {298},
   number = {5594},
   pages = {824-7},
   year = {2002}
}

@article{Cherry2000,
   author = {Cherry, J. L. and Adler, F. R.},
   title = {{How to make a biological switch}},
   journal = {J. Theor. Biol.},
   volume = {203},
   number = {2},
   pages = {117-133},
   year = {2000}
}

@article{Changeux2013,
   author = {Changeux, J. P.},
   title = {{50 years of allosteric interactions: the twists and turns of the models}},
   journal = {Nat. Rev. Mol. Cell Biol.},
   volume = {14},
   number = {12},
   pages = {819-829},
   year = {2013}
}

@article{Monod1963,
   author = {Monod, J. and Changeux, J.P. and Jacob, F.},
   title = {{Allosteric proteins and cellular control systems}},
   journal = {J. Mol. Biol.},
   volume = {6},
   pages = {306-29},
   year = {1963}
}

@article{Lindsley2006,
  author = {Lindsley, J. E. and Rutter, J.},
   title = {{Whence cometh the allosterome?}},
   journal = {Proc. Nat. Acad. Sci. USA},
   volume = {103},
   number = {28},
   pages = {10533-10535},
   year = {2006}
   }

@article{zuniga2014turing,
  title={In {{Turing's}} hands—the making of digits},
  author={Zuniga, A. and Zeller, R.},
  journal={Science},
  volume={345},
  number={6196},
  pages={516--517},
  year={2014},
  publisher={American Association for the Advancement of Science}
}

@article{corson2012geometry,
  title={Geometry, epistasis, and developmental patterning},
  author={Corson, F. and Siggia, E. D.},
  journal={Proc. Nat. Acad. Sci. USA},
  volume={109},
  number={15},
  pages={5568--5575},
  year={2012},
  publisher={National Academy of Sciences}
}

@article{schindlermorphogenesis,
author = {Schindler, A. J. and Sherwood, D. R.},
title = {Morphogenesis of the {{\it Caenorhabditis elegans}} vulva},
journal = {Wiley Interdiscip. Rev. Dev. Biol.},
volume = {2},
number = {1},
pages = {75-95},
year = {2013}
}

@article{liu2008yamanaka,
  title={Yamanaka factors critically regulate the developmental signaling network in mouse embryonic stem cells},
  author={Liu, X. and Huang, J. and Chen, T. and Wang, Y. and Xin, S. and Li, J. and Pei, G. and Kang, J.},
  journal={Cell Res.},
  volume={18},
  number={12},
  pages={1177--1189},
  year={2008},
  publisher={Nature Publishing Group}
}

@inproceedings{laslo2008gene,
  title={Gene regulatory networks directing myeloid and lymphoid cell fates within the immune system},
  author={Laslo, P. and Pongubala, J. M. and Lancki, D. W. and Singh, H.},
  booktitle={Seminars in immunology},
  volume={20},
  number={4},
  pages={228--235},
  year={2008},
  organization={Elsevier}
}

@article{chute2010minireview,
  title={Minireview: nuclear receptors, hematopoiesis, and stem cells},
  author={Chute, J. P. and Ross, J. R. and McDonnell, D. P.},
  journal={Molecular Endocrinology},
  volume={24},
  number={1},
  pages={1--10},
  year={2010},
  publisher={Oxford University Press}
}

@article{jaeger2011gap,
  title={The gap gene network},
  author={Jaeger, J.},
  journal={Cell. Mol. Life Sci.},
  volume={68},
  pages={243--274},
  year={2011},
  publisher={Springer}
}

@book{carroll2001dna,
  title={From {DNA} to Diversity: Molecular Genetics and the Evolution of Animal Design},
  author={Carroll, S. and Grenier, J. and Weatherbee, S.},
  year={2001},
  address={Malden, MA},
  publisher={Blackwell Science},
}

@book{Echols2001,
   author = {Echols, H. and Gross, C.},
   title = {Operators and Promoters: The Story of Molecular Biology and Its Creators},
   publisher = {University of California Press},
   address = {Berkeley},
   year = {2001}
}

@book{Muller-Hill1996,
   author = {B. M\"{u}ller-Hill},
   title = {{The lac Operon: A Short History of a Genetic Paradigm}},
   publisher = {Walter de Gruyter},
   address = {Berlin, New York},
   year = {1996}
}

@book{Judson1996,
   author = {Judson, Horace Freeland},
   title = {The Eighth Day of Creation},
   publisher = {Cold Spring Harbor Laboratory Press},
   address = {New York},
   year = {1996}
}

@article{SantosZavaleta2019,
   author = {Santos-Zavaleta, A. and Salgado, H. and Gama-Castro, S. and Sanchez-Perez, M. and Gomez-Romero, L. and Ledezma-Tejeida, D. and Garcia-Sotelo, J. S. and Alquicira-Hernandez, K. and Muniz-Rascado, L. J. and Pena-Loredo, P. and Ishida-Gutierrez, C. and Velazquez-Ramirez, D. A. and Del Moral-Chavez, V. and Bonavides-Martinez, C. and Mendez-Cruz, C. F. and Galagan, J. and Collado-Vides, J.},
   title = {{RegulonDB} v 10.5: tackling challenges to unify classic and high throughput knowledge of gene regulation in {{\it E. coli}} {K-12}},
   journal = {Nucleic Acids Res.},
   volume = {47},
   number = {D1},
   pages = {D212-D220},
   year = {2019}
}

@article{mangan2006incoherent,
  title={The incoherent feed-forward loop accelerates the response-time of the gal system of {{\it Escherichia coli}}},
  author={Mangan, Shmoolik and Itzkovitz, Shalev and Zaslaver, Alon and Alon, Uri},
  journal={J. Mol. Biol.},
  volume={356},
  number={5},
  pages={1073--1081},
  year={2006},
  publisher={Elsevier}
}

@article{mangan2003coherent,
  title={The coherent feedforward loop serves as a sign-sensitive delay element in transcription networks},
  author={Mangan, Shmoolik and Zaslaver, Alon and Alon, Uri},
  journal={J. Mol. Biol.},
  volume={334},
  number={2},
  pages={197--204},
  year={2003},
  publisher={Elsevier}
}

@article{shen2002network,
  title={Network motifs in the transcriptional regulation network of {{\it {Escherichia coli}}}},
  author={Shen-Orr, Shai S and Milo, Ron and Mangan, Shmoolik and Alon, Uri},
  journal={Nat. Genet.},
  volume={31},
  number={1},
  pages={64--68},
  year={2002},
  publisher={Nature Publishing Group}
}

@article{Huang2007,
   author = {Huang, S. and Guo, Y. P. and May, G. and Enver, T.},
   title = {{Bifurcation dynamics in lineage-commitment in bipotent progenitor cells}},
   journal = {Dev. Biol.},
   volume = {305},
   number = {2},
   pages = {695-713},
   year = {2007}
}

@book{Alon2020,
   author = {Alon, Uri},
   title = {{An Introduction to Systems Biology: Design Principles of Biological Circuits}},
   publisher = {CRC Press, Taylor \& Francis Group},
   address = {Boca Raton},
   edition = {2},
   year = {2020}
}

@book{Goodwin1963,
   author = {Goodwin, Brian C.},
   title = {{Temporal Organization in Cells: A Dynamic Theory of Cellular Control Processes}},
   publisher = {Academic Press},
   address = {London, New York,},
   year = {1963}
}

@book{Covert2015,
   author = {Covert, Markus},
   title = {{Fundamentals of Systems Biology: From Synthetic Circuits to Whole-Cell Models}},
   publisher = {CRC Press, Taylor \& Francis Group},
   address = {Boca Raton},
   year = {2015}
}

@article{Shea1985,
   author = {Shea, M. A. and Ackers, G. K.},
   title = {{The OR control system of bacteriophage lambda. A physical-chemical model for gene regulation}},
   journal = {J. Mol. Biol.},
   volume = {181},
   number = {2},
   pages = {211-30},
   year = {1985}
}

@article{Novick1957,
   author = {Novick, A. and Weiner, M.},
   title = {{Enzyme induction as an all-or-none phenomenon}},
   journal = {Proc. Natl. Acad. Sci. USA},
   volume = {43},
   number = {7},
   pages = {553-566},
   year = {1957}
}

@article{fflnoise_gui2016noise,
  title={Noise decomposition principle in a coherent feed-forward transcriptional regulatory loop},
  author={Gui, Rong and Liu, Quan and Yao, Yuangen and Deng, Haiyou and Ma, Chengzhang and Jia, Ya and Yi, Ming},
  journal={Front. Physiol.},
  volume={7},
  pages={600},
  year={2016},
  publisher={Frontiers Media SA}
}

@article{fflnoise_chakravarty2021systematic,
  title={Systematic analysis of noise reduction properties of coupled and isolated feed-forward loops},
  author={Chakravarty, Suchana and Csik{\'a}sz-Nagy, Attila},
  journal={PLoS Comput. Biol.},
  volume={17},
  number={12},
  pages={e1009622},
  year={2021},
  publisher={Public Library of Science San Francisco, CA USA}
}

@article{fflsoie_ghosh2005noise,
  title={Noise characteristics of feed forward loops},
  author={Ghosh, Bhaswar and Karmakar, Rajesh and Bose, Indrani},
  journal={Phys. Biol.},
  volume={2},
  number={1},
  pages={36},
  year={2005},
  publisher={IOP Publishing}
}

@article{ozbudak2004multistability,
  title={Multistability in the lactose utilization network of {{ \it Escherichia coli}}},
  author={Ozbudak, Ertugrul M and Thattai, Mukund and Lim, Han N and Shraiman, Boris I and Van Oudenaarden, Alexander},
  journal={Nature},
  volume={427},
  number={6976},
  pages={737--740},
  year={2004},
  publisher={Nature Publishing Group UK London}
}

@article{subsoontorn2011bistability,
  title={Bistability of an in vitro synthetic autoregulatory switch},
  author={Subsoontorn, Pakpoom and Kim, Jongmin and Winfree, Erik},
  journal={arXiv preprint arXiv:1101.0723},
  year={2011}
}

@article{siegal2011emergence,
  title={Emergence of switch-like behavior in a large family of simple biochemical networks},
  author={Siegal-Gaskins, Dan and Mejia-Guerra, Maria Katherine and Smith, Gregory D and Grotewold, Erich},
  journal={PLoS Comput. Biol.},
  volume={7},
  number={5},
  pages={e1002039},
  year={2011},
  publisher={Public Library of Science San Francisco, USA}
}

@article{pal2016functional,
  title={Functional characteristics of gene expression motifs with single and dual strategies of regulation},
  author={Pal, Mainak and Ghosh, Sayantari and Bose, Indrani},
  journal={Biomed. Phys. Eng. Express},
  volume={2},
  number={2},
  pages={025009},
  year={2016},
  publisher={IOP Publishing}
}

@article{bintu2005transcriptional,
  title={Transcriptional regulation by the numbers: models},
  author={Bintu, Lacramioara and Buchler, Nicolas E and Garcia, Hernan G and Gerland, Ulrich and Hwa, Terence and Kondev, Jan{\'e} and Phillips, Rob},
  journal={Curr. Opin. Genet. Dev.},
  volume={15},
  number={2},
  pages={116--124},
  year={2005},
  publisher={Elsevier}
}

@article{bintu2005transcriptional2,
  title={Transcriptional regulation by the numbers: applications},
  author={Bintu, Lacramioara and Buchler, Nicolas E and Garcia, Hernan G and Gerland, Ulrich and Hwa, Terence and Kondev, Jane and Kuhlman, Thomas and Phillips, Rob},
  journal={Curr. Opin. Genet. Dev.},
  volume={15},
  number={2},
  pages={125--135},
  year={2005},
  publisher={Elsevier}
}

@article{becskei2001positive,
  title={Positive feedback in eukaryotic gene networks: cell differentiation by graded to binary response conversion},
  author={Becskei, Attila and S{\'e}raphin, Bertrand and Serrano, Luis},
  journal={EMBO J.},
  year={2001},
  publisher={John Wiley \& Sons, Ltd., Chichester, UK}
}

@article{laxhuber2020theoretical,
  title={Theoretical investigation of a genetic switch for metabolic adaptation},
  author={Laxhuber, Kathrin S and Morrison, Muir J and Chure, Griffin and Belliveau, Nathan M and Strandkvist, Charlotte and Naughton, Kyle L and Phillips, Rob},
  journal={PloS One},
  volume={15},
  number={5},
  pages={e0226453},
  year={2020},
  publisher={Public Library of Science San Francisco, CA USA}
}

@article{Rousseau2023,
  author  = {Rousseau, R. J. and Phillips, R.},
  title   = {Bifurcation and multistability in three-gene-driven network models},
  journal = {Bull. Am. Phys. Soc.},
  volume  = {68},
  number = {3},
  year    = {2023},
  note    = {{APS March Meeting, Las Vegas, NV, March 8, 2023}},
  url = {https://meetings.aps.org/Meeting/MAR23/Session/N10.5}
}

@article{Kelvin1901,
   author = {Lord Kelvin},
   title = {{Nineteenth Century Clouds over the Dynamical Theory of
Heat and Light}},
   journal = {Phil. Mag.},
   volume = {2},
   pages = {1-40},
   year = {1901}
}

@article{Barnett2004,
   author = {Barnett, J. A.},
   title = {A history of research on yeasts 7: enzymic adaptation and regulation},
   journal = {Yeast},
   volume = {21},
   number = {9},
   pages = {703-746},
   year = {2004}
}

@book{Ferrell2022,
author = {Ferrell, James E.},
title = {Systems Biology of Cell Signaling: Recurring Themes and Quantitative Models},
publisher = {CRC Press},
address = {Boca Raton, FL},
year = {2022}
}

@article{Monod1949,
   author = {Monod, J.},
   title = {{The Growth of Bacterial Cultures}},
   journal = {Ann. Rev. Microbiol.},
   volume = {3},
   pages = {371-394},
   year = {1949}
}

@article{Ackers1982,
   author = {Ackers, G. K. and Johnson, A. D. and Shea, M. A.},
   title = {{Quantitative model for gene regulation by lambda phage repressor}},
   journal = {Proc. Natl. Acad. Sci. USA},
   volume = {79},
   number = {4},
   pages = {1129-33},
   year = {1982}
}

@article{Buchler2003a,
   author = {Buchler, N. E. and Gerland, U. and Hwa, T.},
   title = {{On schemes of combinatorial transcription logic}},
   journal = {Proc. Natl. Acad. Sci. USA},
   volume = {100},
   number = {9},
   pages = {5136-41},
   year = {2003}
}

@article{Vilar2003b,
   author = {Vilar, J. M. and Guet, C. C. and Leibler, S.},
   title = {{Modeling network dynamics: the {{ \it lac}} operon, a case study}},
   journal = {J. Cell Biol.},
   volume = {161},
   number = {3},
   pages = {471-6},
   year = {2003}
}

@article{Sherman2012,
   author = {Sherman, M. S. and Cohen, B. A.},
   title = {{Thermodynamic state ensemble models of {\it cis}-regulation}},
   journal = {PLoS Comput. Biol.},
   volume = {8},
   number = {3},
   pages = {e1002407},
   year = {2012}
}

@article{Vilar2003a,
   author = {Vilar, J. M. and Leibler, S.},
   title = {{DNA looping and physical constraints on transcription regulation}},
   journal = {J. Mol. Biol.},
   volume = {331},
   number = {5},
   pages = {981-9},
   year = {2003}
}

@article{Kuhlman2007,
   author = {Kuhlman, T. and Zhang, Z. and Saier Jr., M. H. and Hwa, T.},
   title = {{Combinatorial transcriptional control of the lactose operon of \emph{Escherichia coli}}},
   journal = {Proc. Natl. Acad. Sci. USA},
   volume = {104},
   number = {14},
   pages = {6043-8},
   year = {2007}
}

@article{Ullmann2011,
   author = {Ullmann, A.},
   title = {{In memoriam: Jacques Monod (1910-1976)}},
   journal = {Genome Biol. Evol.},
   volume = {3},
   pages = {1025-1033},
   year = {2011}
}

@book{Gamow1985,
   author = {Gamow, George},
   title = {{Thirty Years that Shook Physics: The Story of Quantum Theory}},
   publisher = {Dover Publications},
   address = {New York},
   year = {1985}
}

@book{Longair2020,
   author = {Longair, Malcolm S.},
   title = {{Theoretical Concepts in Physics: An Alternative View of Theoretical Reasoning in Physics}},
   publisher = {Cambridge University Press,},
   year = {2020}
}

@book{Segre2007c,
   author = {Segre, Emilio},
   title = {{From X-rays to Quarks : Modern Physicists and Their Discoveries}},
   publisher = {Dover Publications},
   address = {Mineola, N.Y.},
   year = {2007}
}

@article{Sokolik2015,
   author = {Sokolik, C. and Liu, Y. X. and Bauer, D. and McPherson, J. and Broeker, M. and Heimberg, G. and Qi, L. S. and Sivak, D. A. and Thomson, M.},
   title = {{Transcription Factor Competition Allows Embryonic Stem Cells to Distinguish Authentic Signals from Noise}},
   journal = {Cell Syst.},
   volume = {1},
   number = {2},
   pages = {117-129},
   year = {2015}
}

@book{Phillips2012,
   author = {R. Phillips and J. Kondev and J. Theriot and H. G. Garcia},
   title = {{Physical biology of the cell, 2nd Edition}},
   publisher = {Garland Science},
   address = {New York},
   note = {(Illustrated by N. Orme)},
   year = {2013}
}

@book{Schleif1993Book,
  author = {Schleif, R. F.},
   title = {{Genetics and Molecular Biology}},
   publisher = {Johns Hopkins University Press},
   address = {Baltimore},
   edition = {2nd},
   year = {1993}
}

@article{griffith1968mathematics,
  title={Mathematics of cellular control processes I. Negative feedback to one gene},
  author={Griffith, J. S.},
  journal={J. Theor. Biol.},
  volume={20},
  number={2},
  pages={202--208},
  year={1968},
  publisher={Elsevier}
}

@article{rosenfeld2002negative,
  title={Negative autoregulation speeds the response times of transcription networks},
  author={Rosenfeld, N. and Elowitz, M. B and Alon, U.},
  journal={J. Mol. Biol.},
  volume={323},
  number={5},
  pages={785--793},
  year={2002},
  publisher={Elsevier}
}

@article{mitrophanov2010positive,
  title={Positive autoregulation shapes response timing and intensity in two-component signal transduction systems},
  author={Mitrophanov, A. Y. and Hadley, T. J. and Groisman, E. A},
  journal={J. Mol. Biol.},
  volume={401},
  number={4},
  pages={671--680},
  year={2010},
  publisher={Elsevier}
}

@article{tyson2003sniffers,
  title={Sniffers, buzzers, toggles and blinkers: dynamics of regulatory and signaling pathways in the cell},
  author={Tyson, J. J. and Chen, K. C. and Novak, B.},
  journal={Curr. Opin. Cell Biol.},
  volume={15},
  number={2},
  pages={221--231},
  year={2003},
  publisher={Elsevier}
}

@article{hermsen2011speed,
  title={Speed, sensitivity, and bistability in auto-activating signaling circuits},
  author={Hermsen, R. and Erickson, D. W. and Hwa, T.},
  journal={PLoS Comput. Biol.},
  volume={7},
  number={11},
  pages={e1002265},
  year={2011},
  publisher={Public Library of Science San Francisco, USA}
}

@article{yagil1971relation,
  title={On the relation between effector concentration and the rate of induced enzyme synthesis},
  author={Yagil, G. and Yagil, E.},
  journal={Biophys. J.},
  volume={11},
  number={1},
  pages={11--27},
  year={1971},
  publisher={Elsevier}
}

@article{saiz2012physics,
  title={The physics of protein--DNA interaction networks in the control of gene expression},
  author={Saiz, L.},
  journal={J. Phys.: Condens. Matter},
  volume={24},
  number={19},
  pages={193102},
  year={2012},
  publisher={IOP Publishing}
}

@article{o1980equilibrium,
  title={Equilibrium binding of inducer to lac repressor. operator DNA complex},
  author={O'Gorman, R. B. and Rosenberg, J. M. and Kallai, O. B. and Dickerson, R. E. and Itakura, K. and Riggs, A. D. and Matthews, K. S.},
  journal={J. Biol. Chem.},
  volume={255},
  number={21},
  pages={10107--10114},
  year={1980},
  publisher={Elsevier}
}

@article{DalyMatthews1986exptind,
  title={Characterization and modification of a monomeric mutant of the lactose repressor protein},
  author={Daly, T. J. and Matthews, K. S.},
  journal={Biochemistry},
  volume={25},
  number={19},
  pages={5474 -- 5478},
  year={1986}
}

@article{vilar2011control,
  title={Control of gene expression by modulated self-assembly},
  author={Vilar, J. M. G. and Saiz, L.},
  journal={Nucleic Acids Res.},
  volume={39},
  number={16},
  pages={6854--6863},
  year={2011},
  publisher={Oxford University Press}
}

@article{vilar2013systems,
  title={Systems biophysics of gene expression},
  author={Vilar, J. M. G. and Saiz, L.},
  journal={Biophys. J.},
  volume={104},
  number={12},
  pages={2574--2585},
  year={2013},
  publisher={Elsevier}
}

@article{vilarSaiz2013,
  title={Reliable prediction of complex phenotypes from a modular design in free energy space: an extensive exploration of the lac operon},
  author={Vilar, J. M. G. and Saiz, L.},
  journal={ACS Synth. Biol.},
  volume={2},
  number={10},
  pages={576--586},
  year={2013}
}

@article{saiz2008ab,
  title={Ab initio thermodynamic modeling of distal multisite transcription regulation},
  author={Saiz, L. and Vilar, J. M. G.},
  journal={Nucleic Acids Res.},
  volume={36},
  number={3},
  pages={726--731},
  year={2008},
  publisher={Oxford University Press}
}

@book{TLHill1977,
   author = {Hill, T. L.},
   title = {{Cooperativity Theory in Biochemistry: Steady-State and Equilibrium Systems}},
   publisher = {Springer},
   address = {N.Y.},
   edition = {1},
   year = {1977}
}

@book{TLHill1989,
   author = {Hill, T. L.},
   title = {{Free Energy Transduction and Biochemical Cycle Kinetics}},
   publisher = {Springer},
   address = {N.Y.},
   edition = {1},
   year = {1989}
}

@article{groisman2001pleiotropic,
  title={The pleiotropic two-component regulatory system PhoP-PhoQ},
  author={Groisman, E. A},
  journal={J. Bacteriol.},
  volume={183},
  number={6},
  pages={1835--1842},
  year={2001},
  publisher={American Society for Microbiology}
}

@article{schleif2010arac,
  title={AraC protein, regulation of the l-arabinose operon in Escherichia coli, and the light switch mechanism of AraC action},
  author={Schleif, R.},
  journal={FEMS Microbiol. Rev.},
  volume={34},
  number={5},
  pages={779--796},
  year={2010},
  publisher={Blackwell Publishing Ltd Oxford, UK}
}

@article{hamoen1998competence,
  title={The competence transcription factor of Bacillus subtilis recognizes short A/T-rich sequences arranged in a unique, flexible pattern along the DNA helix},
  author={Hamoen, L. W. and Van Werkhoven, A. F. and Bijlsma, J. J. E. and Dubnau, D. and Venema, G.},
  journal={Genes Dev.},
  volume={12},
  number={10},
  pages={1539--1550},
  year={1998}
}

@article{perron2005dimerization,
  title={Dimerization and DNA binding of the Salmonella enterica PhoP response regulator are phosphorylation independent},
  author={Perron-Savard, P. and De Crescenzo, G. and Moual, H. L.},
  journal={Microbiology},
  volume={151},
  number={12},
  pages={3979--3987},
  year={2005}
}

@article{morelli2009dna,
  title={DNA looping provides stability and robustness to the bacteriophage $\lambda$ switch},
  author={Morelli, M. J. and Ten W., Pieter R. and Allen, R. J.},
  journal={Proc. Natl. Acad. Sci. USA},
  volume={106},
  number={20},
  pages={8101--8106},
  year={2009},
  publisher={National Academy of Sciences}
}

@article{anderson2008dna,
  title={DNA looping can enhance lysogenic CI transcription in phage lambda},
  author={Anderson, L. M. and Yang, H.},
  journal={Proc. Natl. Acad. Sci. USA},
  volume={105},
  number={15},
  pages={5827--5832},
  year={2008},
  publisher={National Academy of Sciences}
}

@article{einav2018theoretical,
  title={Theoretical analysis of inducer and operator binding for cyclic-AMP receptor protein mutants},
  author={Einav, T. and Duque, J. and Phillips, R.},
  journal={PLoS One},
  volume={13},
  number={9},
  pages={e0204275},
  year={2018},
  publisher={Public Library of Science San Francisco, CA USA}
}

@misc{RousseauPhillips2025,
      title={Bifurcations and multistability in inducible three-gene toggle switch networks}, 
      author={Rousseau, R. J. and Phillips, R.},
      year={2025},
      eprint={2509.24926},
      archivePrefix={arXiv},
      primaryClass={q-bio.MN}
}

@article{walczak2010optimizing,
  title={Optimizing information flow in small genetic networks. II. Feed-forward interactions},
  author={Walczak, A. M. and Tka{\v{c}}ik, G. and Bialek, W.},
  journal={Phys. Rev. E},
  volume={81},
  number={4},
  pages={041905},
  year={2010},
  publisher={APS}
}

@article{tkavcik2012optimizing,
  title={Optimizing information flow in small genetic networks. III. A self-interacting gene},
  author={Tka{\v{c}}ik, G. and Walczak, A. M. and Bialek, W.},
  journal={Phys. Rev. E},
  volume={85},
  number={4},
  pages={041903},
  year={2012},
  publisher={APS}
}

@article{tkavcik2011information,
  title={Information transmission in genetic regulatory networks: a review},
  author={Tka{\v{c}}ik, G. and Walczak, A. M.},
  journal={J. Phys.: Condens. Matter},
  volume={23},
  number={15},
  pages={153102},
  year={2011},
  publisher={IOP Publishing}
}

@article{landman2017self,
  title={Self-consistent theory of transcriptional control in complex regulatory architectures},
  author={Landman, J. and Brewster, R. C. and Weinert, F. M. and Phillips, R. and Kegel, W. K.},
  journal={PLoS One},
  volume={12},
  number={7},
  pages={e0179235},
  year={2017},
  publisher={Public Library of Science San Francisco, CA USA}
}

@article{bottani2017hill,
  title={Hill function-based models of transcriptional switches: impact of specific, nonspecific, functional and nonfunctional binding},
  author={Bottani, S. and Veitia, R. A.},
  journal={Biol. Rev.},
  volume={92},
  number={2},
  pages={953--963},
  year={2017},
  publisher={Wiley Online Library}
}

@article{michel2010transcription,
  title={How transcription factors can adjust the gene expression floodgates},
  author={Michel, D.},
  journal={Prog. Biophys. Mol. Biol.},
  volume={102},
  number={1},
  pages={16--37},
  year={2010},
  publisher={Elsevier}
}

@article{daber2009one,
  title={One is not enough},
  author={Daber, R. and Sharp, K. and Lewis, M.},
  journal={J. Mol. Biol.},
  volume={392},
  number={5},
  pages={1133--1144},
  year={2009},
  publisher={Elsevier}
}

@article{daber2011thermodynamic,
  title={Thermodynamic analysis of mutant lac repressors},
  author={Daber, R. and Sochor, M. A. and Lewis, M.},
  journal={J. Mol. Biol.},
  volume={409},
  number={1},
  pages={76--87},
  year={2011},
  publisher={Elsevier}
}

@article{rogers2015synthetic,
  title={Synthetic biosensors for precise gene control and real-time monitoring of metabolites},
  author={Rogers, J. K. and Guzman, C. D. and Taylor, N. D. and Raman, S. and Anderson, K. and Church, G. M.},
  journal={Nucleic Acids Res.},
  volume={43},
  number={15},
  pages={7648--7660},
  year={2015},
  publisher={Oxford University Press}
}

\end{document}